\documentclass[12pt]{article}
\pdfoutput=1

\usepackage[pdftex]{graphicx}
\usepackage{amssymb}
\usepackage{amsmath}
\usepackage{cite}
\usepackage{multirow}

\usepackage{longtable}
\usepackage{booktabs}
\usepackage{array}

\usepackage{anyfontsize}

\usepackage{wrapfig}
\usepackage{color}

\definecolor{tit}{rgb}{0.1,0.2,0.4}

\newcommand{\C}[1]{{\cal C}_{#1}}

\definecolor{DRed}{rgb}{0.8,0,0.1}
\definecolor{DBlue}{rgb}{0,0,0.8}

\begin{document}

\begin{flushright}
{\small
LPT-ORSAY/16-40\\
QFET-2016-09\\
SI-HEP-2016-14 
}
\end{flushright}
$\ $
\vspace{-2mm}
\begin{center}
\fontsize{22}{26}\selectfont
\color{tit} \bf 
Assessing lepton-flavour non-universality\\[2mm]
from $B\to K^*\ell\ell$ angular analyses
\end{center}

\vspace{1mm}
\begin{center}
{\sf Bernat Capdevila}$^{a}$, {\sf S\'ebastien Descotes-Genon}$^b$, {\sf  Joaquim Matias}$^a$ {\sf and Javier Virto}$^c$\\[2mm]
{\em \small
$^a$ Universitat Aut\`onoma de Barcelona, 08193 Bellaterra, Barcelona,\\
Institut de Fisica d'Altes Energies (IFAE), The Barcelona Institute of Science and Technology, Campus UAB, 08193 Bellaterra (Barcelona) Spain
\\[3mm]
$^b$ Laboratoire de Physique Th\'eorique (UMR8627),\\
CNRS, Univ. Paris-Sud, Universit\'e Paris-Saclay,
91405 Orsay, France
\\[3mm]
$^c$ Theoretische Physik 1, Naturwissenschaftlich-Technische Fakult\"at,\\
Universit\"at Siegen, 57068 Siegen, Germany
}
\end{center}

\vspace{1mm}
\begin{abstract}\noindent
\vspace{-5mm}

The $B\to K^*\mu\mu$ decay exhibits deviations with respect to Standard Model expectations and the measurement of the
ratio $R_K$ hints at a violation of lepton-flavour universality in $B\to K\ell\ell$ transitions.
Both effects can be understood in model-independent fits as a short-distance contribution to the Wilson
coefficient $C_{9\mu}$, with some room for similar contributions in other Wilson coefficients for $b\to s\mu\mu$ transitions.
We discuss how a full angular analysis of $B\to K^*ee$ and its comparison with $B\to K^*\mu\mu$ could improve our understanding
of these anomalies and help confirming their interpretation in terms of short-distance New Physics. We discuss several
observables of interest in this context and provide predictions for them within the Standard Model as well as within several
New Physics benchmark scenarios.
We pay special attention to the sensitivity of these observables to hadronic uncertainties from SM contributions with charm loops.

\end{abstract}

\allowdisplaybreaks

\newpage
 
\section{Introduction}

In recent years, several deviations from the Standard Model (SM) have arisen in $B$-physics observables, with
the experimental confirmation of the anomaly~\cite{Descotes-Genon:2013wba} in the $B\to K^*\mu\mu$ observable
$P_5^\prime$~\cite{DescotesGenon:2012zf,Aaij:2013qta,Aaij:2015oid,Abdesselam:2016llu}, several tensions in
branching ratios for $b\to s\mu\mu$ transitions~\cite{Aaij:2013iag,Aaij:2015esa,Aaij:2014tfa} and evidence for
the violation of lepton flavour universality (LFU) in different observables ($R_K$, $R(D)$, $R(D^*)$)~\cite{
Aaij:2014ora,Lees:2013uzd,Huschle:2015rga,Aaij:2015yra,Abdesselam:2016yvr}~\footnote
{
The observable $P_2$~\cite{Becirevic:2011bp,Matias:2012xw} exhibited also a coherent deviation in the bin [2,4.3] with the 1 fb$^{-1}$
dataset~\cite{Aaij:2013qta}. Given the large experimental error in the 3 fb$^{-1}$ dataset in the bin [2.5,4]
due to $F_L^\text{exp}\simeq 1$ in that bin~\cite{Aaij:2015oid}, it was not possible to confirm nor
disprove this deviation. It would be very desirable to collect more data in that bin and in particular to
measure $F_L$ with a higher precision. In fact, a recent analysis by the Babar collaboration~\cite{Lees:2015ymt}
hints at a deviation in the same region.
}.
Global analyses of the deviations in $b\to s\ell\ell$ transitions
point towards a large additional contribution to the Wilson coefficient $C_{9\mu}$ of the semileptonic
operator in the effective Hamiltonian~\cite{Beneke:2001at} for $b\to s\mu\mu$, as initially discussed in
Ref.~\cite{Descotes-Genon:2013wba} and later confirmed by several works~\cite{
Altmannshofer:2013foa,Beaujean:2013soa,Altmannshofer:2014rta,Descotes-Genon:2015uva,Altmannshofer:2015sma,
Hurth:2016fbr}. Even though such a contribution to $C_{9\mu}$ in $b\to s\mu\mu$ appears as a rather economical
way of explaining a large set of deviations with respect to SM expectations, theory predictions for some
$b\to s\mu\mu$ observables may also get a better agreement with data once additional contributions are allowed in
other Wilson coefficients  (such as $C_{9'\mu}$ or $C_{10\mu}$)~\cite{Descotes-Genon:2015uva}. On the other hand,
$B\to K^*ee$ observables and the $R_K$ ratio suggest that $b\to see$ transitions agree well with the
SM~\cite{Aaij:2015dea}, pointing to explanations with New Physics (NP) models with a maximal violation of LFU,
affecting only muon and not electron modes.
These hints of lepton flavour non-universality (LFNU) have triggered a lot of theoretical activity~\cite{1408.1627,1408.4097,1411.4773,1412.7164,1503.09024,
1503.06077,1505.03079,1505.05164,1506.01705,1506.02661,1508.07009,1511.01900,1511.06024,Boucenna:2016wpr,1604.03940}.

As  discussed in several works~\cite{Khodjamirian:2010vf,Beylich:2011aq,0404250,Khodjamirian:2012rm,Lyon:2014hpa,
Ciuchini:2015qxb}, long-distance SM contributions from diagrams involving charm loops enter the computation
of $b\to s\ell\ell$ processes, acting as additional contributions to the Wilson coefficient $C_9$. 
These contributions are process-dependent and they must be estimated through different theoretical methods
according to the dilepton invariant mass $q^2$. The latest estimates of these contributions~\cite{Beylich:2011aq,
Khodjamirian:2010vf} have been included in the global fits for $B\to K^*\mu\mu$~\cite{Descotes-Genon:2015uva,
Altmannshofer:2015sma,Hurth:2016fbr}, providing the consistent picture described above.
In particular, bin-by-bin fits indicate that the data agrees well with a single, process-independent contribution
to $C_{9\mu}$, independent of the dimuon invariant mass, and present only in muon modes,
as expected from a short-distance (NP) flavour-non-universal contribution.
In order to confirm this pattern, it would be very desirable to design observables probing:

\begin{itemize}
\item[\emph{a)}] only the short-distance part of $C_{9\ell}$,
\item[\emph{b)}] other Wilson coefficients, such as $C_{10\ell}$, which do not receive long-distance contributions
from the SM,
\item[\emph{c)}] the amount of lepton-flavour non-universality between electron and muon modes.
\end{itemize}
In all cases, hadronic uncertainties should remain controlled: while non-universality is a smoking-gun-signal
of NP (the SM predictions being very precise), the \emph{measurement} of the effect is affected by the
same hadronic uncertainties as the individual $b\to s\ell\ell$ modes.

The purpose of this article is to investigate which observables can be built that match these criteria,
once a full angular analysis of $B\to K^*ee$, with an accuracy comparable to that of $B\to K^*\mu\mu$,
is available. If the most obvious quantity consists in comparing branching ratios though the ratio $R_{K^*}$
(similar to $R_K$) (see Ref.~\cite{Descotes-Genon:2015uva} for predictions for these ratios for different NP
scenarios), it is also interesting to consider other ratios probing the violation of LFU using the angular
coefficients $J_i$ describing the whole angular kinematics of these decays. In this note, we will discuss
observables that can measure LFNU in $B \to K^*\ell\ell$.
Some of them are variations around the basis of optimised observables introduced in
Refs.~\cite{Matias:2012xw,DescotesGenon:2012zf} and others can be built directly by
combining angular coefficients from muon and electron modes. We will discuss the advantages of these observables
in the context of hadronic uncertainties, and provide predictions in the SM and in several benchmark scenarios
corresponding to the best-fit points obtained in our recent global analysis of $b\to s\ell\ell$
modes~\cite{Descotes-Genon:2015uva}.

We begin with a presentation of the observables of interest in Section~\ref{sec:2}.
In addition to observables naturally derived from
the angular coefficients $J_i$ and the optimised observables~$P^{(\prime)}_i$, we consider other observables,
namely $B_i$ and $M$ (and $\widetilde{B}_i$, $\widetilde M$) which have a reduced sensitivity to charm
contributions in some NP scenarios.
In Section~3 we present our predictions in the SM and in several NP benchmark points,
illustrating how these observables can help in discerning among NP scenarios and how (in)sensitive they
are with respect to hadronic uncertainties.
We present our conclusions in Section~4.
In the appendices we discuss the dependence of $M$ and $\widetilde{M}$ observables on charm contributions,
we recall the definition of binned observables, and we provide further predictions for the various observables
within the different benchmark scenarios.

\section{$B\to K^*\ell\ell$ observables assessing lepton flavour universality}
\label{sec:2}

\subsection{Observables derived from $J_i$, $P_i$ and $S_i$}

We want to exploit the angular analyses of both $B\to K^*\mu\mu$ and $B\to K^*ee$ decays in order to build
observables that will probe the violation of LFU, the short-distance part of $C_{9\mu}$ and/or the other Wilson
coefficients, with limited hadronic uncertainties.
Natural combinations are~\footnote{In the following, we always consider quantities obtained by combining CP-averaged
angular coefficients.}
\begin{equation} \label{obs}
Q_{F_L}=F_L^{\mu}-F_L^{e}\,,\quad
Q_i=P_i^{\mu}-P_i^{e}\,,\quad
T_i=\frac{S_i^{\mu}-S_i^{e}}{S_i^{\mu}+S_i^{e}}\,,\quad
B_i=\frac{J_i^{\mu}}{J_i^{e}}-1\,,\quad
\widetilde{B}_i=\frac{\beta_e^2}{\beta_\mu^2}\frac{J_i^{\mu}}{J_k^{e}}-1\,,
\end{equation}
where $P_i$ should be replaced by $P'_i$ for $Q_{i=4,5,6,8}$. $B_i$ and $\widetilde{B}_i$ differ mostly at
very low $q^2$ and become almost identical for large $q^2$, where $\beta_\ell=\sqrt{1-4m_\ell^2/q^2}\simeq 1$ for both
electrons and muons. The optimised observables $P^{(\prime)}_i$ have already a limited sensitivity to hadronic
uncertainties~\cite{Matias:2012xw,DescotesGenon:2012zf,Descotes-Genon:2013vna,Descotes-Genon:2014uoa,Descotes-Genon:2015uva}, contrary to the
angular averages $S_i$~\cite{Altmannshofer:2008dz,Matias:2012xw,Jager:2012uw,DescotesGenon:2012zf,
Jager:2014rwa,Descotes-Genon:2014uoa}. We thus expect the $Q_i$ observables to exhibit a correspondingly low
sensitivity to hadronic uncertainties.~\footnote{
We also expect a reduced sensitivity to $K\pi$ S-wave contributions
(see \emph{e.g.} \cite{Becirevic:2012dp,Matias:2012qz,Blake:2012mb}).
}
Moreover, these observables are protected from long-distance charm-loop
contributions in the SM.

A measurement of $Q_i$ different from zero would  point to NP  in an unambiguous way, confirming the violation
of LFU observed in $R_K$. A second step  would then consist in identifying the pattern of NP, which requires to
separate the residual hadronic uncertainties (in particular, charm-loop contributions) from the NP contributions.
The set of observables $Q_i$, $T_i$ and $B_k$ ($\widetilde{B}_k$) can be particularly instrumental at this
second stage, with a sensitivity to the various Wilson coefficients depending on the particular angular
coefficients considered.

We have already investigated this sensitivity~\cite{Descotes-Genon:2013vna,Descotes-Genon:2014uoa,
Descotes-Genon:2015uva}, but we would like to highlight the difference of behaviour in the case of two of the
relevant observables $P_4'$ and $P_5'$, directly related to $Q_4$ and $Q_5$ respectively. Both LHCb and Belle
collaborations~\cite{Aaij:2013qta,Aaij:2014tfa,Abdesselam:2016llu} observed the same pattern, i.e., a
significant deviation from the SM for $P_5^\prime$ for $q^2$ between 4 and 8 GeV$^2$ and a result consistent
with the SM within errors for $P_4^\prime$. This behaviour is expected in the presence of NP in  the Wilson
coefficient $C_9$. From the large-recoil expressions of $A_{\perp,_\|,0}^{L,R}$ (see Eqs.~(3.8)-(3.10) of
Ref.~\cite{Kruger:2005ep}) one finds that the right-handed amplitudes
$|A^R_{0,\perp,||}| \propto (C_9^{\rm eff}+C_{10}) + ...$ are suppressed compared to the left-handed ones
in the SM, due to the approximated cancellation  $C_9^{\rm eff}+C_{10} \simeq 0$.
This cancellation is not so effective in the presence of a negative NP contribution to $C_9$,
and $A_{0,\|}^R$, $|A_\perp^R|$  increase while $|A_{0,\|}^L|$, $A_\perp^L$  decrease. Both effects add up
coherently in the numerator of $P_5^\prime \propto {\rm Re}(A_0^L A_{\perp}^{L*} - A_{0}^R A_{\perp}^{R*})$
due to the relative minus sign, and the effect is to reduce the value of $|P_5'|$ in the region far up from the
photon pole, in agreement with the experimental observation.
In $P_4^\prime \propto \text{Re}(A_0^L A_{\|}^{L*} + A_{0}^R A_{\|}^{R*})$, however,
an increase in the right-handed amplitudes will compensate a decrease in the left-handed ones,
due to the relative positive sign.
For this reason, no deviation is expected in $P_4^\prime$ in the presence of NP in $C_9$
(but in the absence of right-handed currents).
The same mechanism is at work for $Q_4$ and $Q_5$.

As discussed in Sec~2.3.1 of Ref.~\cite{Descotes-Genon:2015uva}, LHCb currently determines the polarisation
fraction $F_T$ and $F_L$ using a simplified description of the angular kinematics. This means that these two
quantities are actually measured from $J_{1c}$ rather than $J_{2s}$ and $J_{2c}$ respectively.
Both determinations are equivalent in the massless limit, and therefore this only has a limited impact, apart
from the first bin [0.1,0.98].
In order to interpret the actual measurements more precisely, we define the $\hat{P}_i$ observables involving
$\hat{F}_T$ and $\hat{F}_L$, as measured currently by LHC:
\begin{eqnarray}\label{eq:hatstart}
F_L =\frac{-J_{2c}}{d\Gamma/dq^2}\to  \hat{F}_L=\frac{J_{1c}}{d\Gamma/dq^2}&\qquad&
F_T =\frac{4J_{2s}}{d\Gamma/dq^2}\to  \hat{F}_T=1-\hat{F}_L\\
P_1=\frac{J_3}{2J_{2s}} \to  \hat{P}_1= \frac{J_3}{2\hat{J}_{2s}} &\qquad&
P_2=\frac{J_{6s}}{8J_{2s}}  \to \hat{P}_2=\frac{J_{6s}}{8\hat{J}_{2s}}\\
P_3=-\frac{J_{9}}{4J_{2s}} \to \hat{P}_3=-\frac{J_{9}}{4\hat{J}_{2s}}&\qquad&
P_4'=\frac{J_{4}}{\sqrt{-J_{2s}J_{2c}}}\to \hat{P}_4'=\frac{J_{4}}{\sqrt{\hat{J}_{2s}J_{1c}}}\\
P_5'= \frac{J_{5}}{2\sqrt{-J_{2s}J_{2c}}}\to \hat{P}_5'=\frac{J_{5}}{2\sqrt{\hat{J}_{2s}J_{1c}}}&\qquad&
P_6'=- \frac{J_{7}}{2\sqrt{-J_{2s}J_{2c}}}\to \hat{P}_6'=- \frac{J_{7}}{2\sqrt{\hat{J}_{2s}J_{1c}}}\\
P_8'=- \frac{J_{8}}{\sqrt{-J_{2s}J_{2c}}}\to \hat{P}_8'=- \frac{J_{8}}{\sqrt{\hat{J}_{2s}J_{1c}}}
&\qquad& {\rm with}\ \hat{J}_{2s}=\frac{1}{16}(6J_{1s} - J_{1c} - 2J_{2s} - J_{2c})\label{eq:hatend}
\end{eqnarray}
and we will provide predictions for both $Q_i$ and $\hat{Q}_i$ observables, in order to illustrate the
differences in the first bin, as well as the insensitivity of the effect in higher bins.

In the case of the $S_i$, the consideration of the $T_i$ ratio is also natural, but unfortunately these
quantities are quite sensitive to hadronic uncertainties. They depend on soft form factors even in the large
recoil limit  due to lepton mass effects at very low $q^2$, related to differences between muon and electron
contributions in the normalization. 
Finally, the ratios $B_i$ that are soft-form-factor independent at leading order in the large-recoil limit
will be shown to complement the observables $Q_i$ in an interesting way.

\subsection{Observables with reduced sensitivity to charm effects}\label{sec:obsBM}

In the presence of NP, all observables $Q_i, T_i$ and $B_i$ are in principle affected by long-distance charm loop
contributions in $C_9$, both transversity-independent and transversity-dependent.
We define these two terms in the following way:
transversity-independent long-distance charm corresponds to an identical contribution to all $B\to K^*\ell\ell$
transversity amplitudes, whereas transversity-dependent contributions differ for each amplitude.
Both of them are expected to exhibit a $q^2$-dependence in general. The explicit computation of charm-loop
contributions performed in Ref.~\cite{Khodjamirian:2010vf} using light-cone sum rules indicates that they are
transversity-dependent, in agreement with general expectations that such hadronic contributions are different
for different external hadronic states (including different $K^*$ helicities).
It is interesting to investigate these issues by considering specific observables
with different sensitivity to transversity-dependent and independent long-distance charm
contributions, as well as to LFNU New Physics.

One can think of exploiting the angular coefficients in electron and muon modes in order to build observables
only sensitive to some of the Wilson coefficients, and in some cases, insensitive to transversity-independent
long-distance charm contributions. It is easy to check that in the large-recoil limit and in the absence of
right-handed or scalar operators, four angular coefficients exhibit a linear sensitivity to $C_9$. Taking the
results from Refs.~\cite{Kruger:2005ep,Matias:2012xw} we have:
\begin{eqnarray}
\beta_\ell J_{6s}-2iJ_9 &=& 
    16\beta_\ell^2 N^2m_B^2(1-\hat{s})^2 C_{10}^\ell\left[2C_7\frac{\hat{m}_b}{\hat{s}}+C_9^\ell\right ]\xi_\perp^2+\ldots
\label{eq:largerec}\\
\beta_\ell J_{5}-2iJ_8 &=& 
8\beta_\ell^2 N^2m_B^2(1-\hat{s})^3 \frac{\hat{m}_{K^*}}{\sqrt{\hat{s}}}C_{10}^\ell\left[C_7\hat{m}_b\left(\frac{1}{\hat{s}}+1\right)+C_9^\ell\right]\xi_\perp\xi_{||}+\ldots
\label{eq:largerec2}
\end{eqnarray}
where $\hat{s}=q^2/m_B^2$ and $\hat{m}_b=m_b/m_B$, $\xi_\perp$ and $\xi_{||}$ correspond to the soft form
factors~\cite{Beneke:2001at}, and the ellipses indicate terms suppressed in the large-recoil limit (including terms of order
$m_\ell^2/q^2$).
If we limit ourselves to real NP contributions, it is interesting to  consider $B_5$ and
$B_{6s}$ (and $\widetilde{B}_5$ and $\widetilde{B}_{6s}$) in Eq.~(\ref{obs}), as well as a combination of them
in the form~\footnote{The definitions of $B_{5,6s}$ ($\widetilde{B}_{5,6s}$) and $M$ ($\widetilde{M}$) could be
adapted to the imaginary contributions $J_{8,9}$. However the latter vanish in the case of real NP contributions.
Since current data does not indicate any need for complex NP contributions, we will not include these
additional observables here.}
\begin{equation}
M=\frac{(J_5^{\mu} - J_5^{e}) (J_{6s}^{\mu} - J_{6s}^{e})}{J_{6s}^{\mu} J_5^{e} - J_{6s}^{e} J_5^{\mu}},
\quad {\widetilde M}=\frac{(\beta_e^2 J_5^{\mu} - \beta_\mu^2 J_5^{e}) (\beta_e^2 J_{6s}^{\mu} -
\beta_\mu^2 J_{6s}^{e})}{\beta_e^2 \beta_\mu^2 (J_{6s}^{\mu} J_5^{e} - J_{6s}^{e} J_5^{\mu})}\,.
\end{equation}
By construction, $B_5$ and $\widetilde{B}_5$ have a pole at the position of the zero of $J_5^{e}$ in the SM
(around $q^2 = 2~\text{GeV}^2$) and $B_{6s}$, $\widetilde{B}_{6s}$ have a pole at the position of the zero
of $A_{\rm FB}$ in the SM (around $q^2 = 4~\text{GeV}^2$). We  expect large
uncertainties for these observables in the corresponding bins. On the contrary, $M$ is well behaved in the same
bins, but it will have large uncertainties when $B_5 \simeq B_{6s}$. In this sense, the observable $M$ is well
suited for NP scenarios and energy regions that yield very different contributions to $B_5$ and $B_{6s}$.
While the $B_i$ have a value in the SM slightly different from zero (specially the first bin) due to
$\beta_\mu/\beta_e$ kinematic effects, the ${\widetilde B}_i$ observables vanish by construction in the
SM.~\footnote{
The measurement of ${\widetilde B}_i$  requires the measurement of the quantities
$\langle J_i^{\ell}/\beta_{\ell}^2 \rangle$.
Experimentally, this can be done by assigning a $\beta_{\ell}^2$ factor to the data on an event-by-event
basis~\cite{nico}.
}

Even more interesting is the case of $\widetilde{M}$, constructed in the same spirit as $\widetilde B_i$,
\emph{i.e.} to cancel the dependence of the angular coefficients on $\beta_\ell$.
Its first bin can be accurately predicted even in the
presence of NP, while its $M$ counterpart suffers from large uncertainties in that bin.
In the next section we will discuss some NP scenarios and show how these set of observables can become
instrumental to disentangle them.

Let us write $C_{ie}=C_i$ and $C_{i\mu}=C_i+\delta C_i$ for $i\neq 9$, so that $\delta C_i$ measure the LFU
violation, whereas $C_{ie}$ can include LFU NP effects. Furthermore,  for $i=9$ we take
$C_{9e}=C_9+ \Delta C_9$ and $C_{9\mu}=C_9 + \delta C_9+ \Delta C_9$ where  $\Delta C_9$ is a long-distance
charm contribution. In order to illustrate the relevant aspects of the various observables,
within this Section we will give analytic formulas assuming the contribution $\Delta C_9$ is transversity
independent and neglecting imaginary parts. But all our numerical evaluations will be based on
complete expressions, as computed in Ref.~\cite{Descotes-Genon:2015uva} where transversity-dependent charm
contributions are included following Ref.~\cite{Khodjamirian:2010vf}, and imaginary parts are properly
accounted for.
We see that $\delta C_{7,7'}=0$~\footnote{$C_7$ includes both the SM $C_7^{\rm eff}$ plus possible LFU NP
(the same applies to $C_9$).} and $\delta C_9$ are directly related to short-distance physics,
while $\Delta C_9$ comes from long-distance contributions from $c\bar{c}$ loops where the lepton pair
is created by an electromagnetic current, and thus identical for $C_{9e}$ and $C_{9\mu}$.
Any $\delta C_i\neq 0$ indicates the presence of LFNU New Physics.

In the large-recoil limit and in the absence of right-handed or scalar operators, we have:
\begin{eqnarray}
B_5&=&\frac{\beta_\mu^2-\beta_e^2}{\beta_e^2}+ \frac{\beta_\mu^2}{\beta_e^2} \frac{\delta C_{10}}{C_{10}}
+\frac{\beta_\mu^2}{\beta_e^2} \frac{(C_{10} + \delta C_{10})
\delta C_{9}\hat{s}}{C_{10} (C_{7} \hat{m}_b (1+\hat{s})+ ( C_{9}+ \Delta C_9) \hat{s})}+\ldots
\label{obsi1}\\
B_{6s}&=&\frac{\beta_\mu^2-\beta_e^2}{\beta_e^2}+\frac{\beta_\mu^2}{\beta_e^2}\frac{\delta C_{10}}{C_{10} }
+ \frac{\beta_\mu^2}{\beta_e^2} \frac{(C_{10} + \delta C_{10}) \delta C_{9} \hat{s}}
{C_{10} (2 C_{7} \hat{m}_b + ( C_{9}+ \Delta C_9) \hat{s})}+\ldots
\label{obsi2} \\
M&=&{\widetilde M}+\Delta M+ \mathcal{A}\Delta C_9+\mathcal{B} \Delta C_9^2+\ldots \label{obsM} \\
{\widetilde M}&=&{\widetilde M}_0 + {\cal A}^\prime \delta C_{10}  \Delta C_9 + {\cal B}^\prime
\delta C_{10}^2  \Delta C_{9}^2 + \ldots  
\end{eqnarray}
where $\widetilde{M}_0$, $\Delta M$, ${\cal A}^{(\prime)}$ and ${\cal B}^{(\prime)}$ are defined in
App.~\ref{sec:mobservable}, and the ellipsis denote again terms neglected in Eqs.~(\ref{eq:largerec}) and~(\ref{eq:largerec2}) and suppressed in the large-recoil limit.
The difference between the muon and electron masses relative to $q^2$, induces a non-vanishing SM value
for the $B_i$ observables at low $q^2$.
$\widetilde B_i$ are exactly zero in the SM, and can be obtained from Eqs.~(\ref{obsi1}),~(\ref{obsi2}) in the limit
$\beta_\ell \to 1$. Note that the $B_i$ observables always have a residual charm dependence
$\Delta C_9$ in the denominator in the presence of NP.

From Eq.~(\ref{obsM}), $M$ appears sensitive to the muon-electron mass difference via $\Delta M$, ${\cal A}$
and ${\cal B}$, and the last two terms introduce a sensitivity to charm effects through $\Delta C_9$.
Moreover, the first bin of $M$ is very sensitive to this mass difference and will be affected by very large
uncertainties in some NP scenarios.
On the contrary, $\widetilde{M}$ is blind to such mass effects. In addition, if there is no NP in
$\delta C_{10}$ then $\widetilde {M}$ becomes also insensitive to transversity-independent charm effects at
leading order and at large recoil. This means that $\widetilde{M}$ is particularly clean  at low $q^2$
(where large-recoil expressions are relevant), especially in the presence of NP in $\delta C_9$. For larger
values of $q^2$ and/or in the presence of NP in $C_{10}$, subleading charm effects are present and will enlarge
the uncertainties, even though the impact of NP on this observable remains very large. $\widetilde{M}$ at low
$q^2$ will turn out to be very efficient to disentangle NP scenarios.

We have the following behaviour for $\delta C_9=0$:
\begin{equation}
B_5=B_{6s}= \frac{\beta_\mu^2-\beta_e^2}{\beta_e^2}+\frac{\beta_\mu^2}{\beta_e^2}\frac{\delta C_{10}}{C_{10} }\ .  
\end{equation}
For $B_5$ and $B_{6s}$, the limit of very small $q^2$ is equivalent to $\delta C_9=0$, and $M$ is not well
predicted in this limit (subleading effects dominate the computation). This is however not a problem
in the current context where global analyses point towards a large NP contribution to $C_9$.
On the other hand, if $\delta C_{10}=0$, we have~\footnote{
The corresponding expressions for $\widetilde{B}_{5,6s}$ when $\delta C_{10}=0$ can be easily obtained from Eqs.(\ref{limiteq1})-(\ref{limiteq2}) by taking $\beta\to 1$  and the one of $M$ can be obtained from App.\ref{sec:mobservable}.}
\begin{eqnarray} \label{limiteq1}
B_5&=&\frac{\beta_\mu^2-\beta_e^2}{\beta_e^2} +\frac{\beta_\mu^2}{\beta_e^2} \frac{ \delta C_{9} \hat{s}}{ (C_{7} \hat{m}_b (1+\hat{s})+ ( C_{9}+ \Delta C_9) \hat{s})}+\ldots \\ \label{limiteq2}
B_{6s}&=&\frac{\beta_\mu^2-\beta_e^2}{\beta_e^2}+ \frac{\beta_\mu^2}{\beta_e^2} \frac{ \delta C_{9} \hat{s}}{ (2 C_{7} \hat{m}_b +  (C_{9}+ \Delta C_9) \hat{s})}+\ldots \\
\widetilde{M}&=&-\frac{\delta C_9 \hat{s}}{C_7 \hat{m}_b (1-\hat{s})}+\ldots
\end{eqnarray}
$B_5$ and $B_{6s}$ contain then a residual charm sensitivity through $\Delta C_9$,
while $\widetilde {M}$ is totally free from this transversity-independent long-distance charm at leading order.
This is a very specific property of $\widetilde{M}$ which is independent of transversity-independent charm
contributions in the presence of New Physics in $C_9$ only. Transversity-dependent charm effects are
kinematically suppressed at very low $q^2$ in these observables as it will be shown later on.

In the case where both $\delta C_9$ and $\delta C_{10}$ are non-zero, a precise interpretation of these
observables requires a more detailed study (including an assessment of all $c\bar{c}$ contributions to $C_9$). 
We see therefore that some of these observables will have a limited sensitivity to charm-loop contributions
in some cases (SM, NP only in $C_{9\mu}$), but not in other cases (NP also in $C_{10,\mu}$ for instance).

As a conclusion, the behaviour of $B_5$ ($\widetilde{B}_5$), $B_{6s}$ ($\widetilde{B}_{6s}$) and $M$ ($\widetilde{M}$) in specific $q^2$-regions should provide  powerful tests of physics beyond the SM, with a limited sensitivity to hadronic uncertainties.

\section{Predictions in the SM and in typical NP benchmark scenarios}

\subsection{Observables and scenarios}

The above discussion assumed that one can determine exactly the value of the angular coefficients $J_i$ differentially in $q^2$. This is in principle possible using the method of amplitudes in Ref.~\cite{Egede:2015kha} even if for electrons it could be particularly difficult. The other methods (likelihood fit and method of moments) lead to binned observables, where the cancellations advocated above hold only in an approximate way, for bins small enough so that the angular coefficients do not exhibit steep variations. The modifications due to binning for the predictions of observables were described in detail in Ref.~\cite{Descotes-Genon:2013vna}, and are also recalled in App.~B for the observables described above. They will obviously have an impact on the previous discussion concerning the cancellation of hadronic uncertainties, which will then be only approximate.

\begin{figure}
 \begin{center}
  \includegraphics[scale=0.28]{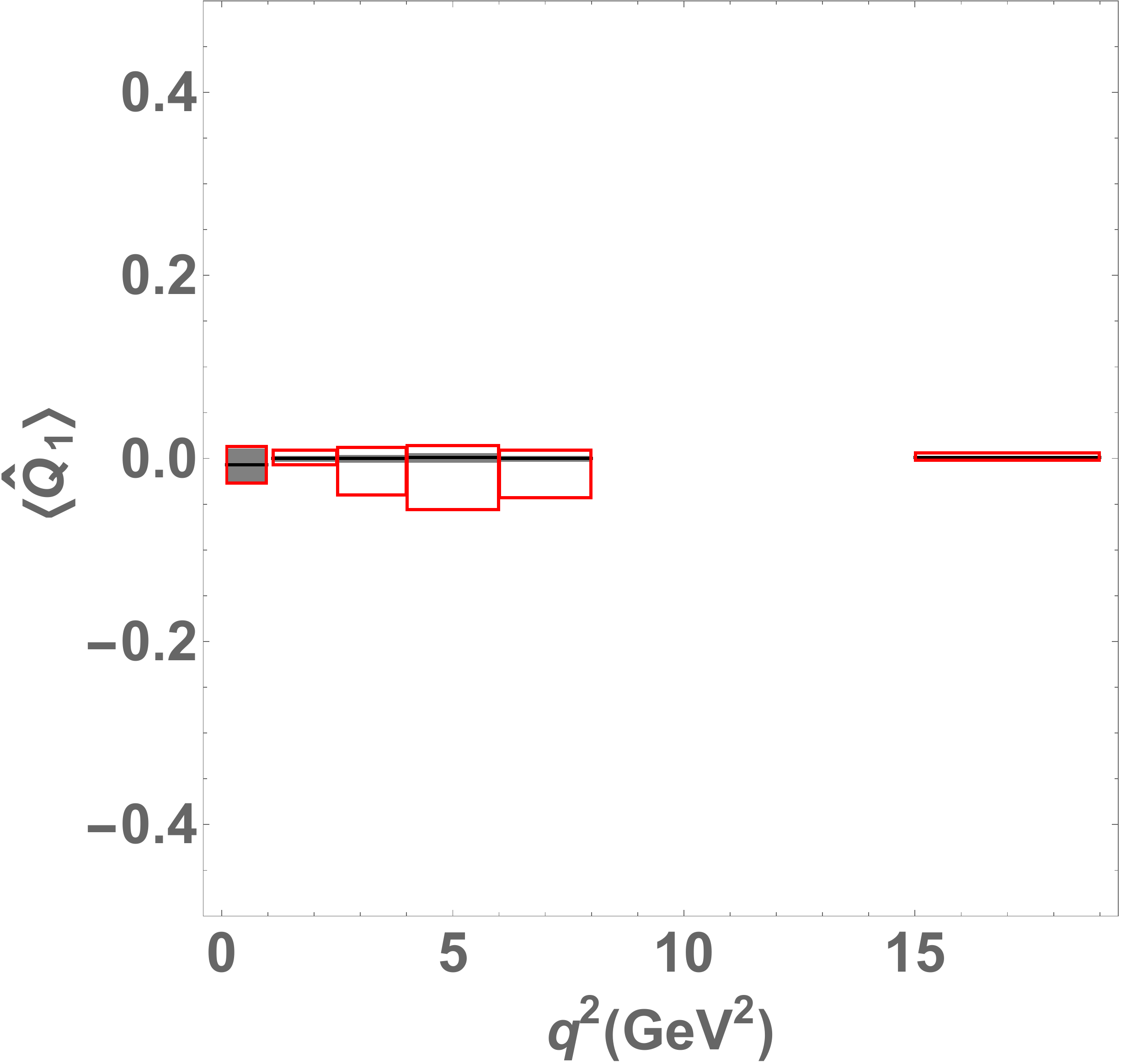}\hspace{7mm}
    \includegraphics[scale=0.28]{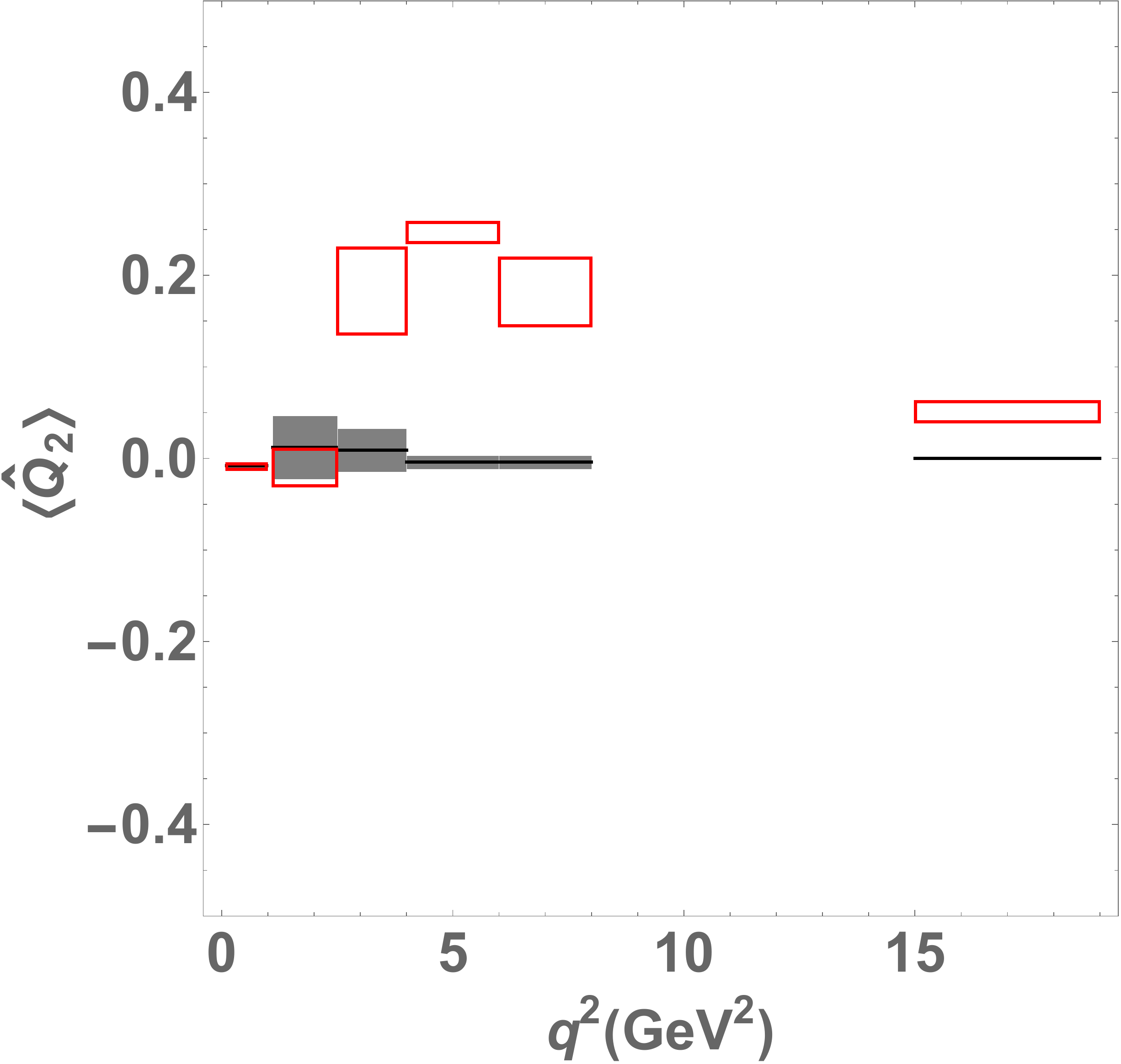}\bigskip
    
   \includegraphics[scale=0.28]{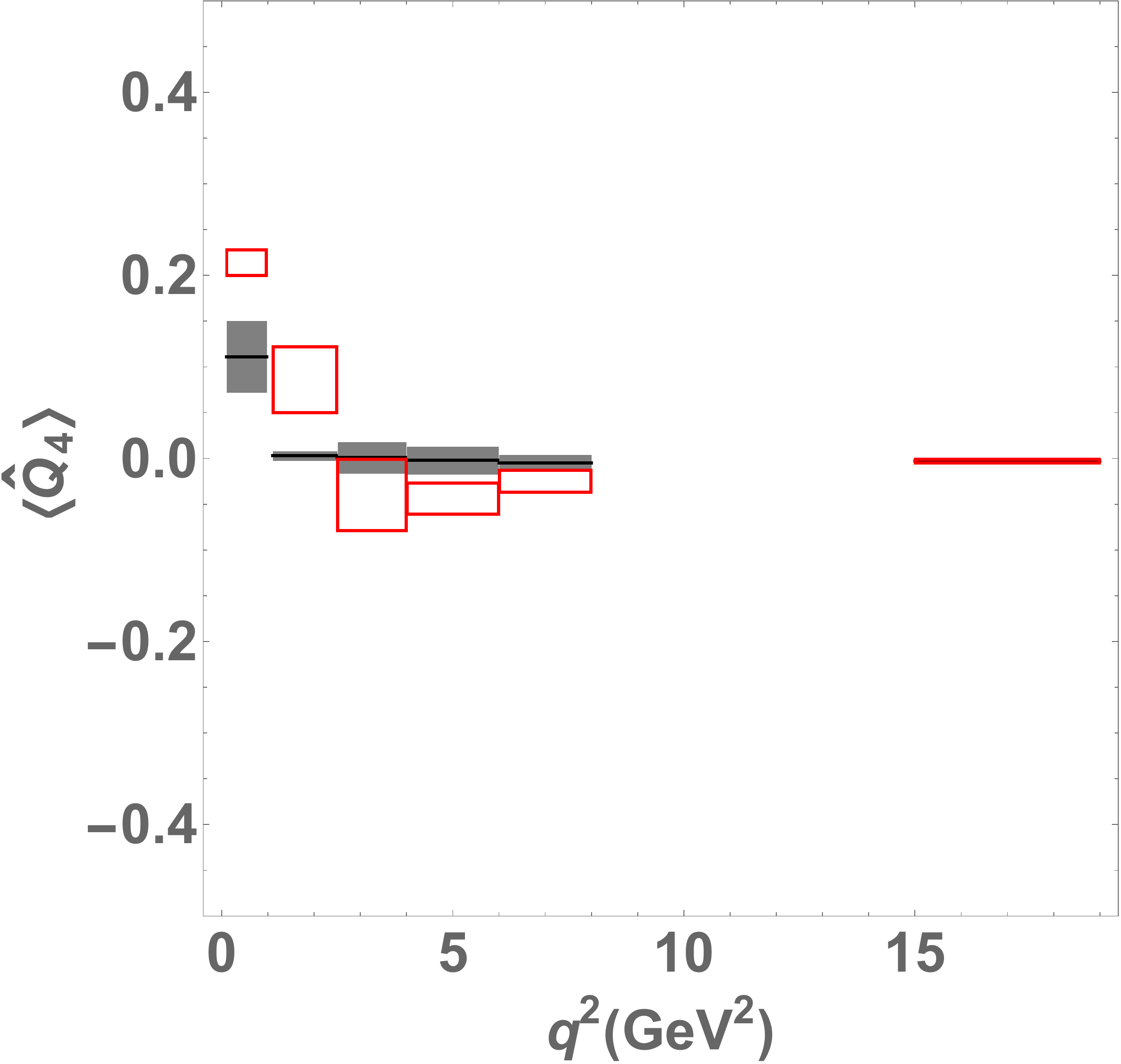}\hspace{7mm}
    \includegraphics[scale=0.28]{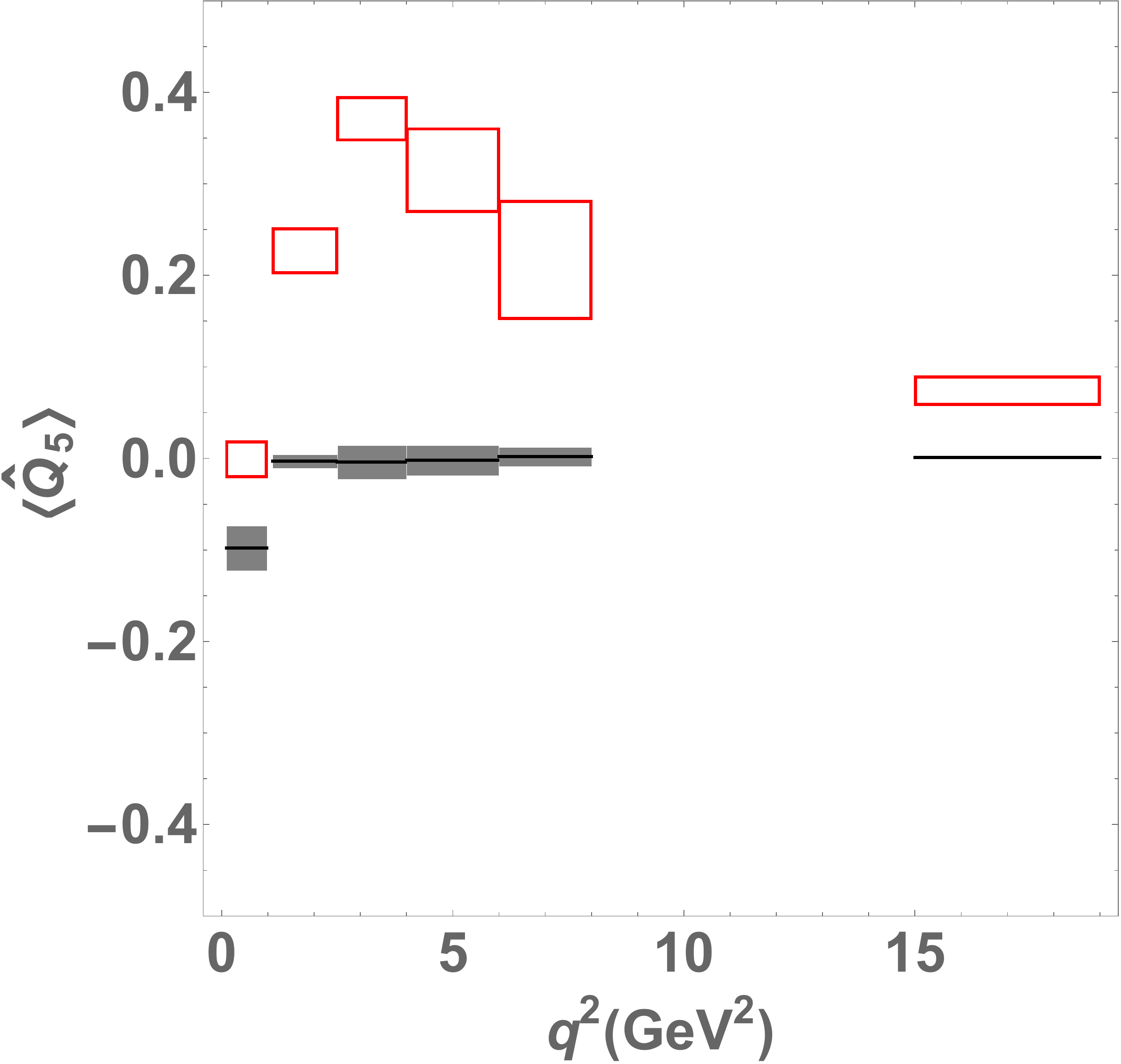}  \bigskip 
       \includegraphics[scale=0.28]{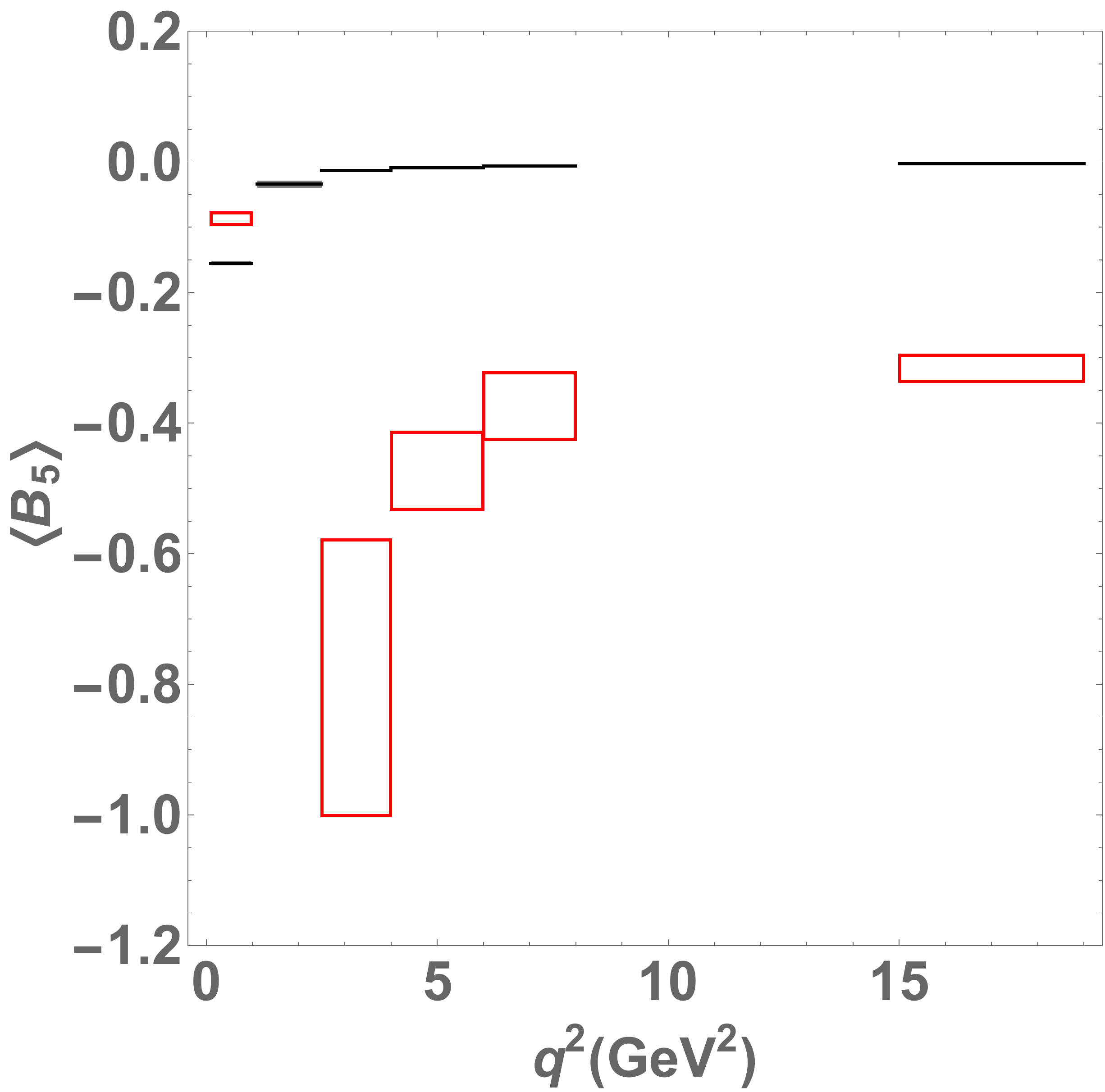}\hspace{7mm}
    \includegraphics[scale=0.28]{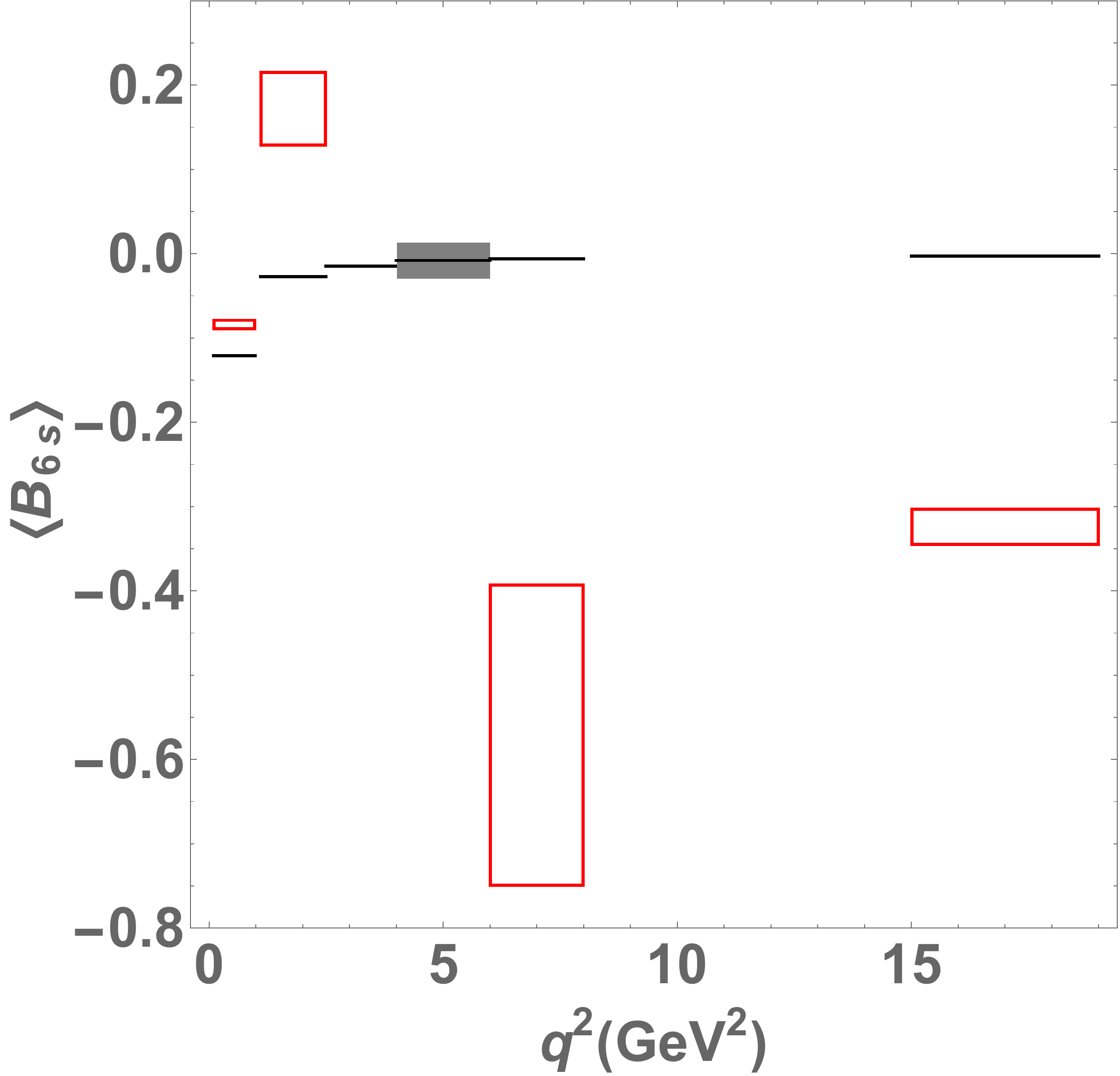}     
       \vspace{0.0cm}
  \caption{\textit{Scenario 1}.  SM predictions (grey boxes) and NP predictions (red boxes), assuming $C_{9\mu}^{\rm NP}=-1.11$.}
  \label{fig:scenario1}
 \end{center}
\end{figure}

\begin{figure}
 \begin{center}
        \includegraphics[scale=0.28]{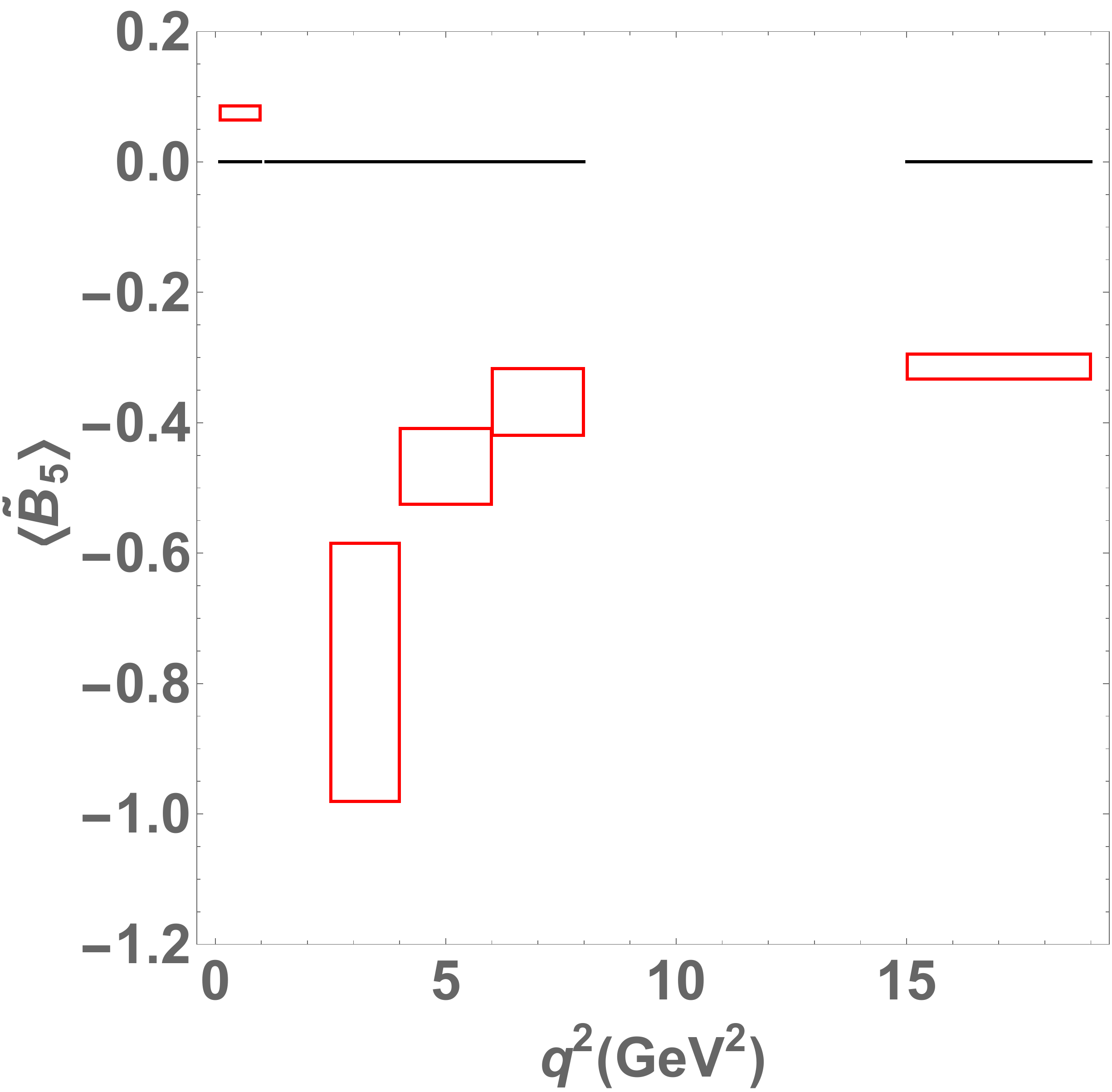}\hspace{7mm}
    \includegraphics[scale=0.28]{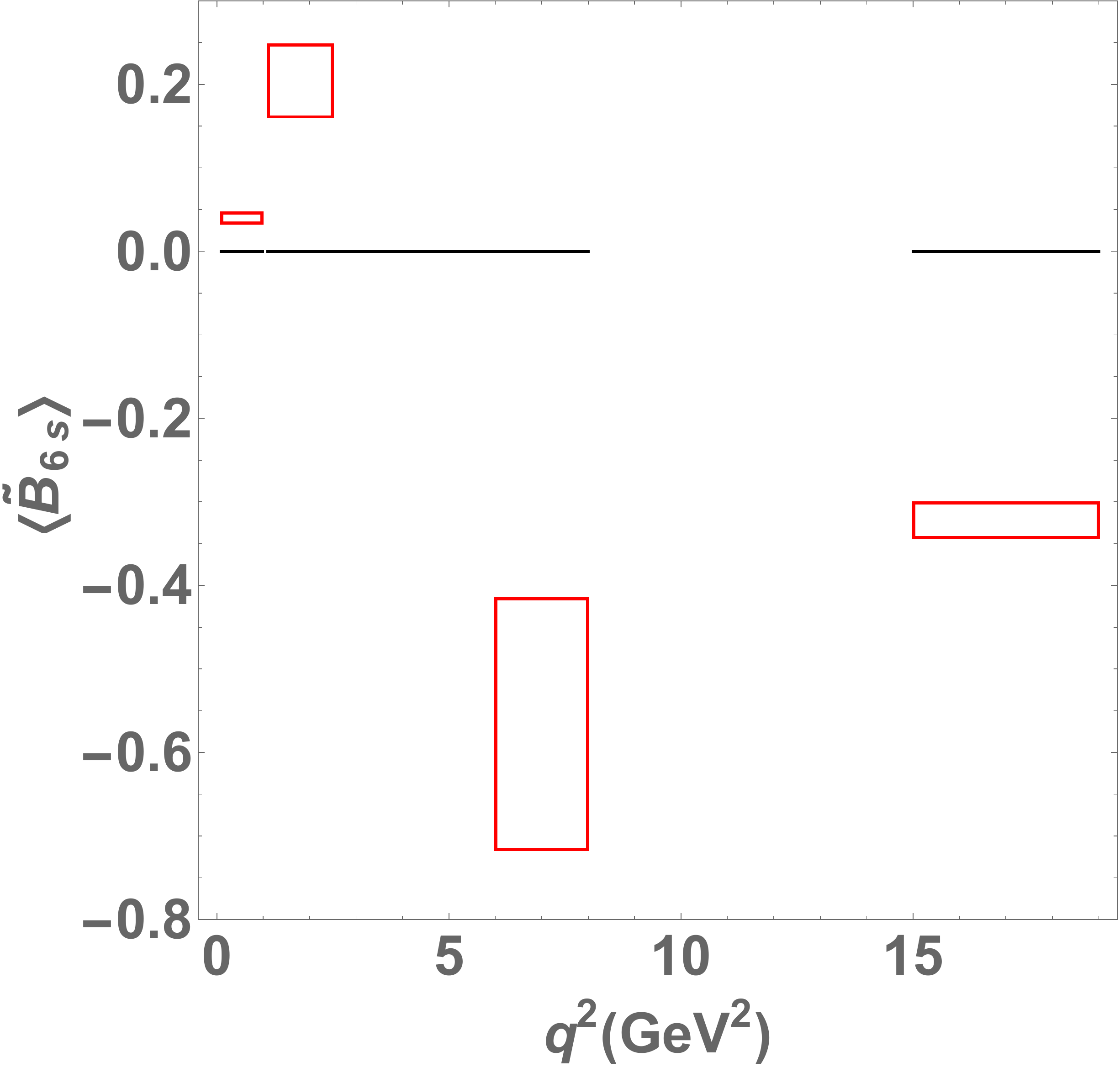} \\[8mm]
  \includegraphics[scale=0.28]{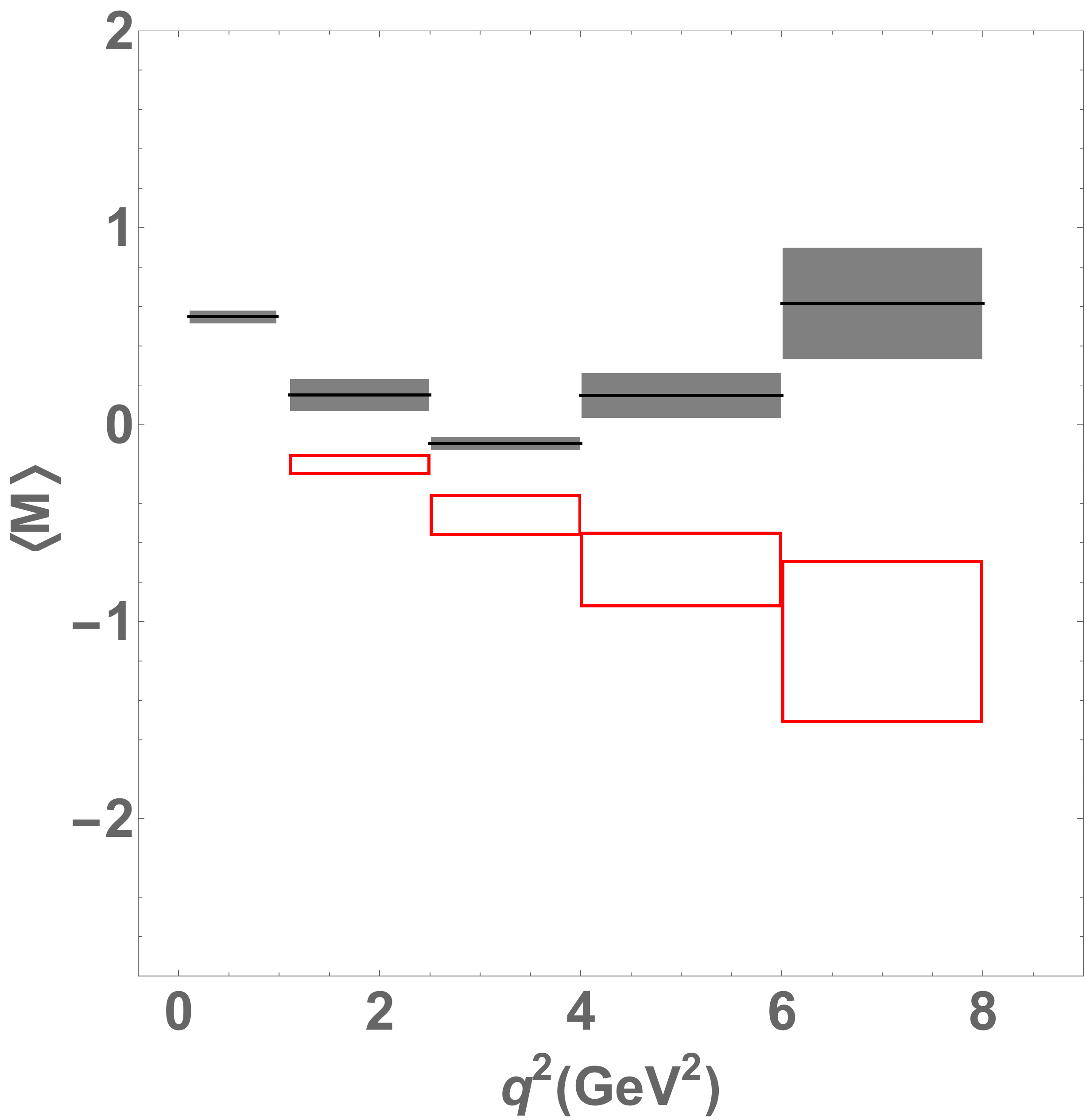}\hspace{7mm}
    \includegraphics[scale=0.29]{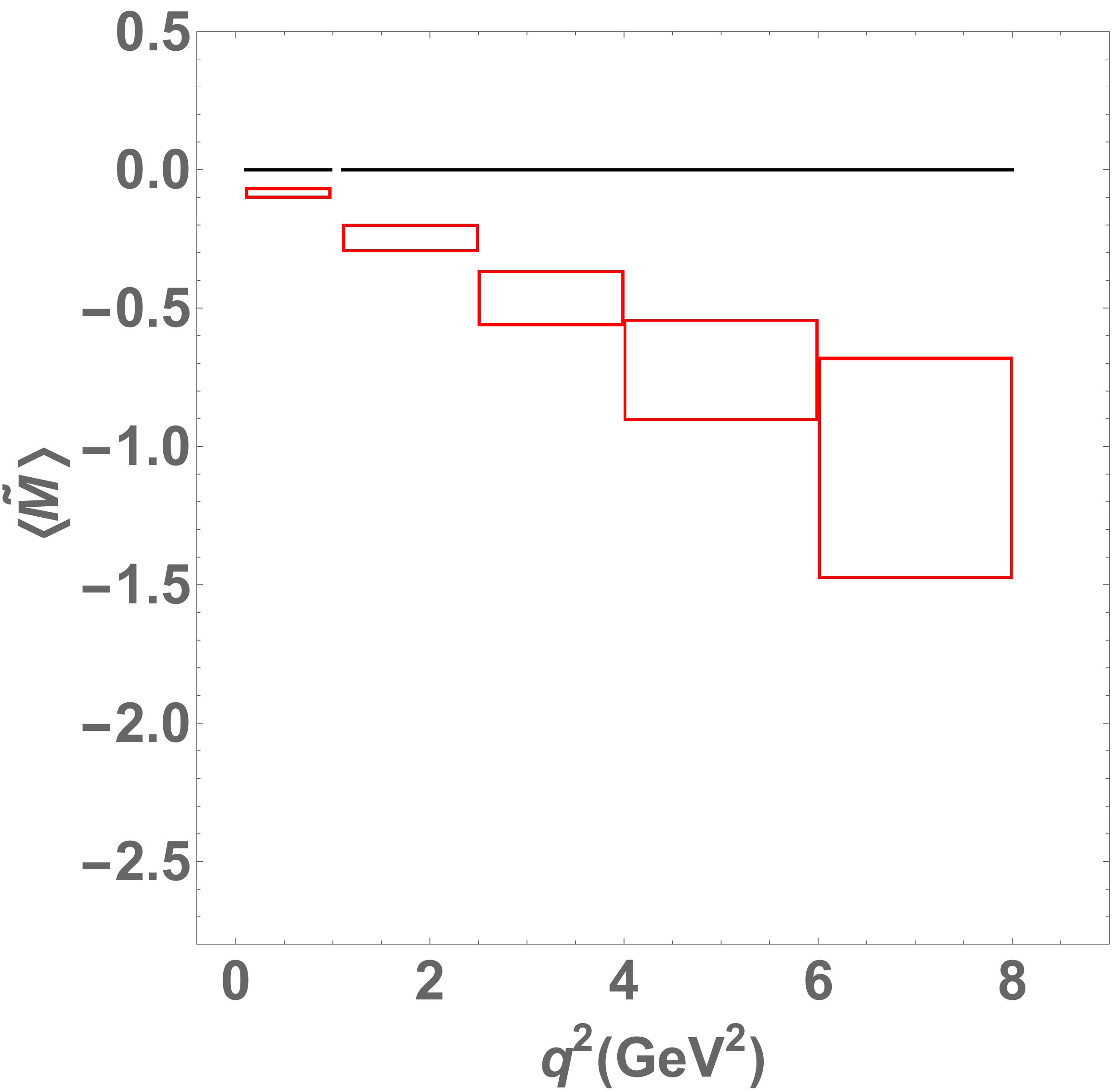}\bigskip
       \vspace{0.0cm}
  \caption{\textit{Scenario 1}.  SM predictions (grey boxes) and NP predictions (red boxes), assuming $C_{9\mu}^{\rm NP}=-1.11$.}
  \label{fig:scenario1b}
 \end{center}
\end{figure}

\begin{figure}
 \begin{center}
  \includegraphics[scale=0.28]{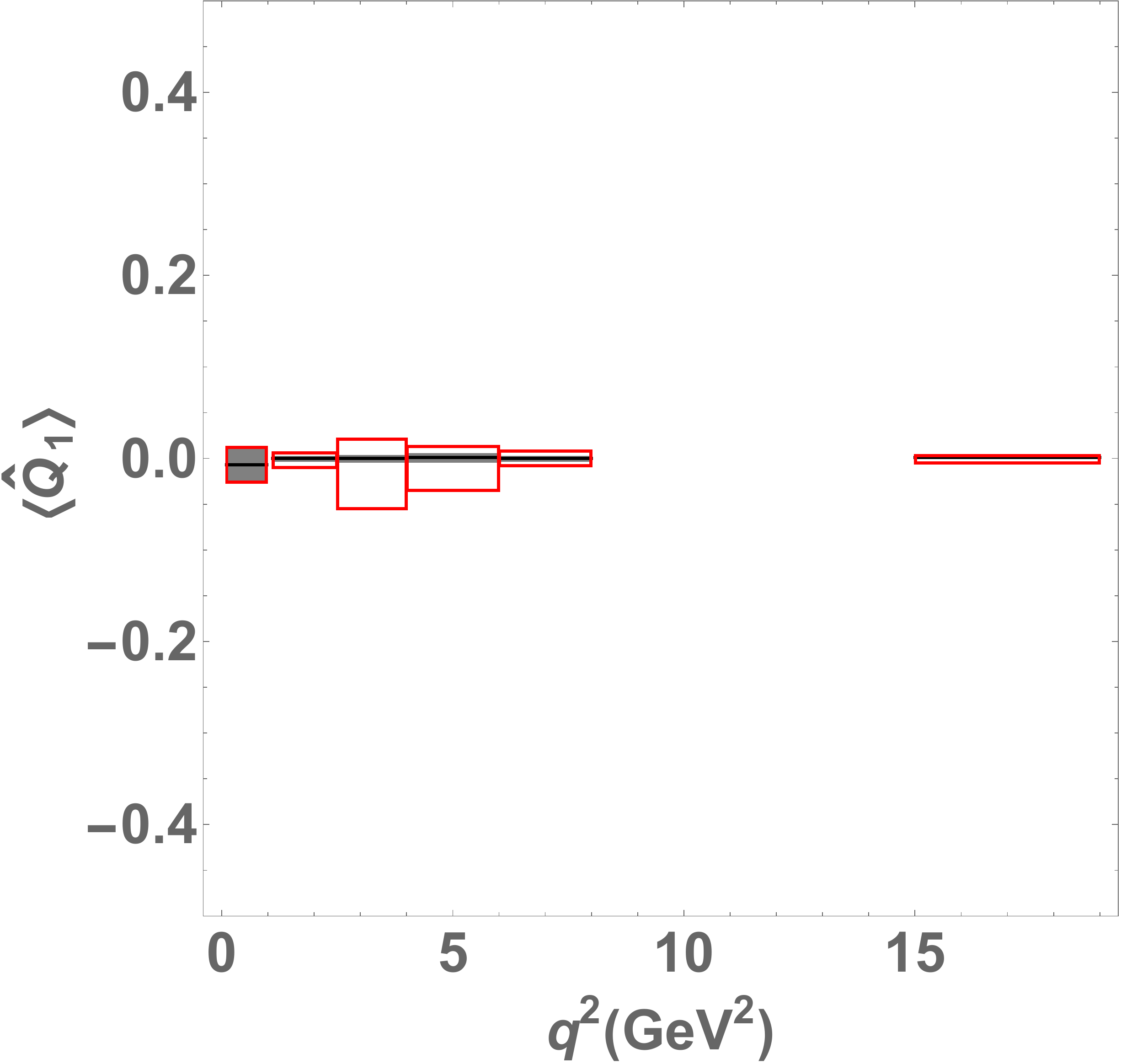}\hspace{7mm}
    \includegraphics[scale=0.28]{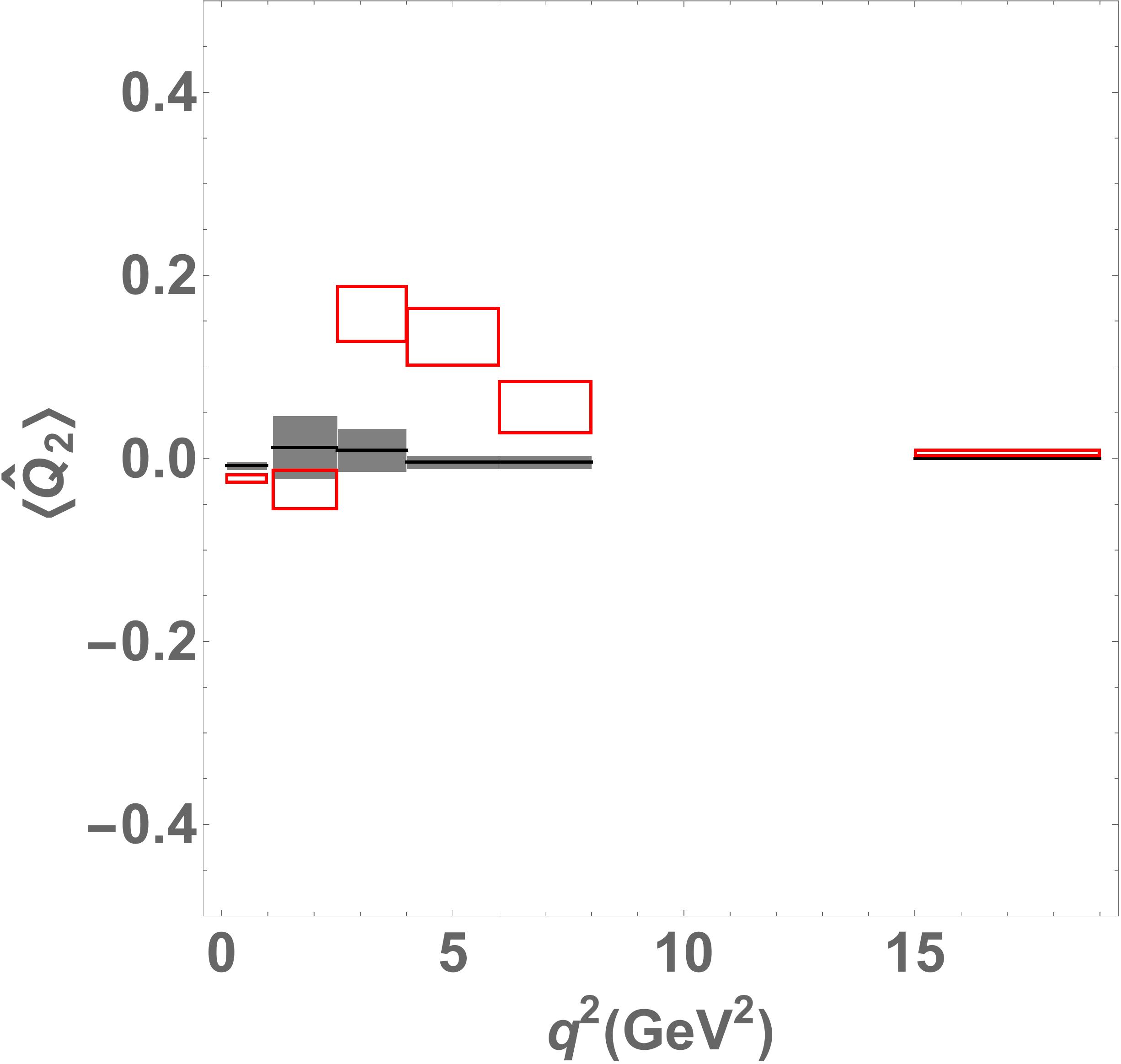}\bigskip
    
   \includegraphics[scale=0.28]{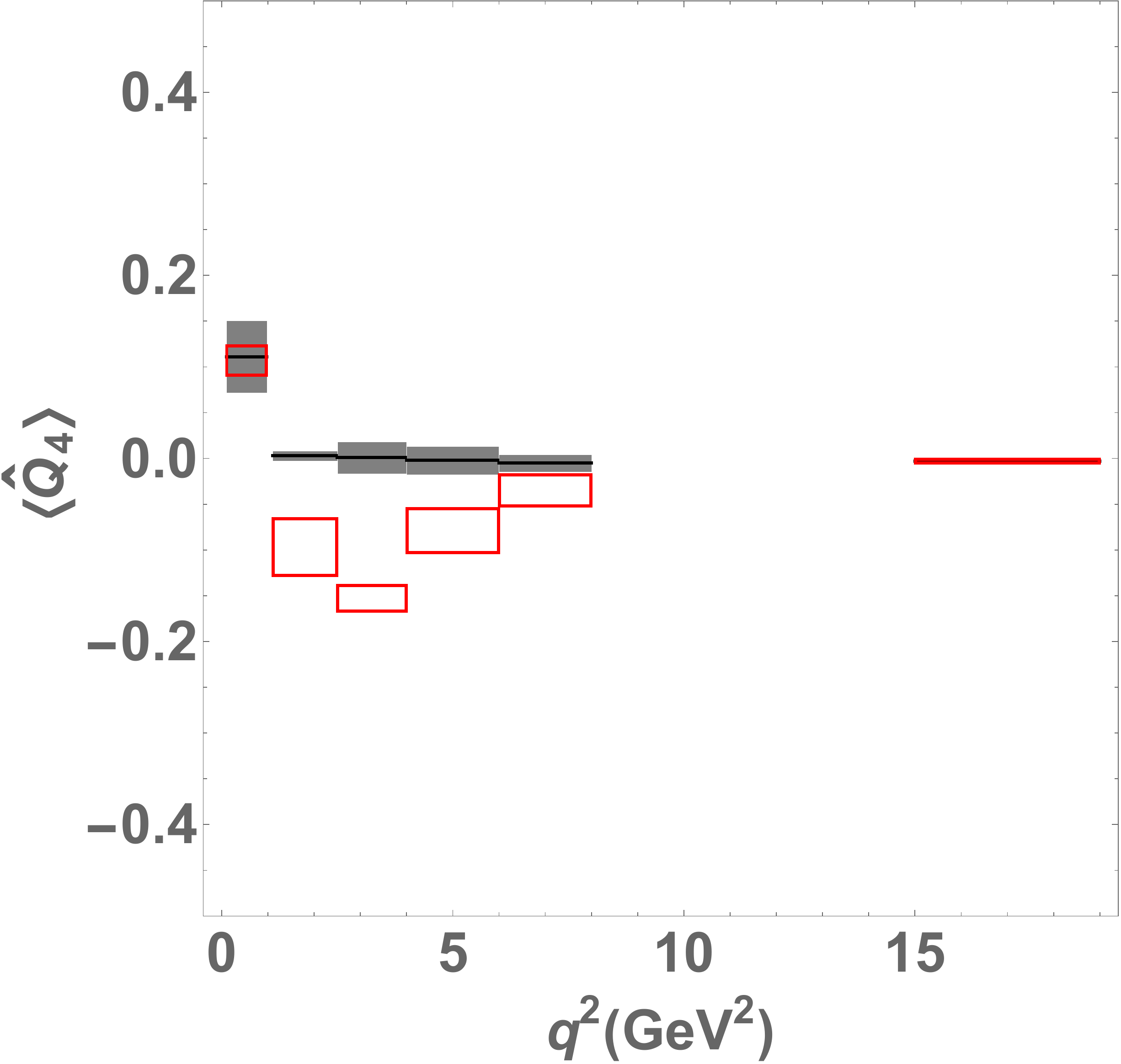}\hspace{7mm}
    \includegraphics[scale=0.28]{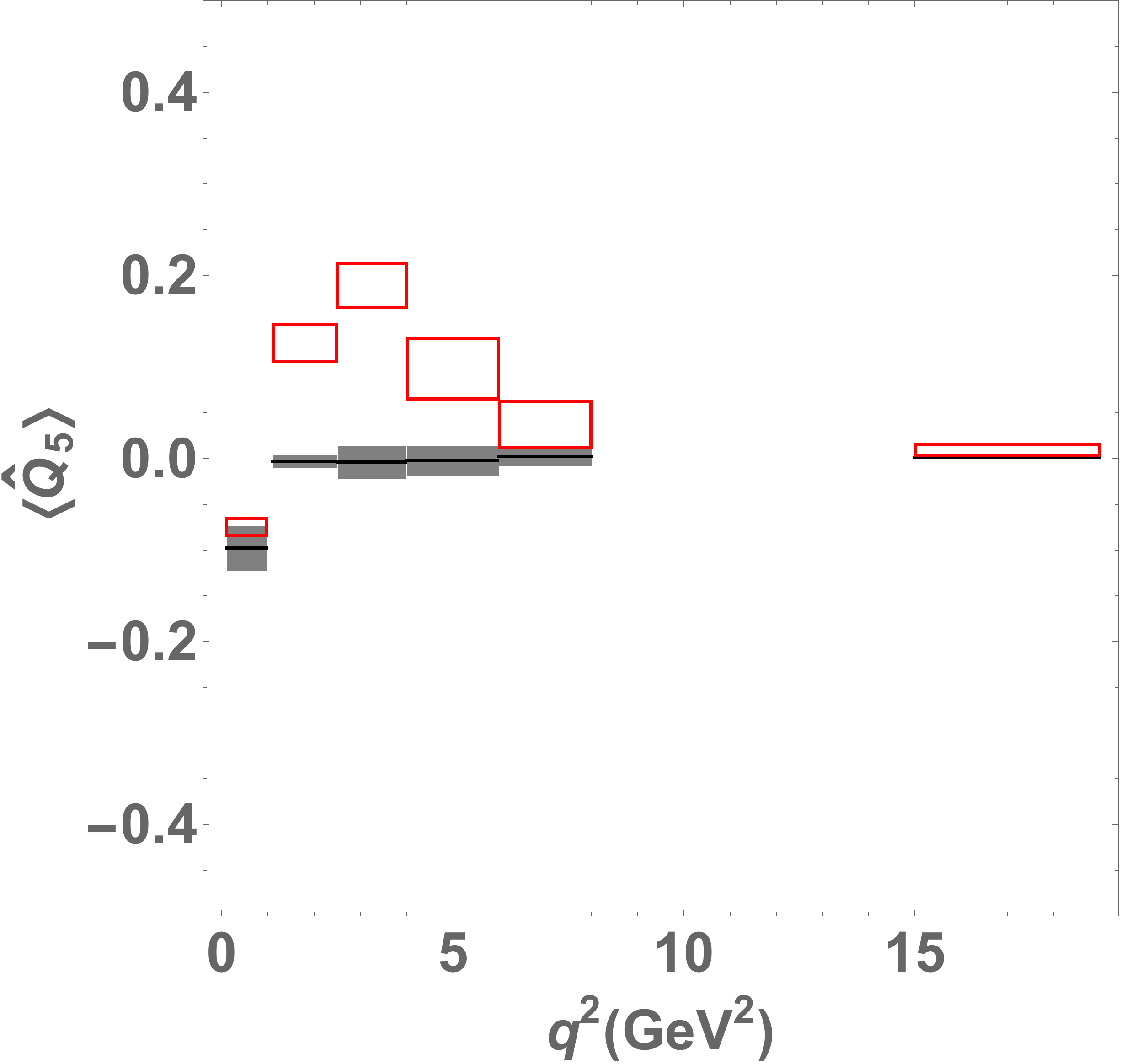}  \bigskip
           \includegraphics[scale=0.28]{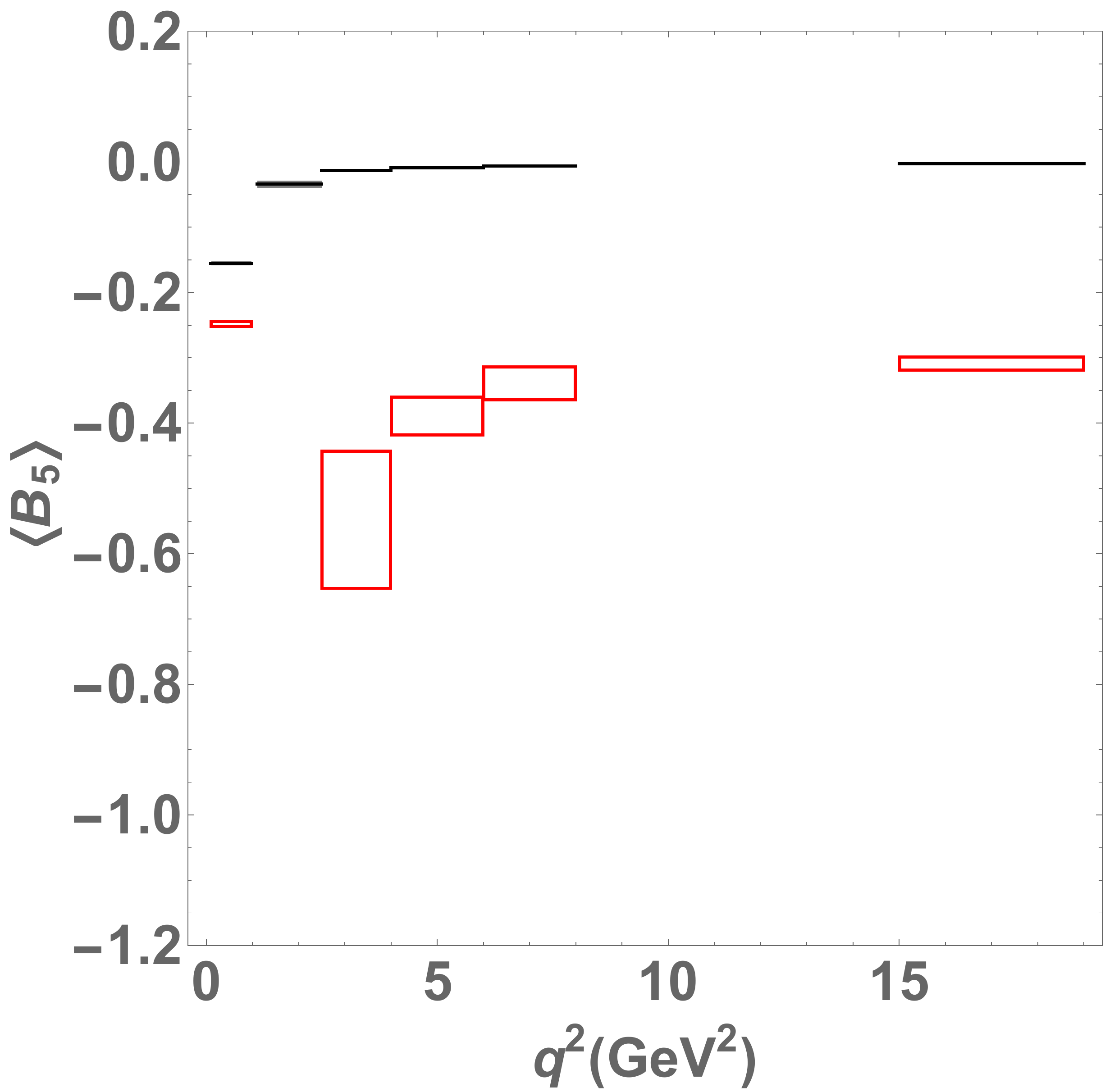}\hspace{7mm}
    \includegraphics[scale=0.28]{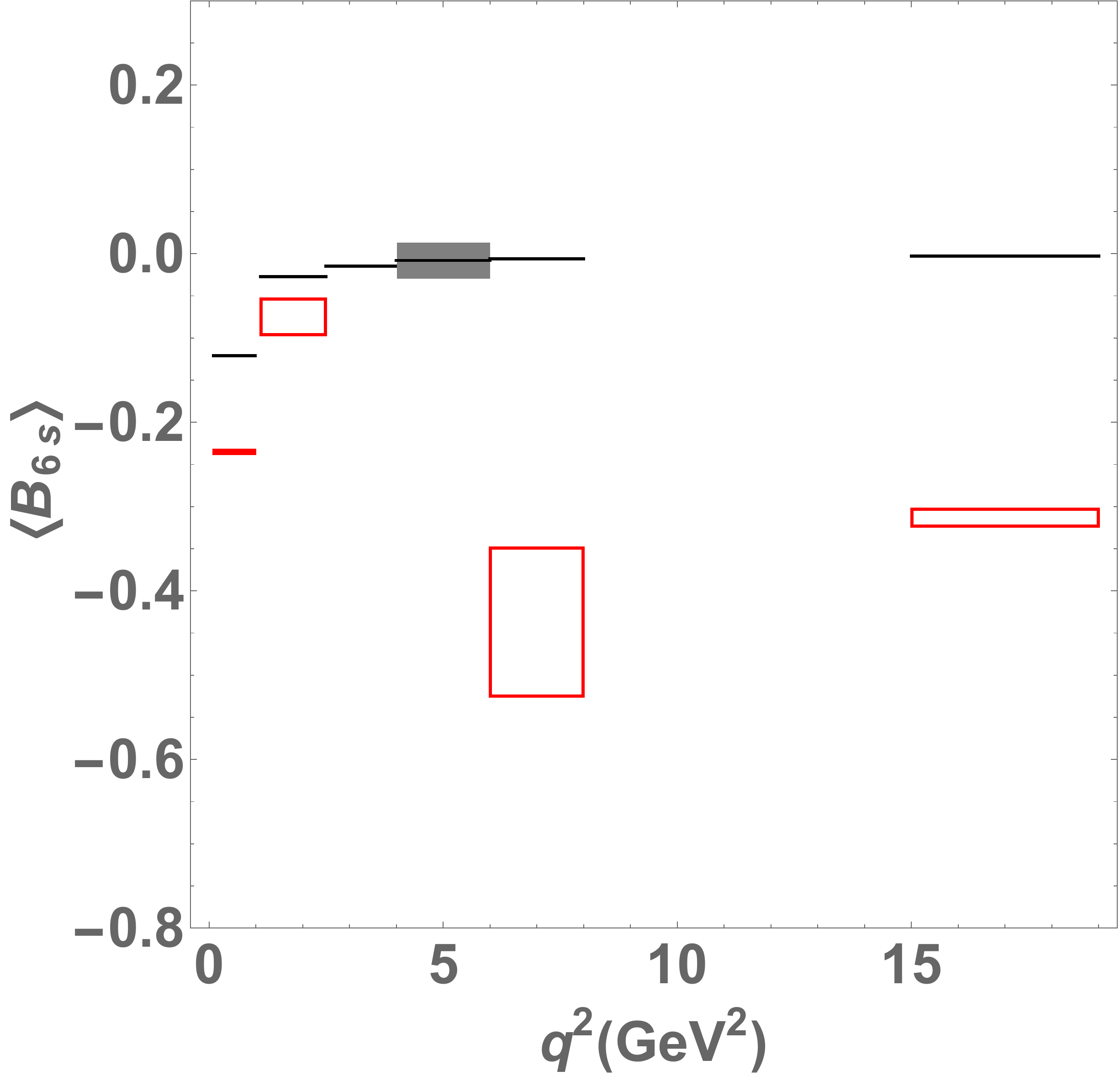} 
           \vspace{0.0cm}
        \caption{\textit{Scenario 2}. SM predictions (grey boxes) and NP predictions (red boxes), assuming $C_{9\mu}^{\rm NP}=-C_{10\mu}^{\rm NP}=-0.65$.}
  \label{fig:scenario2}
 \end{center}
\end{figure}

\begin{figure}
 \begin{center}
        \includegraphics[scale=0.28]{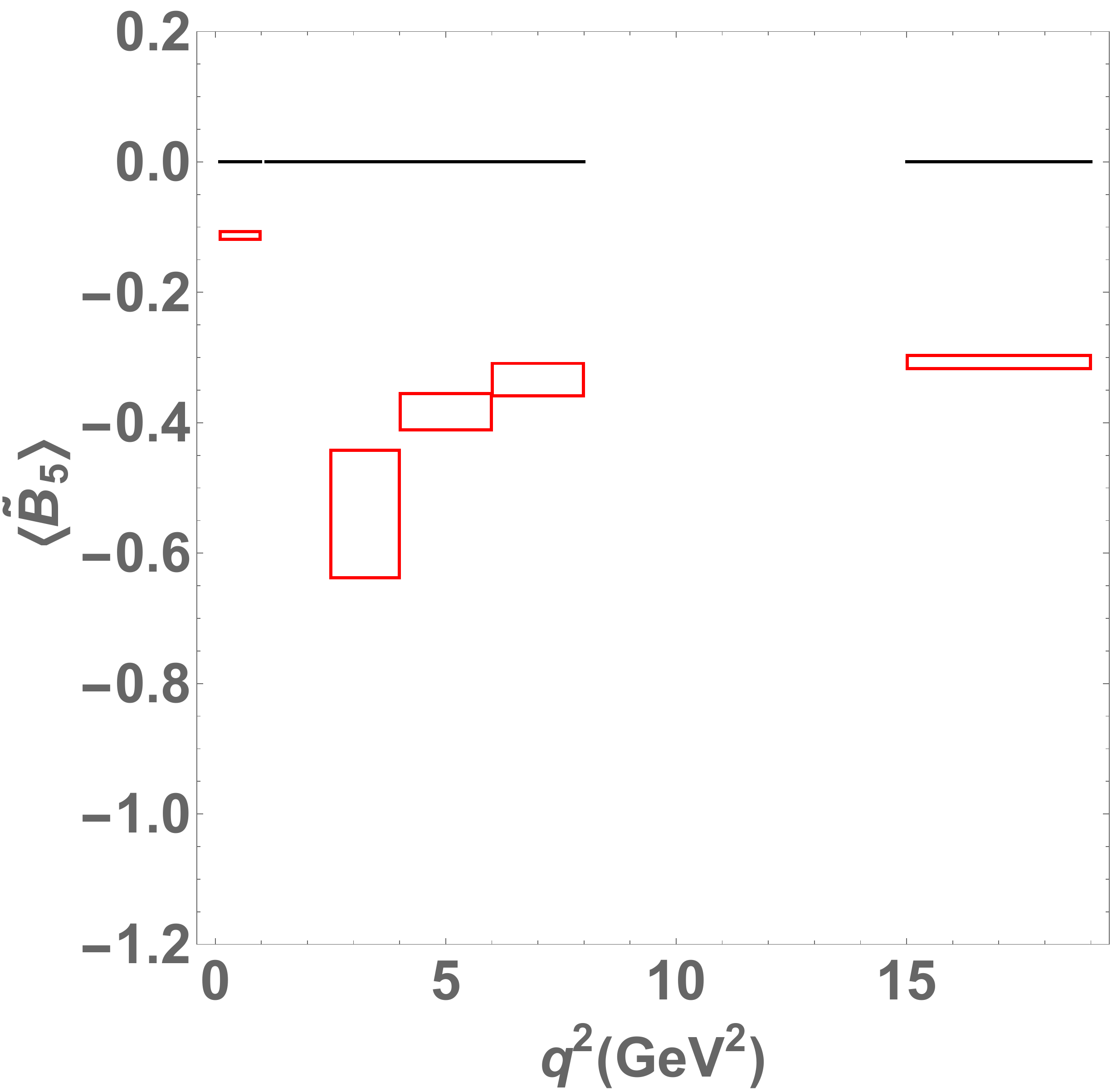}\hspace{7mm}
    \includegraphics[scale=0.28]{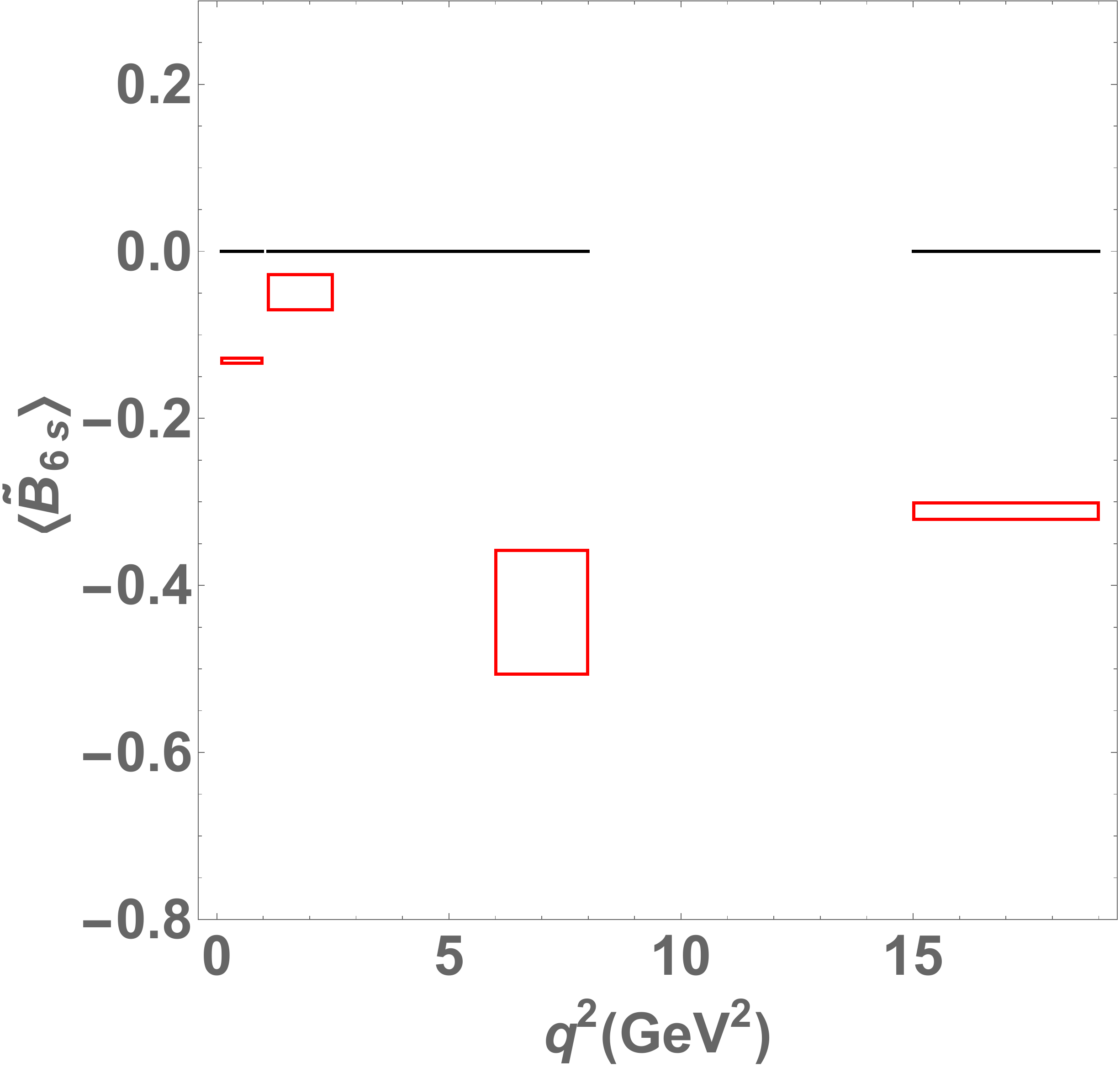} \\[10mm]
  \includegraphics[scale=0.28]{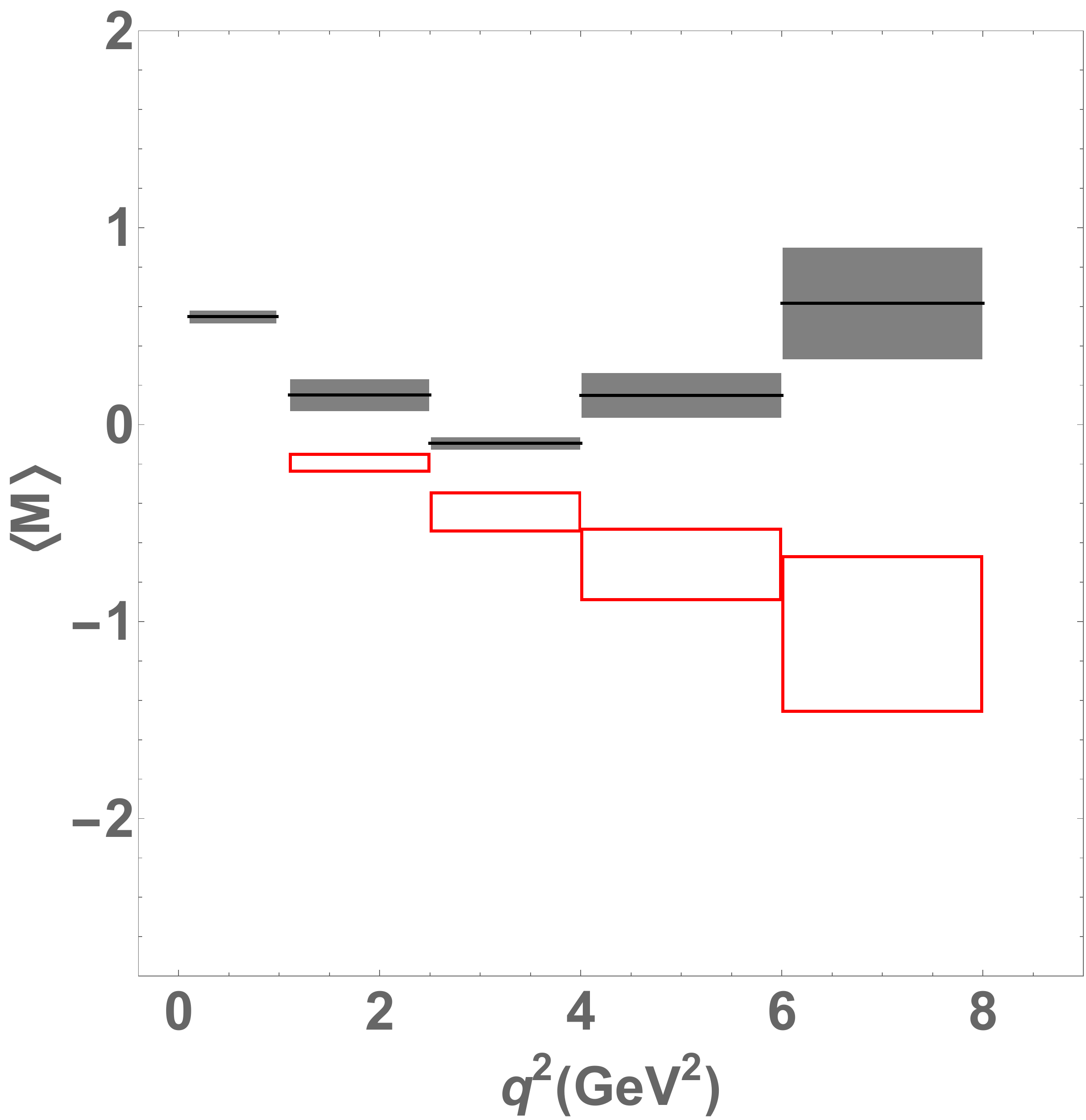}\hspace{7mm}
    \includegraphics[scale=0.28]{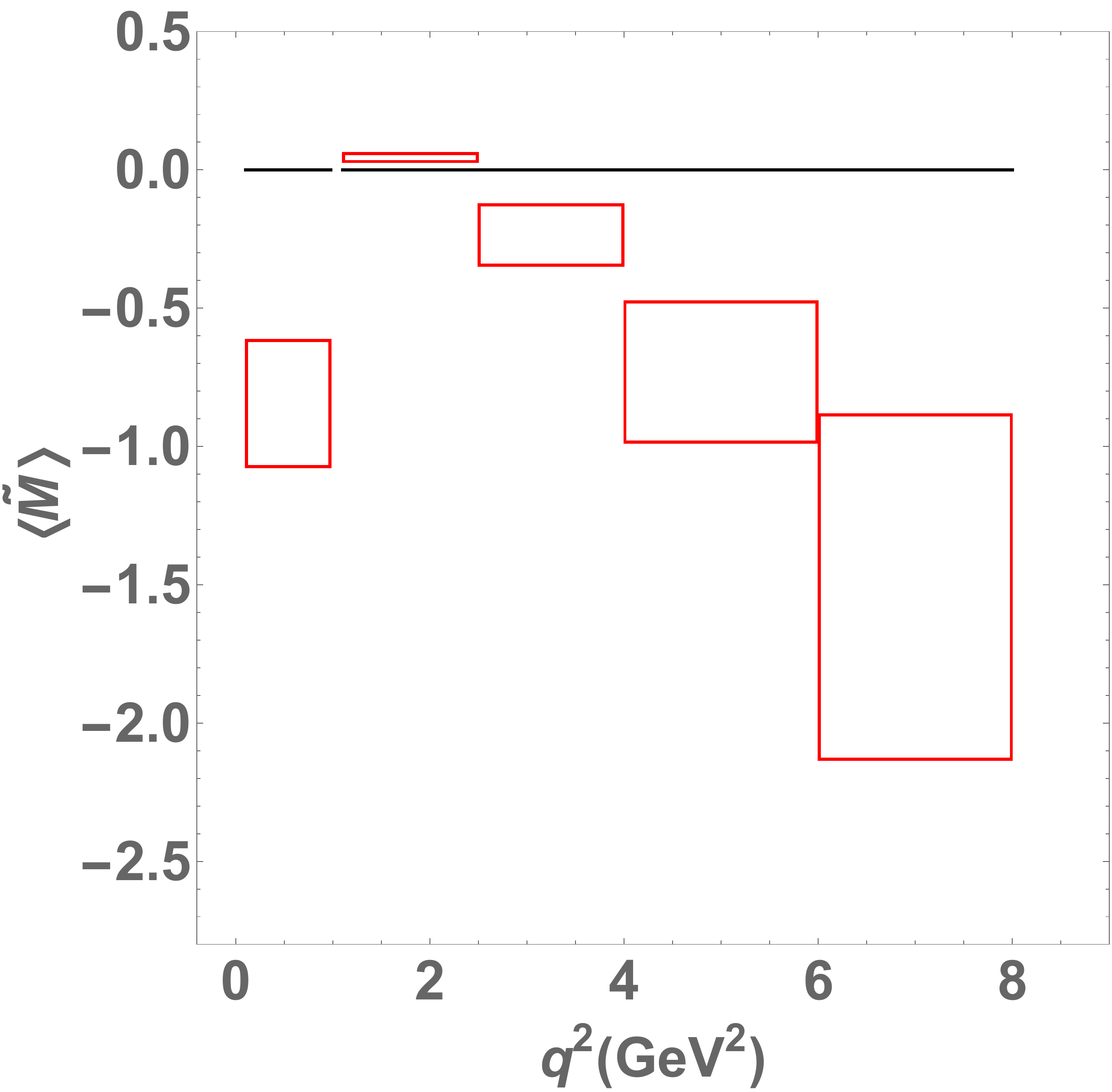}\bigskip
       \vspace{0.0cm}
        \caption{\textit{Scenario 2}. SM predictions (grey boxes) and NP predictions (red boxes), assuming $C_{9\mu}^{\rm NP}=-C_{10\mu}^{\rm NP}=-0.65$.}
  \label{fig:scenario2b}
 \end{center}
\end{figure}

\begin{figure}
 \begin{center}
  \includegraphics[scale=0.28]{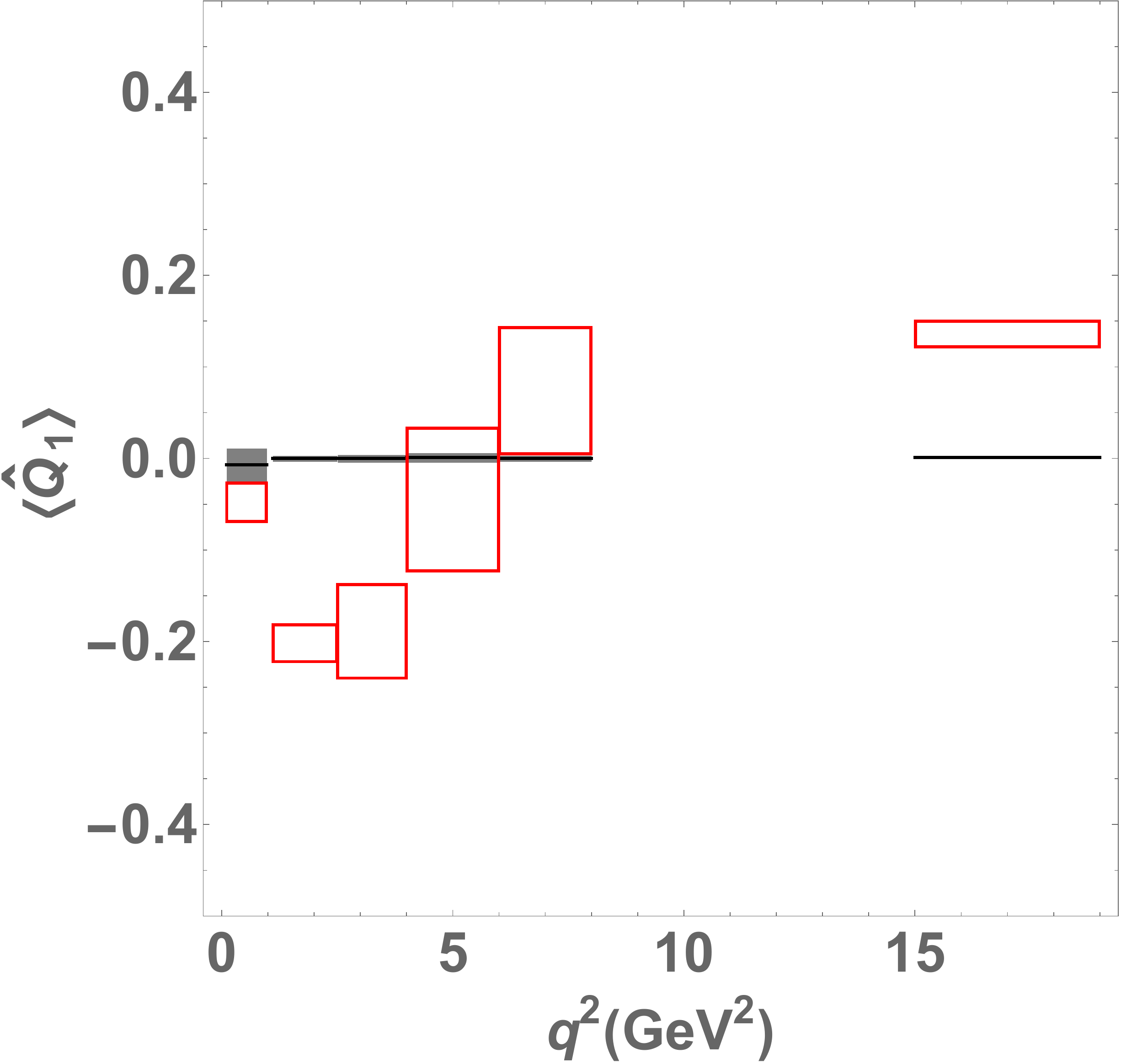}\hspace{7mm}
    \includegraphics[scale=0.28]{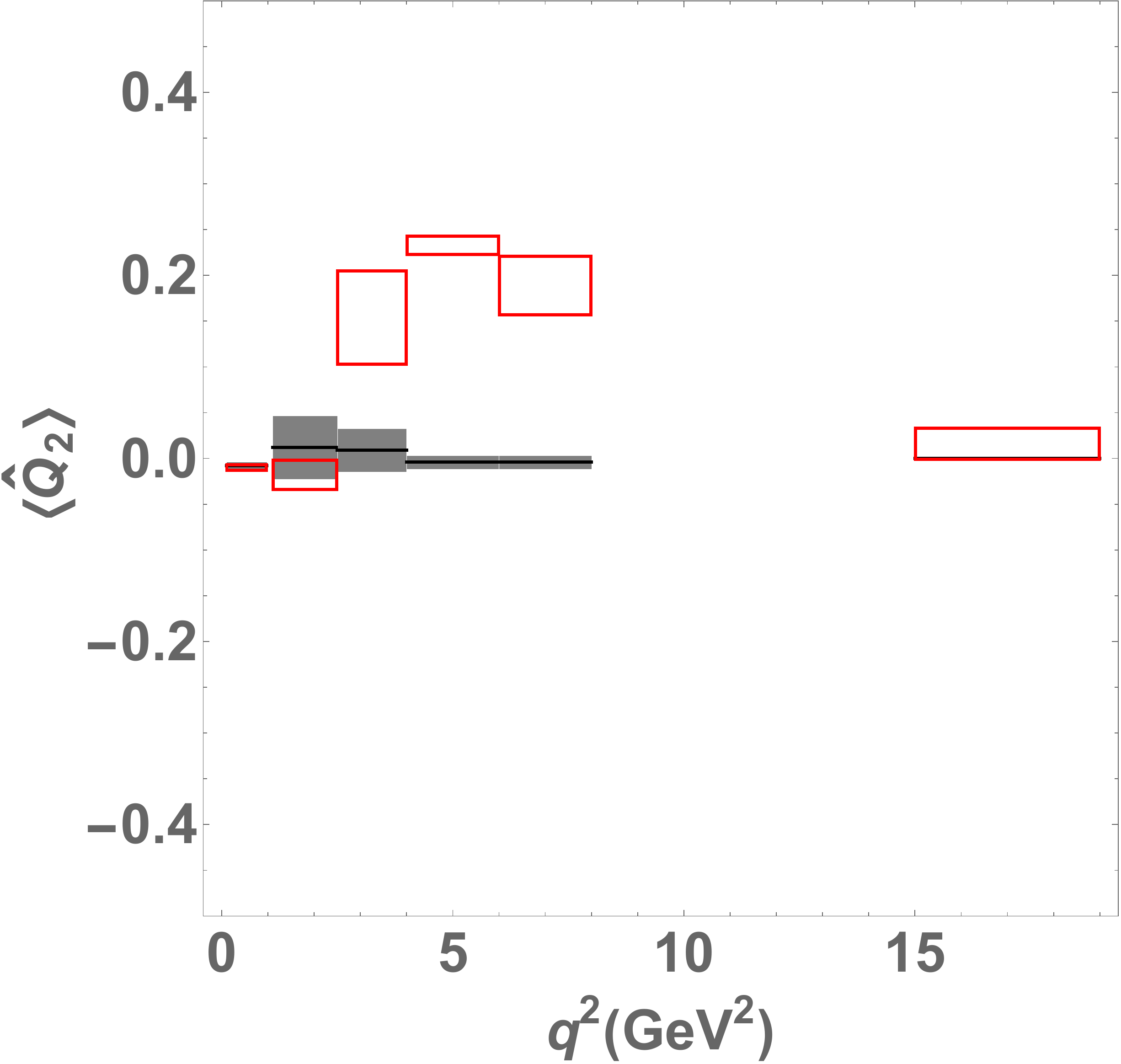}\bigskip
    
   \includegraphics[scale=0.28]{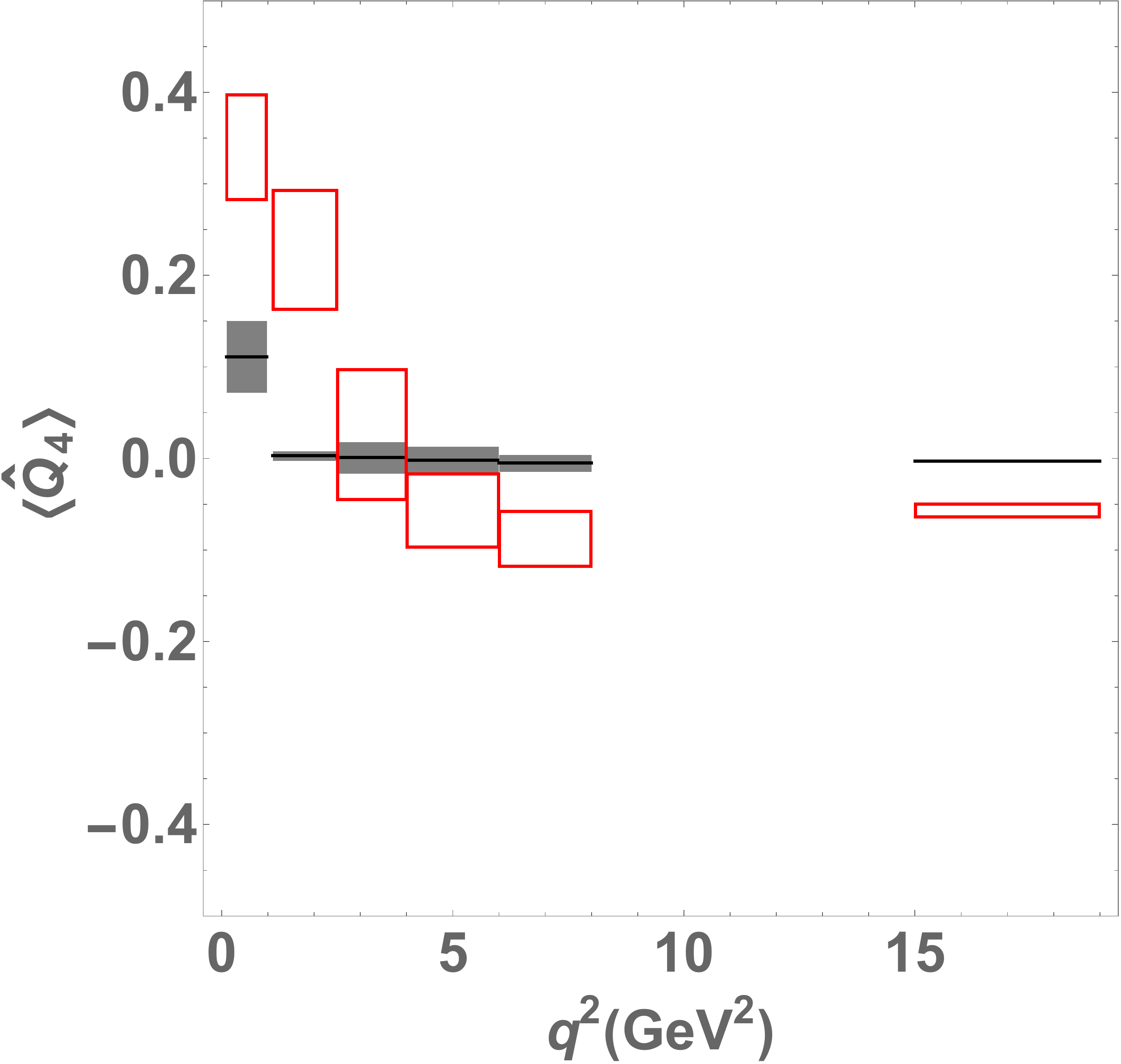}\hspace{7mm}
    \includegraphics[scale=0.28]{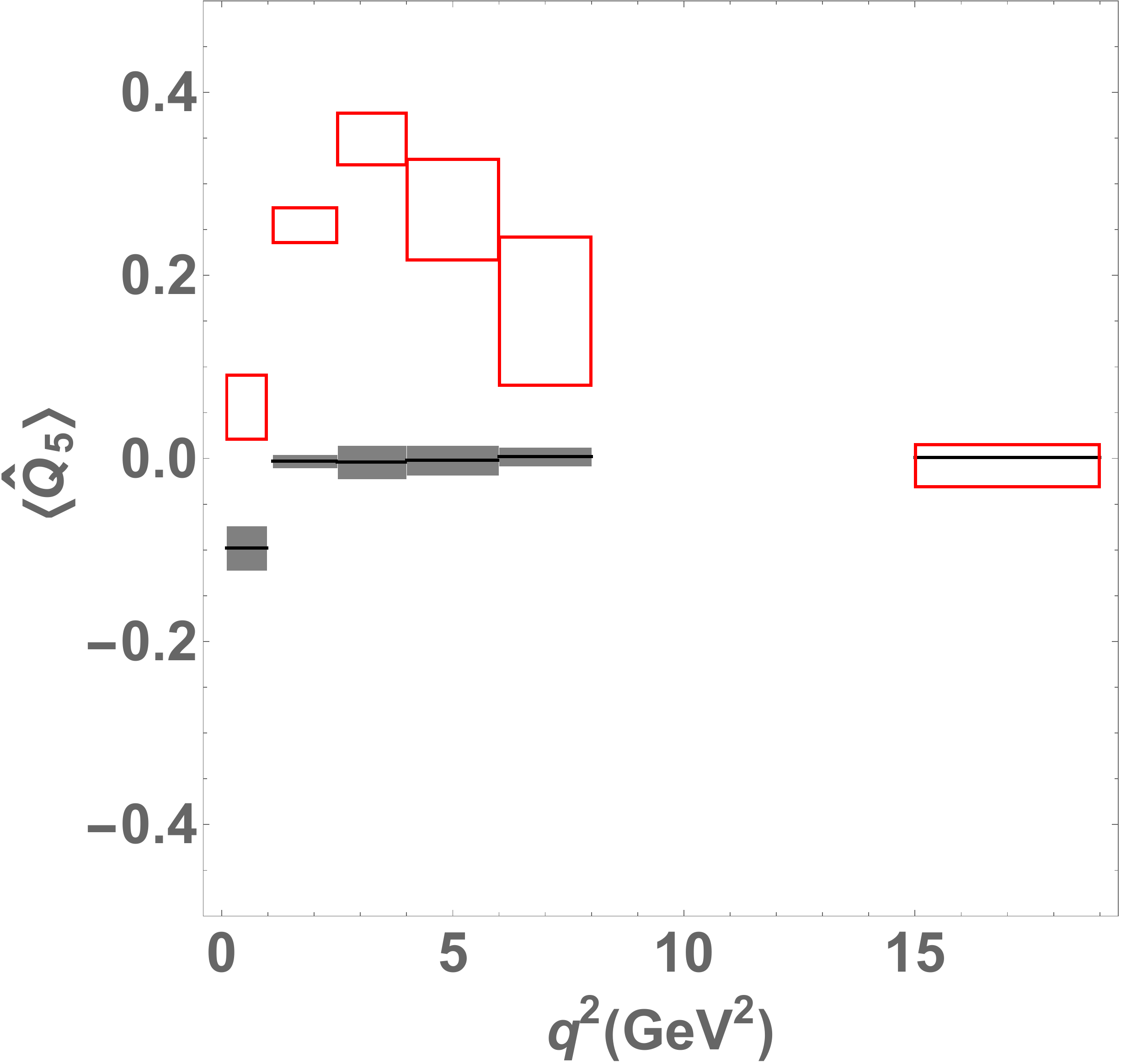}  \bigskip
           \includegraphics[scale=0.28]{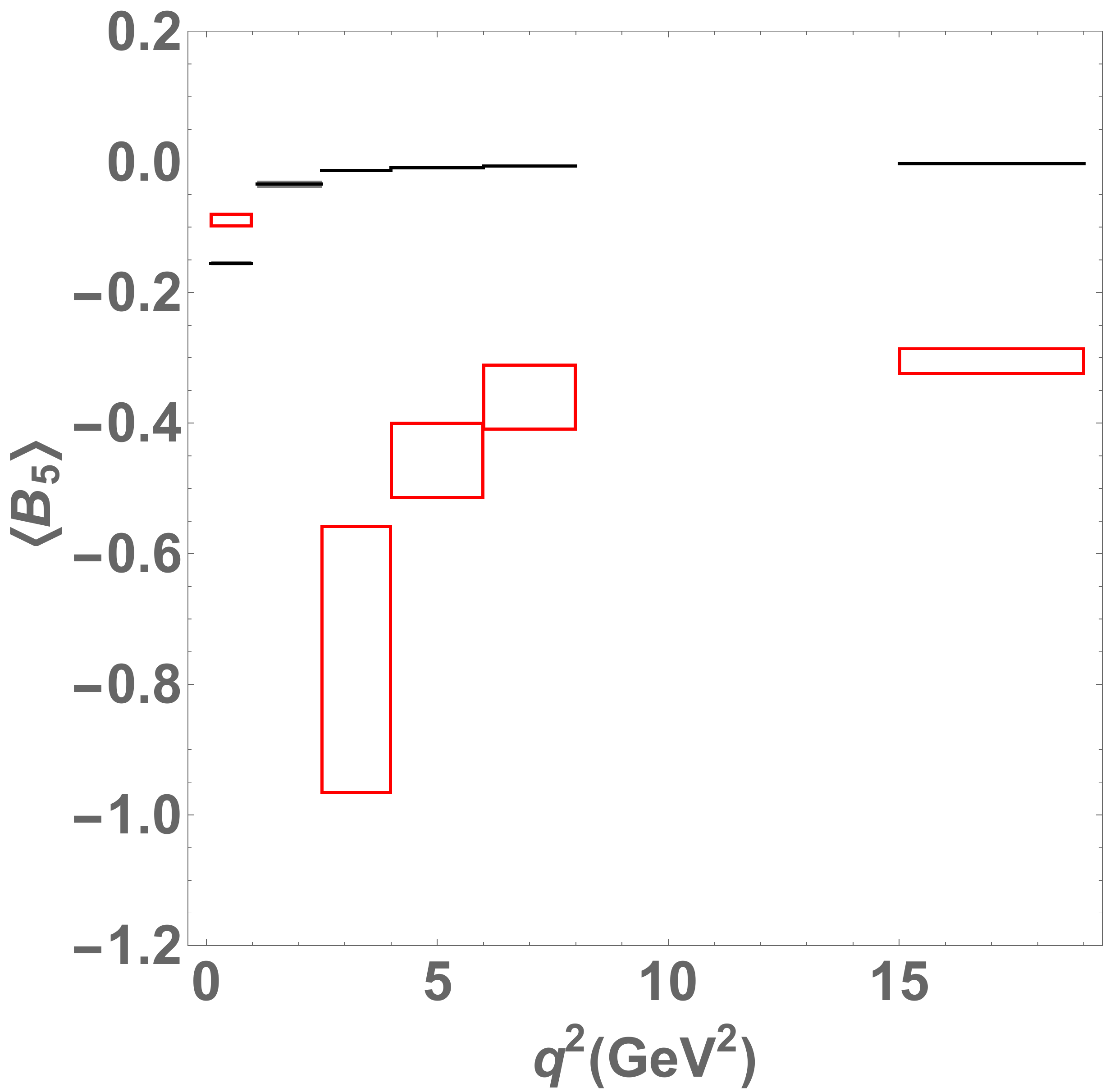}\hspace{7mm}
    \includegraphics[scale=0.28]{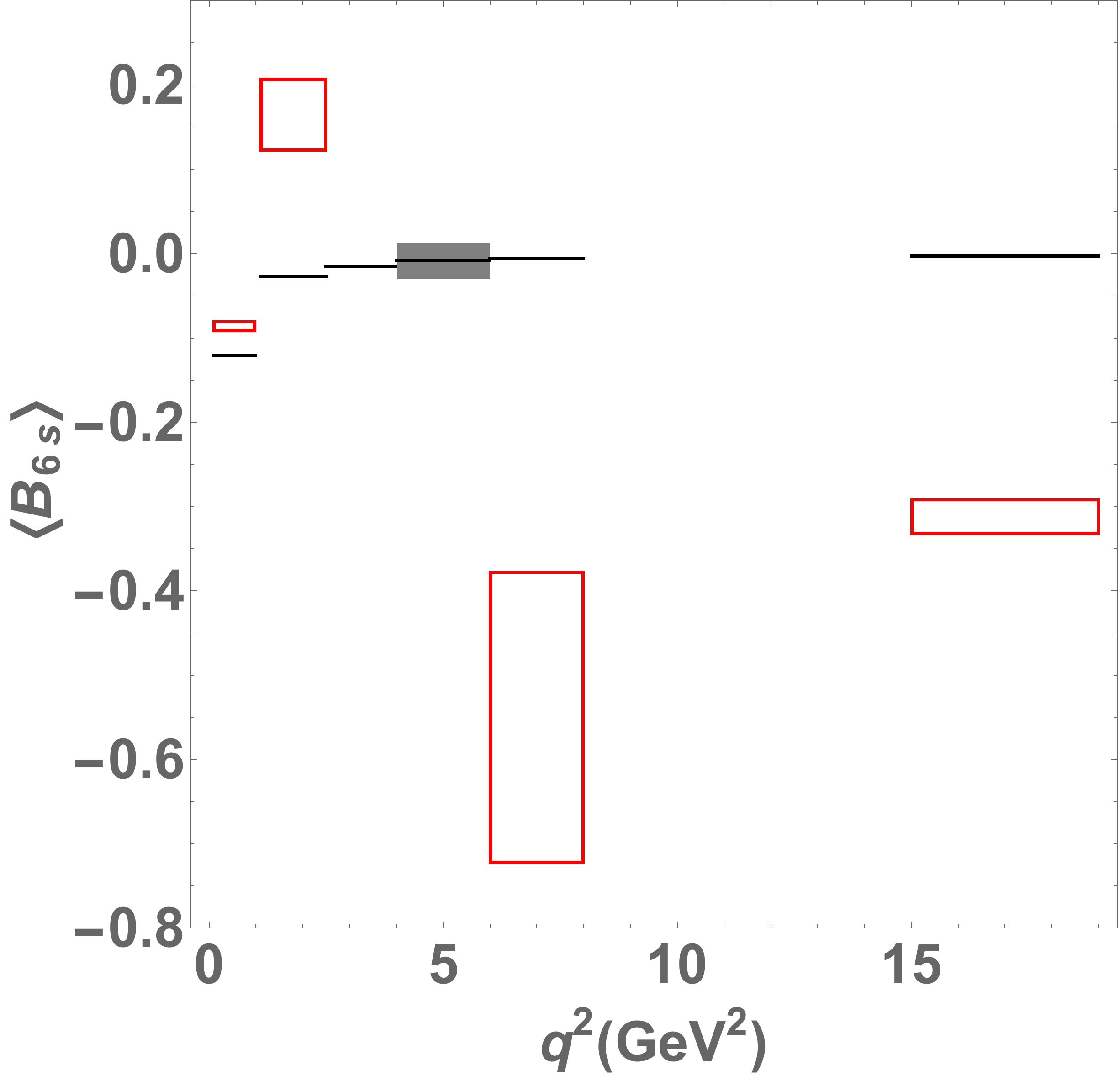}     
       \vspace{0.0cm}
       
        \caption{\textit{Scenario 3}.  SM predictions (grey boxes) and NP predictions (red boxes), assuming $C_{9\mu}^{\rm NP}=-C_{9'\mu}^{{\rm NP}}=-1.07$}
  \label{fig:scenario3}
 \end{center}
\end{figure}

\begin{figure}
 \begin{center}
        \includegraphics[scale=0.28]{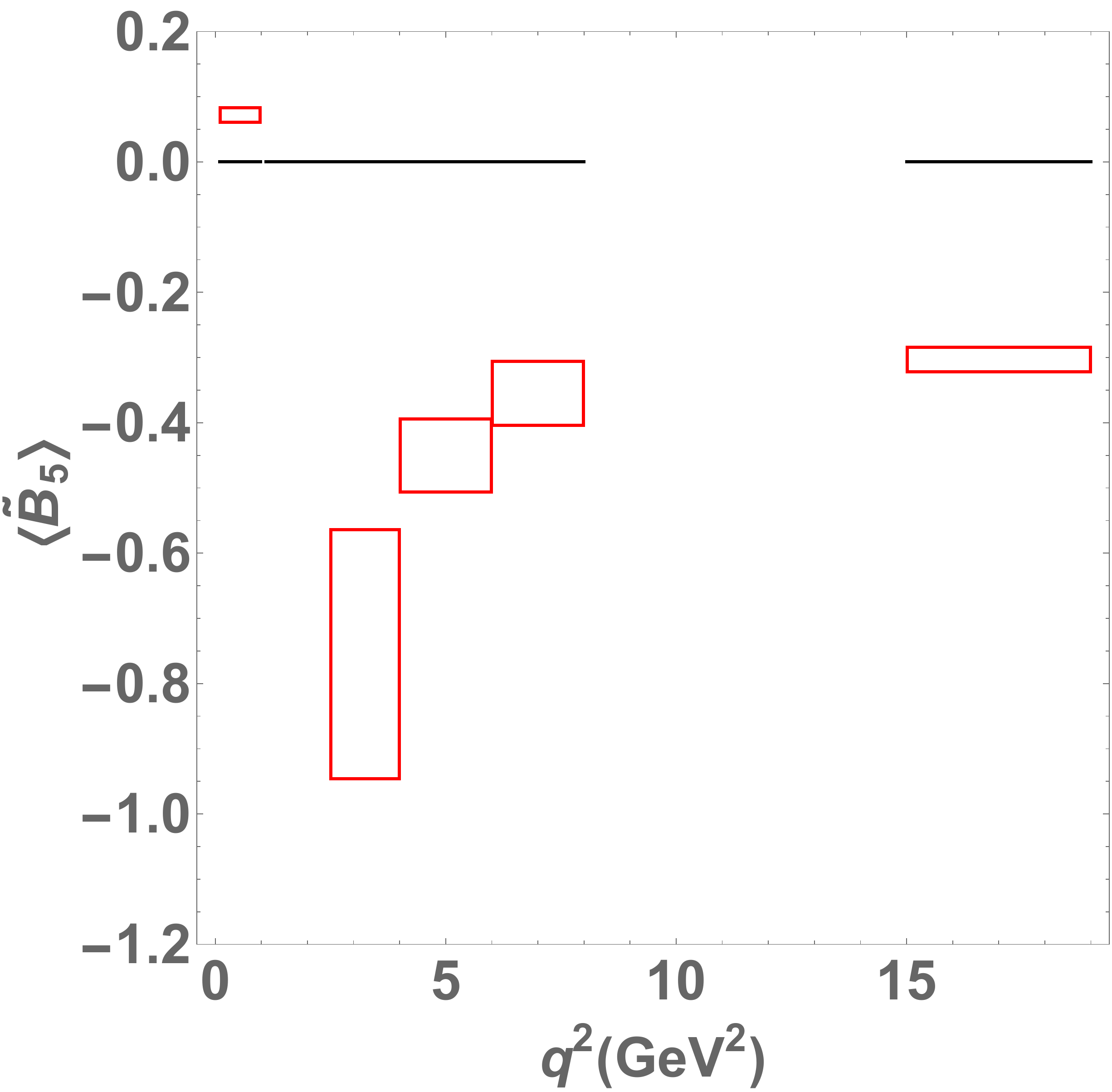}\hspace{7mm}
    \includegraphics[scale=0.28]{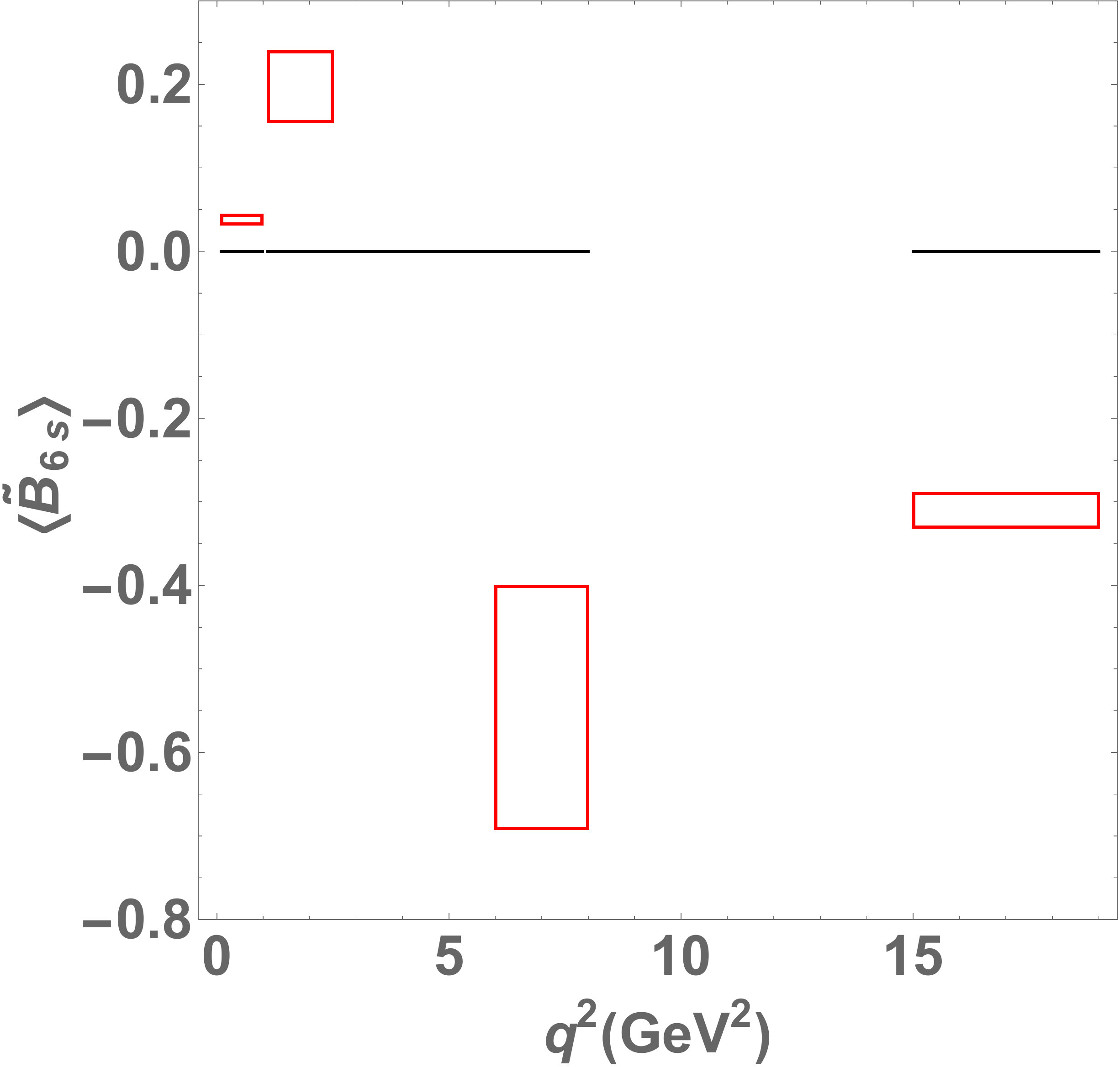}\\[10mm]
  \includegraphics[scale=0.28]{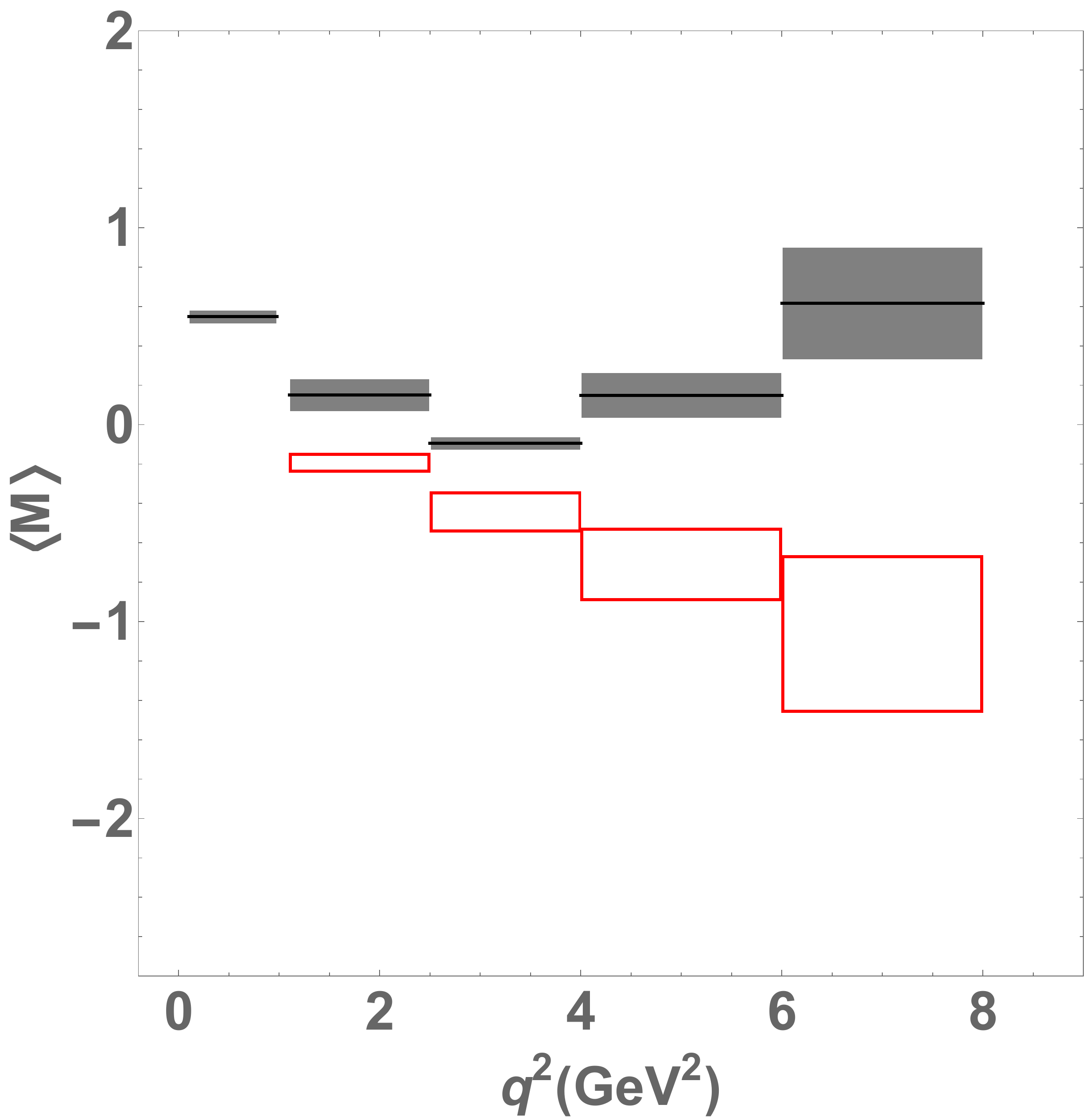}\hspace{7mm}
    \includegraphics[scale=0.29]{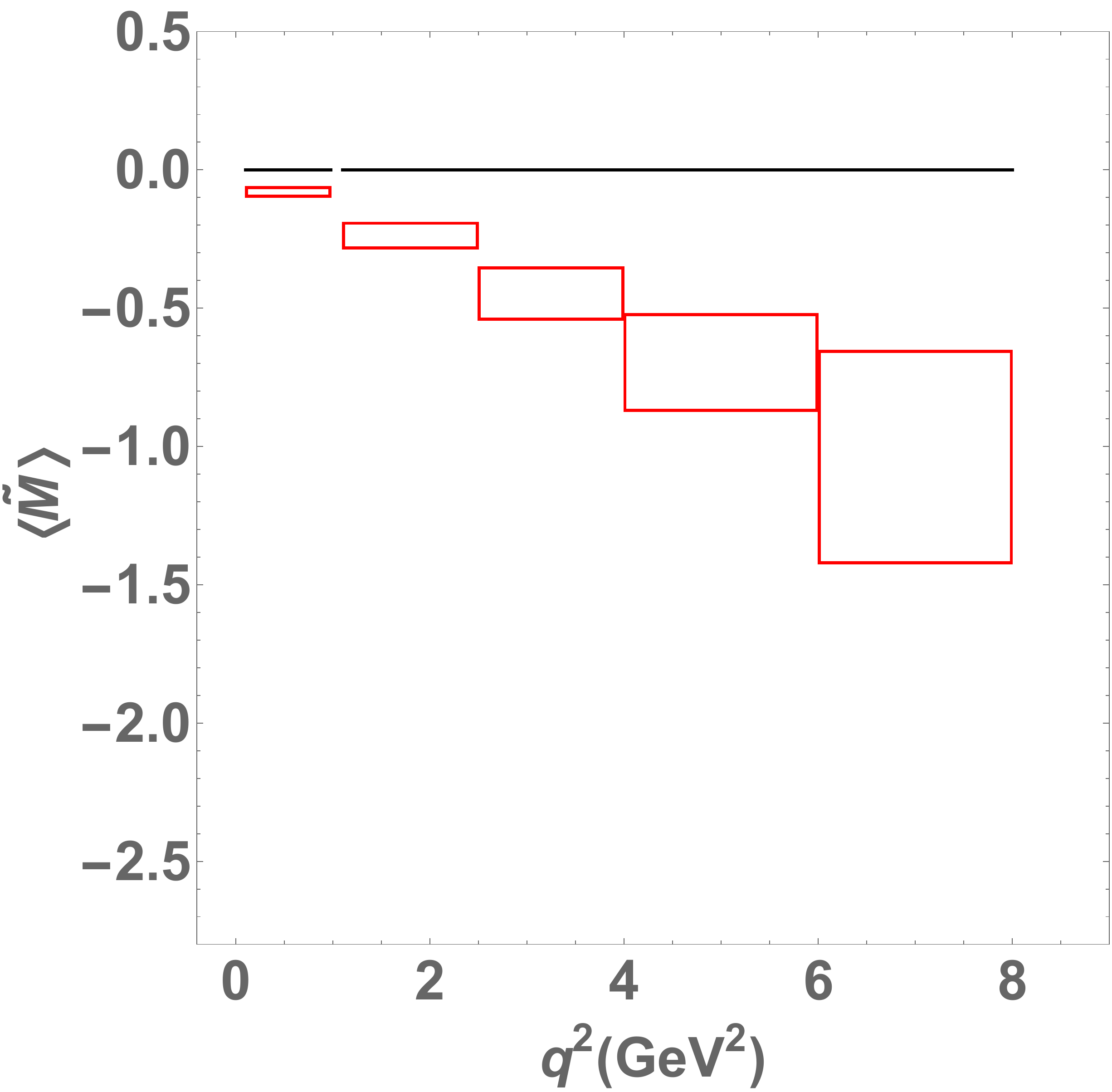}\bigskip
       \vspace{0.0cm}
        \caption{\textit{Scenario 3}.  SM predictions (grey boxes) and NP predictions (red boxes), assuming $C_{9\mu}^{\rm NP}=-C_{9'\mu}^{{\rm NP}}=-1.07$}
  \label{fig:scenario3b}
 \end{center}
\end{figure}

\begin{figure}
 \begin{center}
  \includegraphics[scale=0.28]{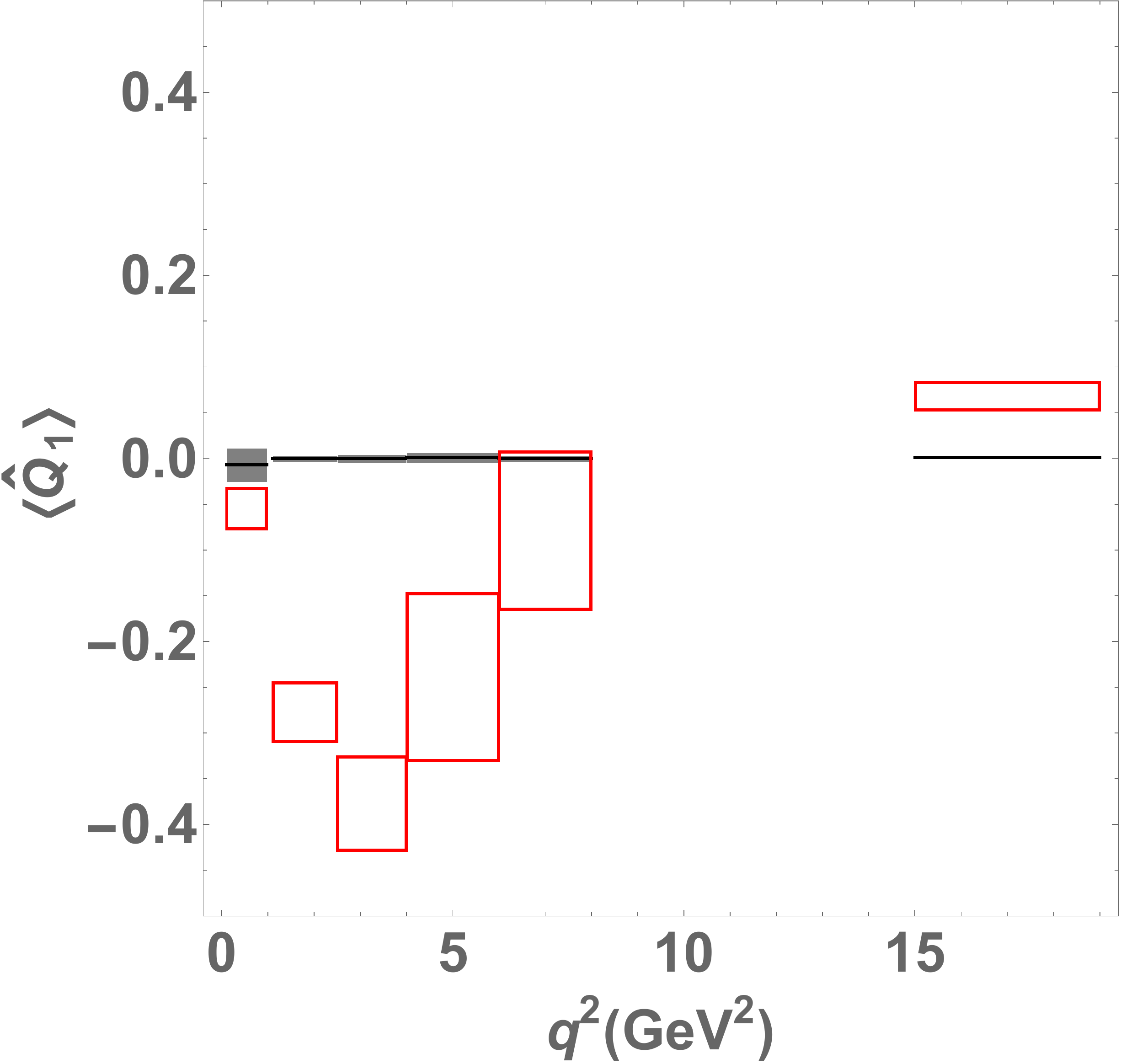}\hspace{7mm}
    \includegraphics[scale=0.28]{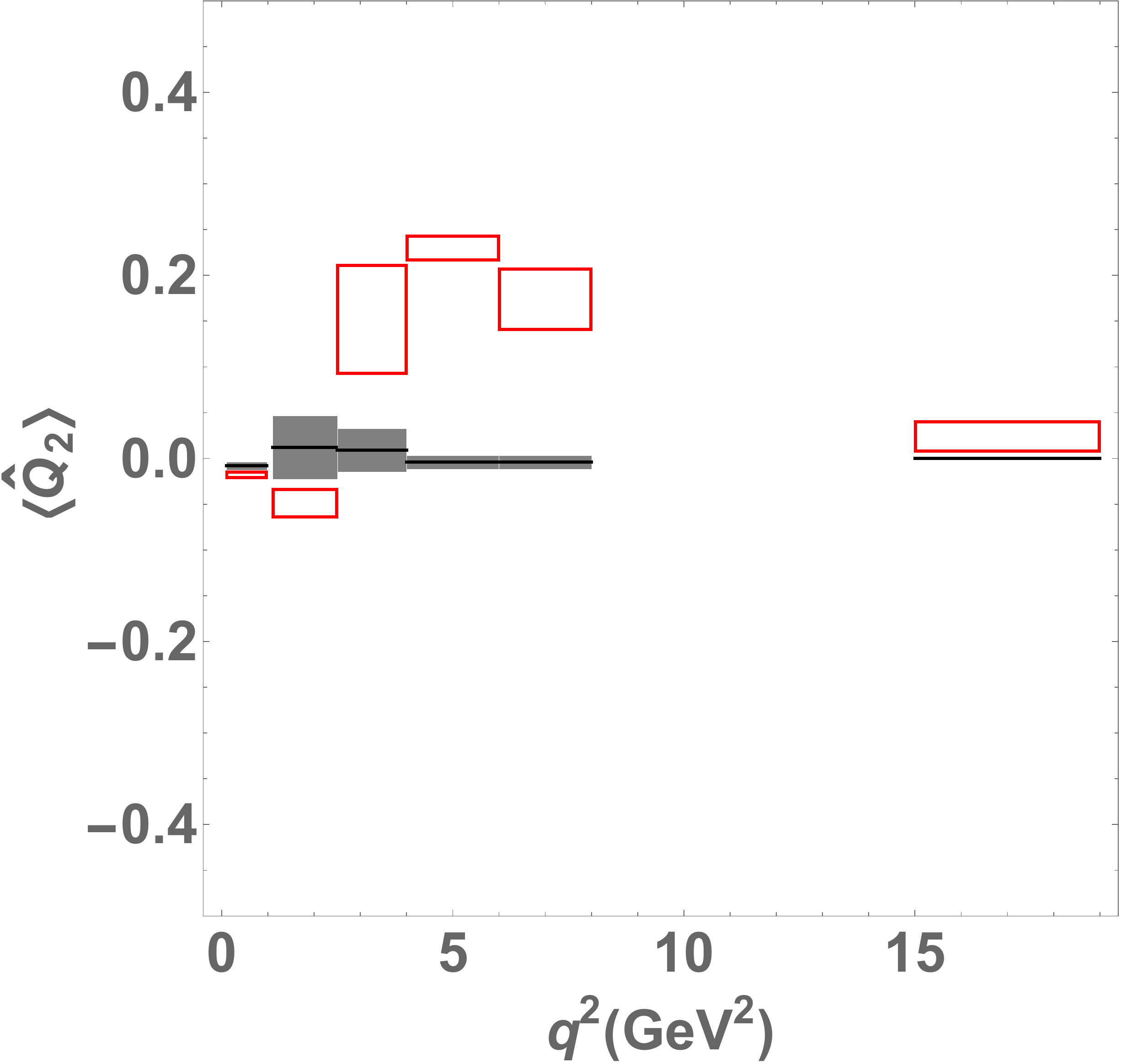}\bigskip
    
   \includegraphics[scale=0.28]{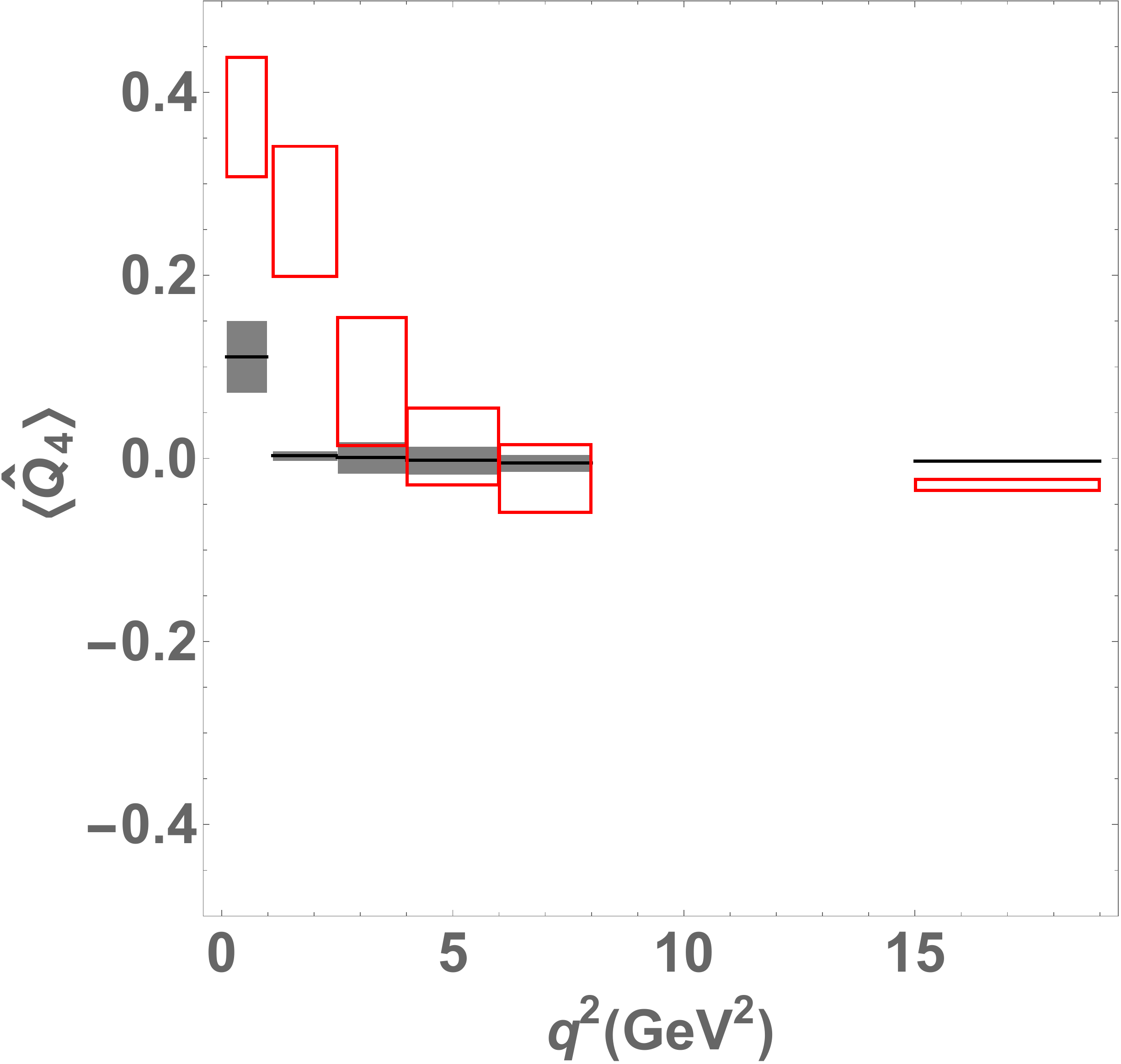}\hspace{7mm}
    \includegraphics[scale=0.28]{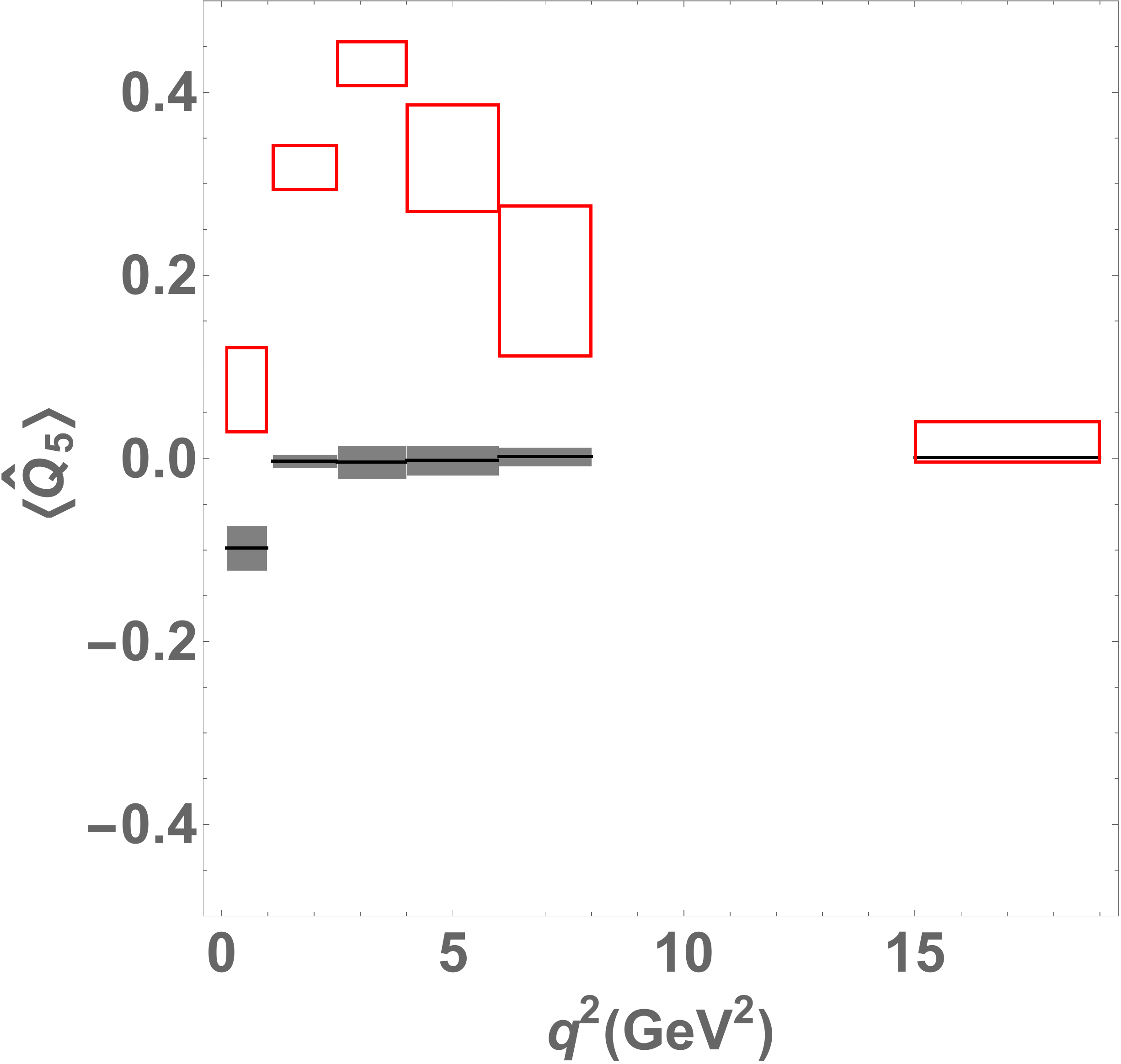}  \bigskip
           \includegraphics[scale=0.28]{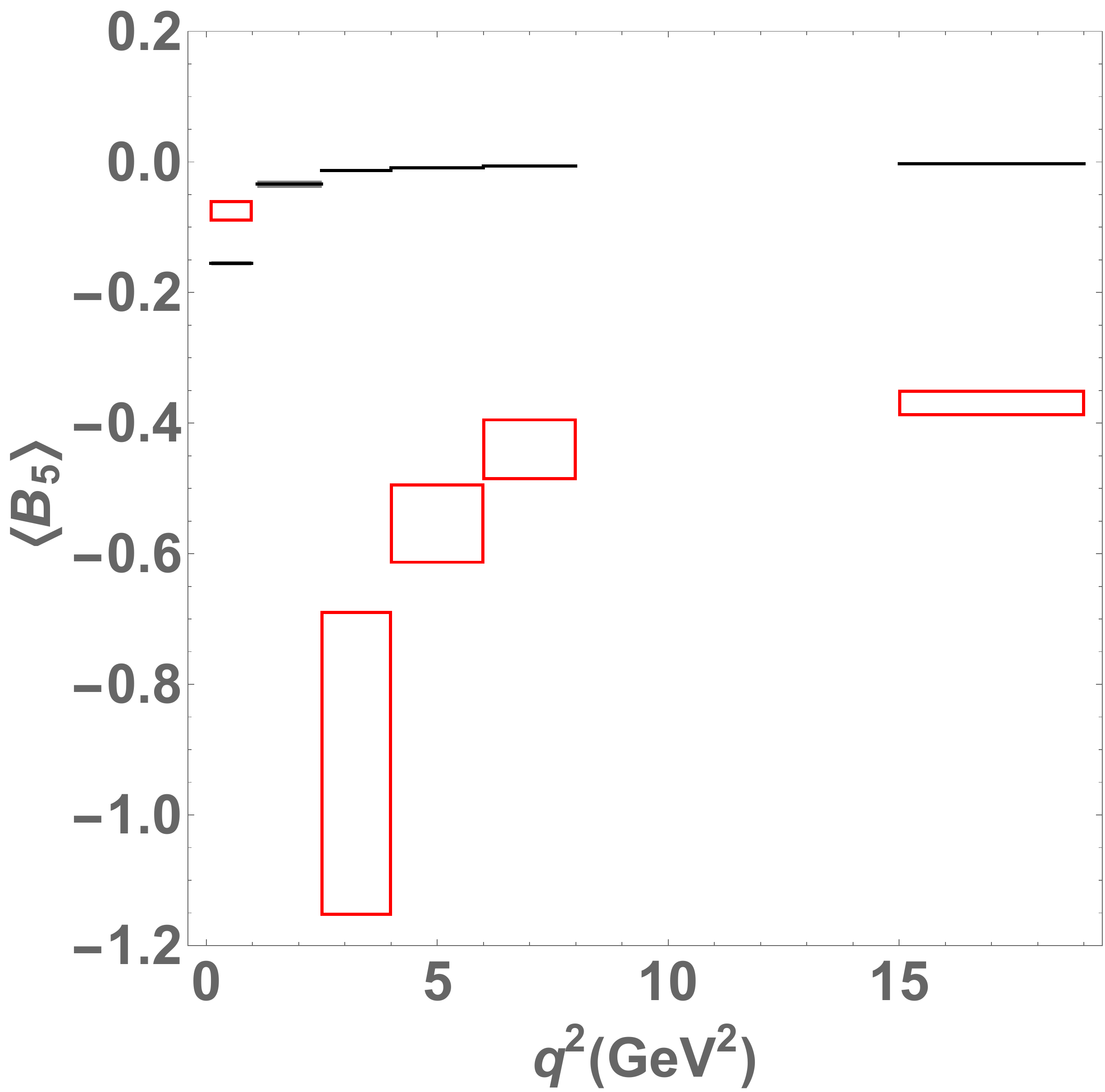}\hspace{7mm}
    \includegraphics[scale=0.28]{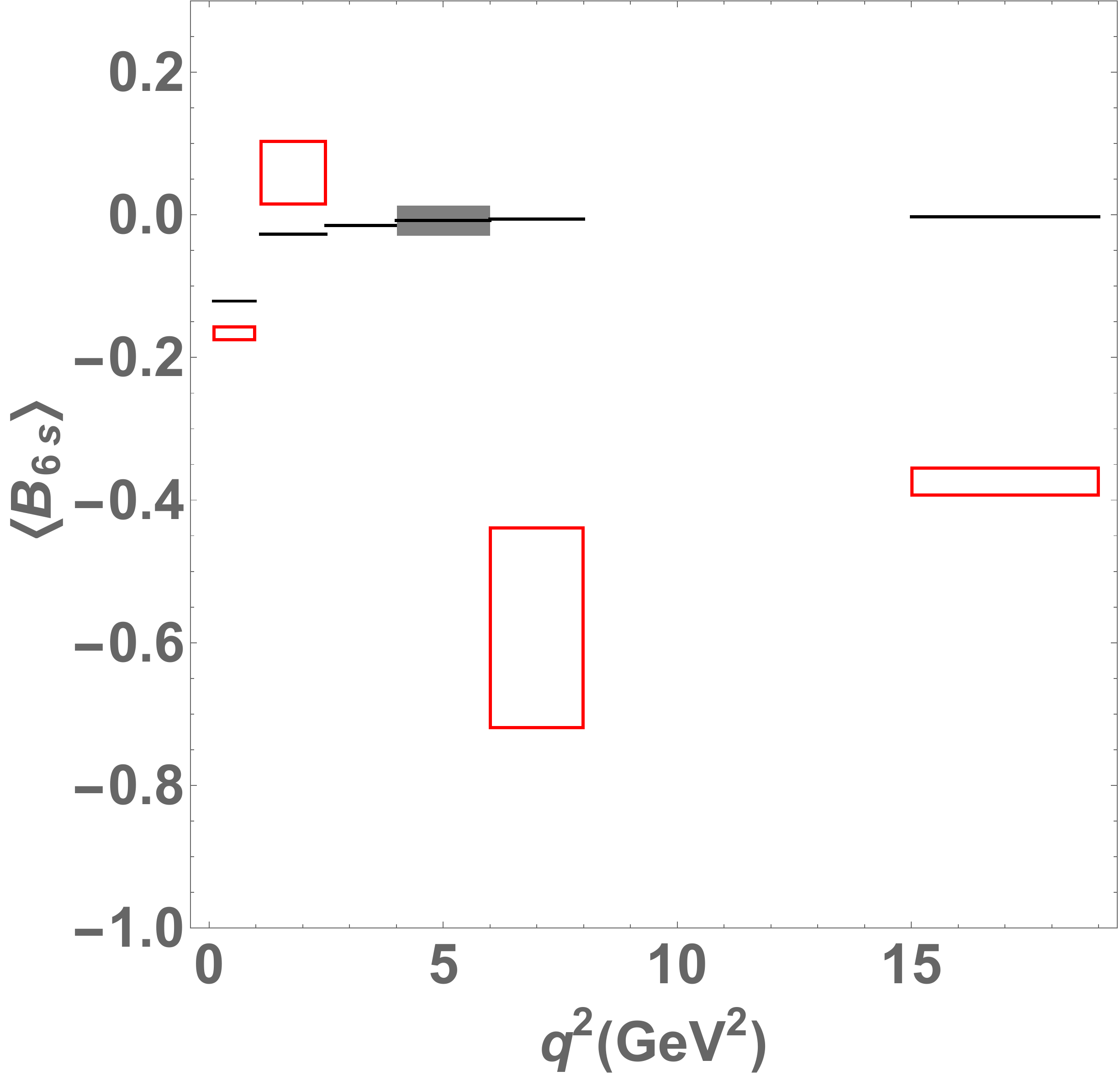}     
       \vspace{0.0cm}
         \caption{\textit{Scenario 4}.  SM predictions (grey boxes) and NP predictions (red boxes), assuming $C_{9\mu}^{\rm NP}=-C_{9'\mu}^{{\rm NP}}=-1.18$ and $C_{10\mu}^{\rm NP}=C_{10'\mu}^{{\rm NP}}=0.38$}       
  \label{fig:scenario4}
 \end{center}
\end{figure}

\begin{figure}
 \begin{center}
        \includegraphics[scale=0.28]{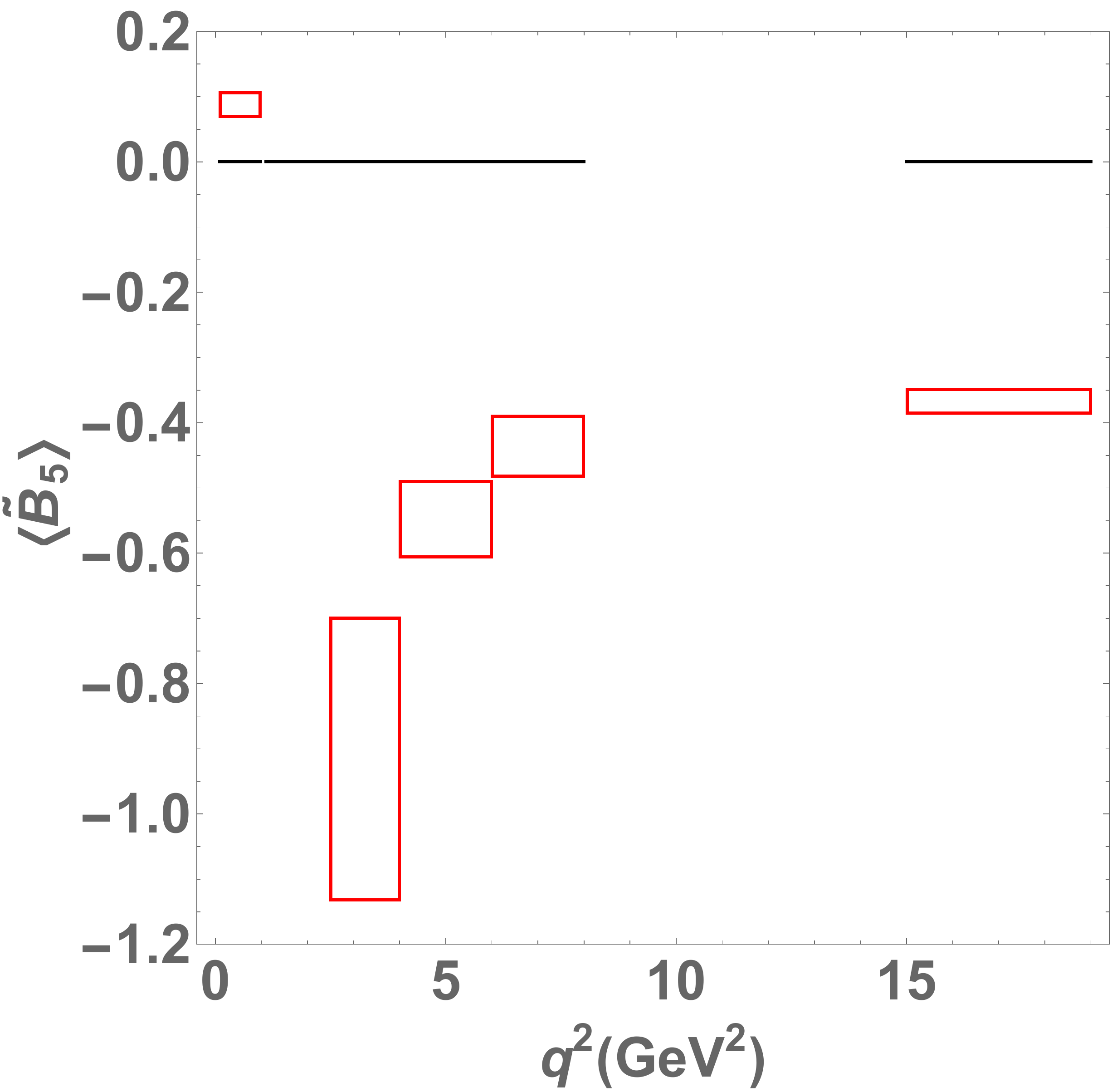}\hspace{7mm}
    \includegraphics[scale=0.28]{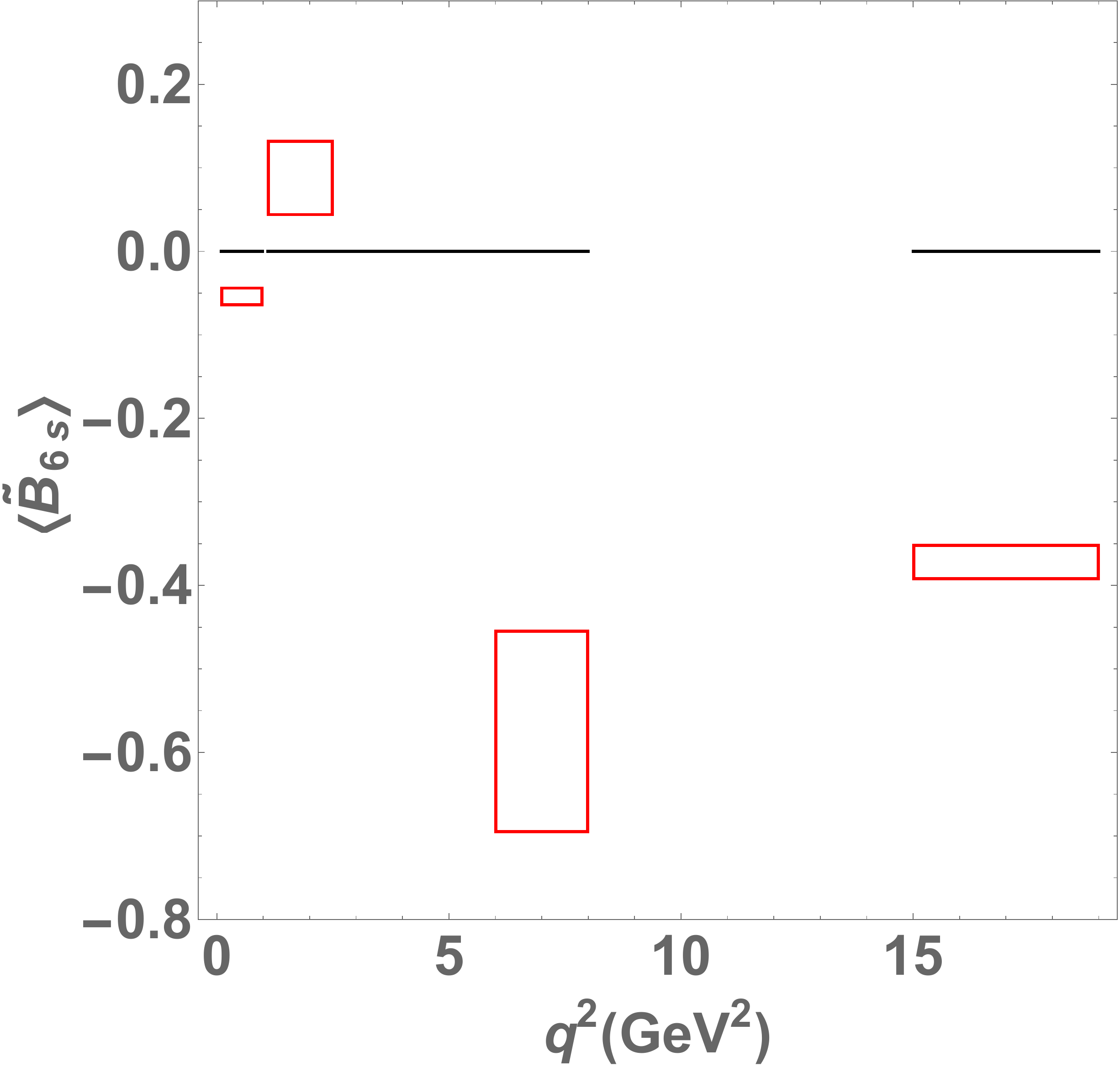} \\[10mm]
  \includegraphics[scale=0.28]{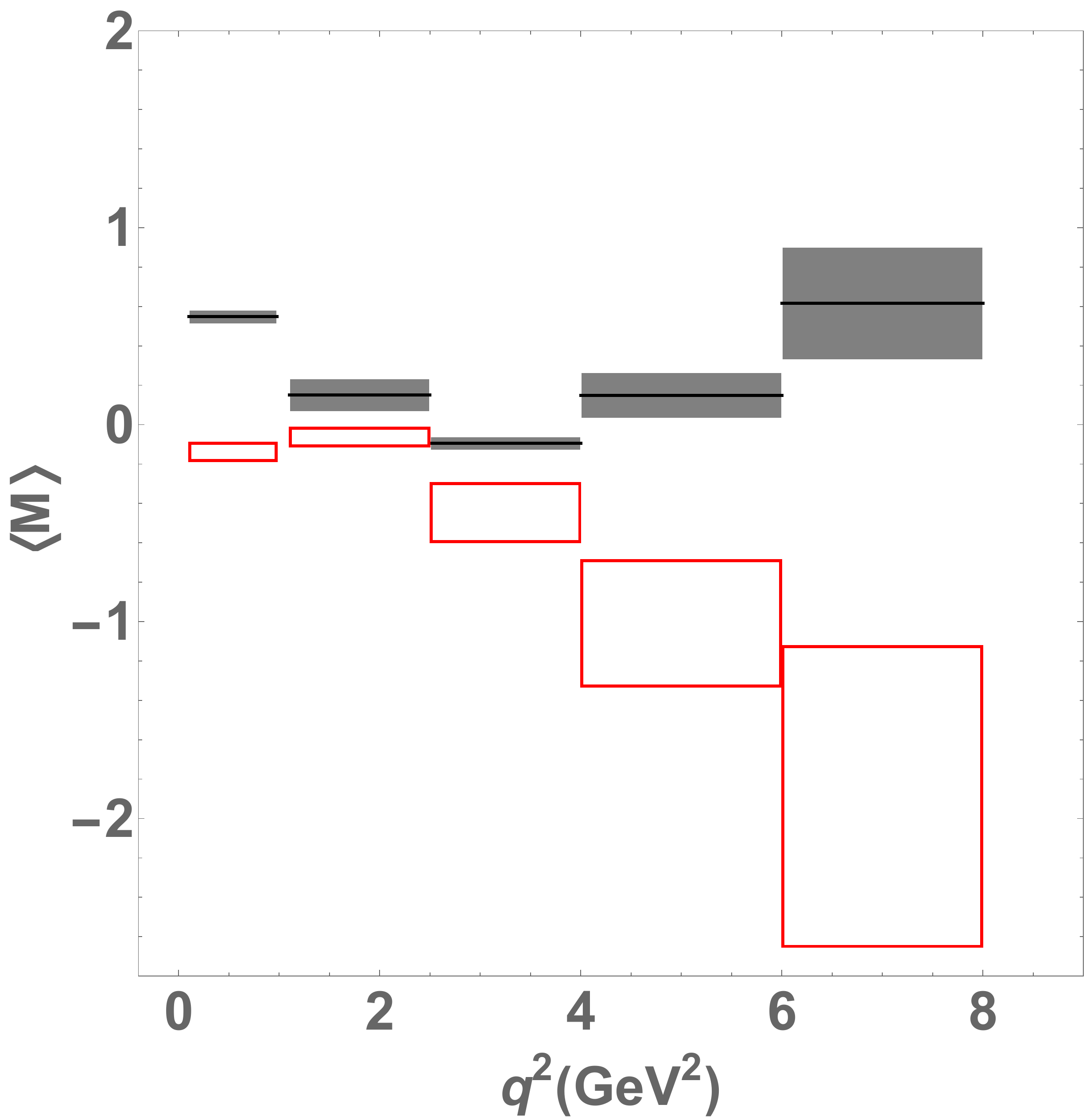}\hspace{7mm}
    \includegraphics[scale=0.29]{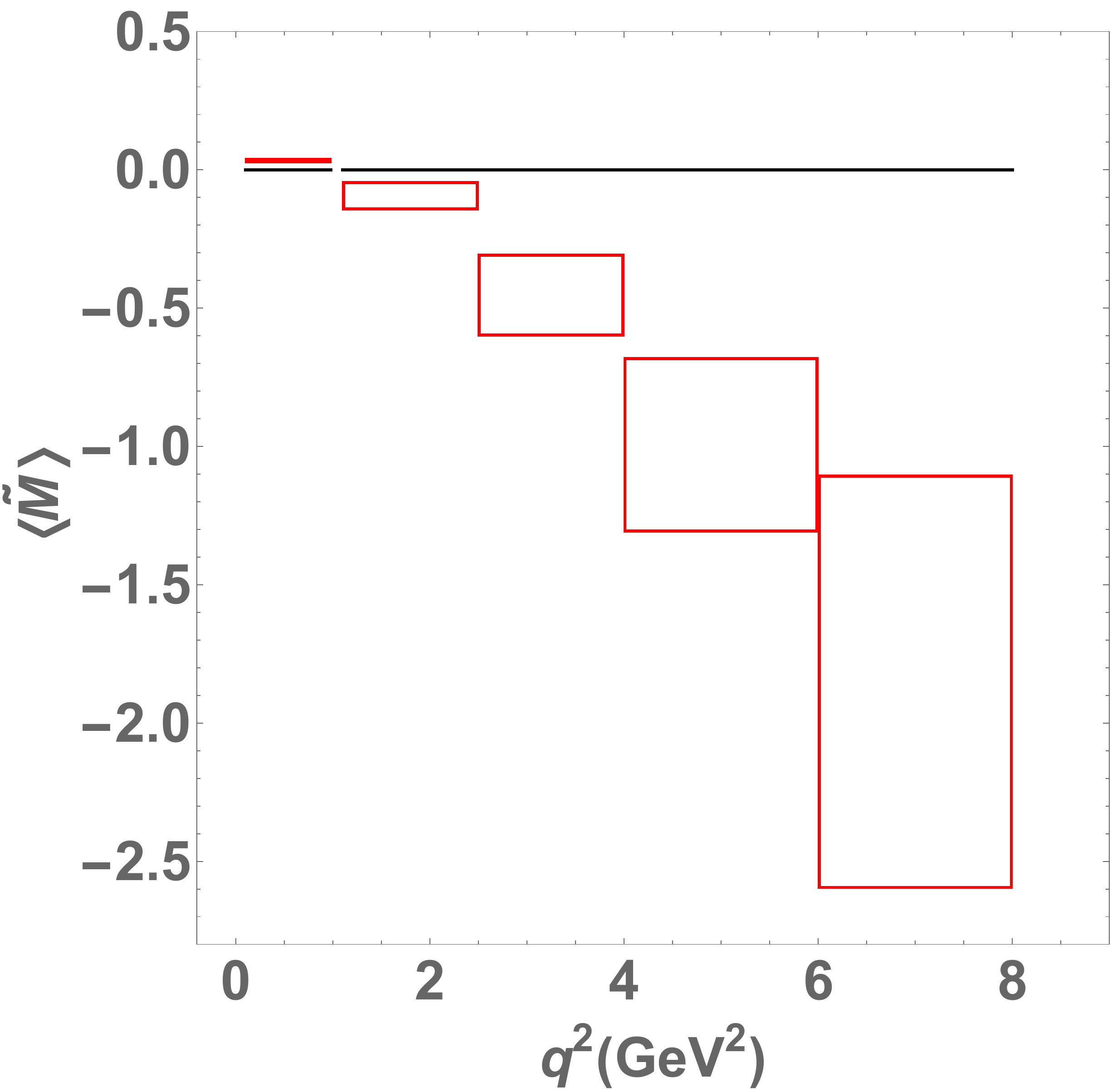}\bigskip
       \vspace{0.0cm}
         \caption{\textit{Scenario 4}.  SM predictions (grey boxes) and NP predictions (red boxes), assuming $C_{9\mu}^{\rm NP}=-C_{9'\mu}^{{\rm NP}}=-1.18$ and $C_{10\mu}^{\rm NP}=C_{10'\mu}^{{\rm NP}}=0.38$}       
  \label{fig:scenario4b}
 \end{center}
\end{figure}

In order to illustrate the interest of the various observables, in addition to the SM, we consider several NP benchmark scenarios corresponding to the best-fit points for hypotheses with a large pull in the global analysis of Ref.~\cite{Descotes-Genon:2015uva} (with NP contributions in $b\to s\mu\mu$ but not in $b\to s ee$). We follow the same approach as in Ref.~\cite{Descotes-Genon:2015uva} and compute the various observables following the definition of binned observables in App.~\ref{sec:binned}. The results are shown in App.~\ref{sec:pred} and in Figs.~\ref{fig:scenario1}-\ref{fig:scenario4b}.

In the SM, $Q_i$, $T_i$ and $B_i$ are expected to be close to zero, as shown in App.~\ref{sec:pred}.
The binned observables $B_5$ and $B_{6s}$ are actually different from zero due to the kinematic factors $\beta_\mu^2$ and $\beta_e^2$ in the transversity amplitudes -- one could imagine measuring the binned values of $J_{5,6s}^\ell/\beta_\ell^2$ and checking that the values for both lepton flavours are indeed identical. The difference between $\beta_\mu$ and $\beta_e$ becomes less relevant for large $q^2$ (above 2.5 GeV$^2$), leading to $B_5$ and $B_{6s}$ decreasing in magnitude and getting closer to each other.
In the same region, $M$  becomes larger as it involves the difference $B_5-B_{6s}$
in the denominator. In the presence of NP affecting differently $C_{9\mu}$ and $C_{9e}$, $B_5$ and $B_{6s}$ are different over the whole kinematic range. In the SM, the binned version of $M$ is charm dependent  due to $\beta_\mu/\beta_e$ terms. In the presence of LFNU in $C_9$, it is interesting to
focus instead on the observable $\widetilde M$, which is not affected by lepton-mass effects and is essentially charm independent at very low-$q^2$. If there are NP contributions in other Wilson coefficients, the situation becomes more complicated concerning the charm dependence of the observables. In the remainder of this Section we will identify patterns based on the set of $Q_i$ and $\hat Q_i$, and we will describe a very promising test based on $B_5$, $B_{6s}$ and $M$.

The observables $\hat{Q_i}$ (see Figs.~\ref{fig:scenario1}-\ref{fig:scenario4b}) show specific patterns for the different scenarios considered here:
\begin{itemize}

\item {\bf Scenario 1: $C_{9\mu}^{\rm NP}=-1.1$.} Both $\hat{Q_2}$ and $\hat{Q_5}$ are affected significantly,
especially the latter. The most interesting region is $q^2 \gtrsim 6 \text{GeV}^2$, taking
into account that these observables receive essentially no charm contributions in the SM. No deviation should be observed in $\hat{Q_1}$ or $\hat{Q_4}$ in the same region within this scenario (see the discussion in 
Section~\ref{sec:2} concerning the sensitivity of $P_4^\prime$ to $C_9$).

\item {\bf Scenario 2: $C_{9\mu}^{\rm NP}=-C_{10\mu}^{\rm NP}=-0.65$.} Within this scenario $\hat{Q_2}$ and $\hat{Q_5}$ show milder deviations, especially in the bin 6-8 GeV$^2$ where they are expected to be SM-like (contrary to Scenario 1). Indeed, the constraint from $B_s \to \mu\mu$ on $C_{10\mu}$  reduces the allowed size of the deviation in $C_{9\mu}$ in this particular scenario. On the contrary, $\hat{Q_4}$ could be particularly interesting in the region below 6 GeV$^2$ with a $q^2$-dependence rather different from Scenario 1. No deviation is expected in $\hat{Q_1}$.

\item {\bf Scenarios 3 and 4: $C_{9\mu}^{\rm NP}=-C_{9\mu}^{\prime}=-1.07$ and $C_{9\mu}^{\rm NP}=-C_{9\mu}^\prime=-1.18$, $C_{10\mu}^{\rm NP}=C_{10\mu}^\prime=0.38$ respectively.} Both scenarios are quite difficult to distinguish using these observables. They have implications in all four relevant observables $\hat{Q}_{1,2,4,5}$. The behaviour of $\hat{Q_2}$ and $\hat{Q_5}$ is similar to Scenario 1, making the three scenarios difficult to disentangle when looking only to these observables. $\hat{Q_1}$, which is designed to test the presence of right-handed currents, is affected significantly. Finally, $\hat{Q_4}$ both at very low- and large-$q^2$ (but within the large recoil region) could be useful if accurate measurements are obtained. In particular, above 6 GeV$^2$ this observable  is only sensitive to right-handed currents~\cite{inprep}.
\end{itemize}

The same discussion applies to the observables ${Q_i}$.
We note that $\hat Q_i$ ($Q_i$) in  the bin [6-8],  which have no charm uncertainties in the SM, may play a central role in disentangling the first two scenarios. 

These observables are quite complementary to $R_{K^*}$, for which we provide predictions in App.~\ref{sec:pred}. Indeed, the value of $R_{K^*}$ is very similar (within uncertainties) in the first two scenarios, whereas a larger suppression is expected for the other scenarios at moderately large $q^2$,
illustrating the complementarity with the $\hat{Q_i}$ ($Q_i$) observables.
For completeness we also present predictions for the observables $T_i$ in the same appendix. 

\subsection{$B$ and $\widetilde{B}$ observables}

We also give predictions for the $B_i$ observables in App.~\ref{sec:pred} and in
Figs.~\ref{fig:scenario1}-\ref{fig:scenario4b} within each scenario. In the plots we have not shown the
predictions in the bins where $B_5$ or $B_{6s}$ have a pole ([1.1,2.5] for $B_5$, [2.5,4] and [4,6] for
$B_{6s}$) and cannot be predicted accurately. All scenarios give very similar predictions, apart from the
first bin of $B_5$ and the two first bins of $B_{6s}$.

The first bin of these observables is predicted accurately both in the SM and in the presence of NP. 
Not only it is  insensitive to form factors in the large-recoil limit at leading order,
but it is also protected from long-distance charm contributions due to a kinematical suppression
of the charm-dependent contribution at low~$q^2$ (see also Ref.~\cite{inprep}). 
The analysis of this bin in the SM and in the scenarios presented above is particularly interesting.
As explained in the previous section, the SM predictions $B_5^{\rm SM}=-0.155 \pm 0.003$ and
$B_{6s}^{\rm SM}=-0.121 \pm 0.001$ are only different from zero due to $\beta_\mu/\beta_e$ effects integrated
over the bin. This can be checked through the corresponding prediction for the $\widetilde B_i$ observables,
which are free from these effects and equal to zero in the SM. In the case of a negative NP contribution to
$C_{9\mu}$, both $B_5$ and $B_{6s}$ receive a positive contribution that pushes them towards zero in the first
bin. If there is a  positive NP contribution in $C_{10\mu}$, the contribution to both observables is negative
and large (of size  $C_{10\mu}^{\rm NP}/C_{10\mu}$).
In summary, a contribution close to zero will favour a scenario with NP only in $C_{9\mu}<0$,
whereas values of $B_5$ and $B_{6s}$ lower than the SM will signal NP in $C_{10\mu}$
(NP in $C_{9\mu}$ is better discriminated by other observables). In both cases $B_5$ and $B_{6s}$ are almost equal,
while a contribution to $C_{10\mu}^\prime$ would break this degeneracy.
The second bin of $B_{6s}$ exhibits a similar pattern (above the SM in Scenario~1, below in Scenario~2). 

The same discussion applies to $\widetilde{B}_i$, which have a similar behaviour in those bins,
the only difference being that they are centered around zero (SM prediction). For instance, the first bin of
$\widetilde B_5$ and $\widetilde B_{6s}$ in the Scenario 1 (Scenario 2) receives a positive (negative)
contribution. The second bin of $\widetilde B_{6s}$ follows the same rules as $B_{6s}$.

The low-recoil behaviour of the $B_i$ and $\widetilde B_i$ observables is particularly interesting because it points to large deviations that cannot be seen easily in the $Q_i$ observables. Unfortunately, they are not useful in distinguishing Scenarios~1 and~2, except if compared together with the corresponding $Q_i$ at low recoil, which show a slightly different behaviour in that region.

\subsection{$M$  and $\widetilde{M}$ observables}

$M$ is also an interesting observable to get information on the existence of NP contributions and
identifying their nature. This can be seen from the results in App.~C and Figs.~\ref{fig:scenario1}-\ref{fig:scenario4b}
by looking at the third bin, where it can be noted that this observable can help to disentangle Scenario 2 from Scenarios 1 and 3,
thus testing for the presence of NP in $C_{10\mu}$.

However in the first bin, where $B_5 \simeq B_{6s}$, $M$ is poorly predicted.
In these region it proves instead very useful to exploit the alternative  observable $\widetilde{M}$, where effects related to
$\beta_\ell$ are removed. This observable then gives additional information in discerning between Scenario 2 and Scenarios 1 and 3.
The effects in this first bin can also be confirmed by looking at the second bin (notice that $\widetilde{M}$
is well defined in its second bin even if $\widetilde{B_5}$ has a pole  in its second bin).

\subsection{Hadronic uncertainties}

The observables presented  here, specially $Q_i$, $B_i$ and $\widetilde{B}_i$, are built to be very accurate
in the SM, and almost insensitive to long-distance charm contributions. Moreover, whether NP is present or not,
these observables are built to have no dependence on soft form factors at leading order in the large-recoil
limit. In the presence of NP, these observables become again sensitive to charm-loop contributions,
but in a very specific way that we discuss now.

Let us first recall that we introduced the observables $\hat{Q}_i$ in order to provide predictions taking
into account how LHC measures $F_L$ currently.
Here the cancellation of soft form factors between numerator and denominator is not fully operative and
these observables are thus sensitive to soft form factors arising in $J_{1c}$ but suppressed by powers of
$m_\ell^2/q^2$.
This explains why the errors of $\hat{Q}_i$ are larger (but still small in most of the bins) than for~$Q_i$.
The observables $T_i$ exhibit a residual sensitivity to soft form factors in most of the bins. 
Finally, the observable $M$ suffers from large uncertainties when $B_5\simeq B_{6s}$, even though
it is designed to have no dependence on soft form factors at leading order in the large-recoil limit.

Concerning long-distance charm-loop contributions, the most interesting observables are $B_i$ ($\widetilde B_i$)
and $M$ ($\widetilde{M}$). In the analytic expressions provided in Section~\ref{sec:obsBM}, we have assumed
that the charm contribution $\Delta C_9$ entered all transversity amplitudes in the same way.
One can generalize the expressions for $B_{5,6s}$ and $\widetilde M$ in Eqs.~(\ref{obsi1},\ref{obsi2}) and
allow for transversity-dependent charm contributions $\Delta C_9^{\perp,\|,0}(q^2)$ associated to each
amplitude:

\begin{equation}
B_5=\dfrac{\beta_\mu^2-\beta_e^2}{\beta_e^2}+\dfrac{\beta_\mu^2}{\beta_e^2}\dfrac{\delta C_{10}}{C_{10}}+\dfrac{\beta_\mu^2}{\beta_e^2}\dfrac{2(C_{10}+\delta C_{10})\delta C_9\hat{s}}{C_{10}\left(2C_7\hat{m}_b(1+\hat{s})+(2C_9+\Delta C_{9,0}+\Delta C_{9,\perp})\hat{s}\right)}
\end{equation}

\begin{equation}
B_{6s}=\dfrac{\beta_\mu^2-\beta_e^2}{\beta_e^2}+\dfrac{\beta_\mu^2}{\beta_e^2}\dfrac{\delta C_{10}}{C_{10}}+\dfrac{\beta_\mu^2}{\beta_e^2}\dfrac{2(C_{10}+\delta C_{10})\delta C_9\hat{s}}{C_{10}\left(4C_7\hat{m}_b+(2C_9+\Delta C_{9,\perp}+\Delta C_{9,\parallel})\hat{s}\right)}
\end{equation}

\begin{eqnarray}
\label{Mtilde}
\widetilde{M}=&\dfrac{\left(2C_{10}\delta C_9\hat{s}+\delta C_{10}\left(2C_7\hat{m}_b(1+\hat{s})+(2C_9+2\delta C_9+\Delta C_{9,\perp}+\Delta C_{9,0})\hat{s}\right)\right)}{2C_{10}(C_{10}+\delta C_{10})\delta C_9\left(2C_7\hat{m}_b(\hat{s}-1)+(\Delta C_{9,0}-\Delta C_{9,\parallel})\hat{s}\right)\hat{s}}\nonumber \\
&\times \left(2C_{10}\delta C_9\hat{s}+\delta C_{10}\left(4C_7\hat{m}_b+(2C_9+2\delta C_9+\Delta C_{9,\perp}+\Delta C_{9,\parallel})\hat{s}\right)\right)
\end{eqnarray}

The corresponding expressions for the $\widetilde{B}_{5,6s}$ are obtained in the limit $\beta_\ell \to 1$.
In the case of NP only in $\delta C_{9}$ they simplify to
\begin{eqnarray} \label{limiteq}
B_5&=&\frac{\beta_\mu^2-\beta_e^2}{\beta_e^2} +\frac{\beta_\mu^2}{\beta_e^2} \frac{ \delta C_{9} \hat{s}}{ (C_{7} \hat{m}_b (1+\hat{s})+ ( C_{9}+ (\Delta C_9^0+\Delta C_9^\perp)/2) \hat{s})}+\ldots \\
B_{6s}&=&\frac{\beta_\mu^2-\beta_e^2}{\beta_e^2}+ \frac{\beta_\mu^2}{\beta_e^2} \frac{ \delta C_{9} \hat{s}}{ (2 C_{7} \hat{m}_b +  (C_{9}+ (\Delta C_9^{\|}+\Delta C_9^\perp)/2) \hat{s})}+\ldots \\
\widetilde{M}&=&-\frac{\delta C_9 \hat{s}}{C_7 \hat{m}_b (1-\hat{s}) - (\Delta C_9^0 - \Delta C_9^\|) \hat{s} /2}+\ldots
\end{eqnarray}
The observable $\widetilde{M}$ was designed to cancel exactly a transversity-independent charm contribution
$\Delta C_9$ at leading order in the large recoil limit, which occurs in the denominator of the $B_i$
observables. The above expressions indicate that for $B_i$, all the long-distance charm dependence is contained
in the denominator, and its numerical impact is somehow reduced by a large $C_9$, which explains their reduced
sensitivity to $\Delta C_9$ (this is even more efficient at very low $q^2$ due to the photon pole).
In the case of $\widetilde{M}$, $C_9$ cancels, leaving only the photon pole to tame the sensitivity to
transversity-dependent charm-loop contributions. For this reason at higher $q^2$ values, where the photon
pole contribution is smaller, the sensitivity to this transversity-dependent charm contribution is maximal
in $\widetilde{M}$ as can be seen in App.~\ref{sec:pred} and in Figs.~\ref{fig:scenario1}-\ref{fig:scenario4b}.
In addition, looking at Eq.~(\ref{Mtilde}), it is interesting to note that $\widetilde{M}$ is sensitive to charm contributions only if \emph{a)}
there is LFNU New Physics in $C_{10}$ or right-handed operators, or \emph{b)} there are transversity-dependent charm-loop
contributions (such that $\Delta C_9^0 \ne \Delta C_9^\|$).

We should finally comment on the fact that our predictions do not include any evaluation of Bremsstrahlung
effects.  Naively one expects these effects to be of order
$\alpha\log(m_e^2/m_\mu^2)\sim 8$\%~\cite{Boucenna:2016wpr}. Part of these effects are taken into account
at the level of the experimental analysis by means of a Montecarlo simulation with PHOTOS~\cite{Barberio:1993qi},
which accounts for soft-photon emission from the leptons.
Other contributions (e.g., real emission from the mesons, virtual photons)
should still be estimated by separating in the theoretical computations the radiative corrections already
implemented experimentally and those to be estimated theoretically
(see Refs.~\cite{Stoffer:2013sfa,Bernard:2015vqa} for a discussion of this issue in the context of
$K_{\ell4}$ decays). Such a work goes far beyond the present note, but the impact of such effects should be
expected to be of a few percent.

\section{Discussion and conclusion}

The recent LHCb and Belle results on $b\to s\ell\ell$ transitions, with the anomalies observed in some
angular observables such as $P_5'(B\to K^*\mu\mu)$, and the hints of LFNU in $B\to K\ell\ell$ have raised
a considerable interest for these processes.
In the present article we have discussed how angular analyses of $B\to K^*ee$ and $B\to K^*\mu\mu$
decay modes can be combined to understand better the pattern of anomalies observed
and to get a solid handle on the size of some SM long-distance contributions.

We have proposed different sets of observables comparing $B\to K^*ee$ and $B\to K^*\mu\mu$,
discussing their respective merits. A first set of observables is obtained directly from the observables
that have been introduced for $B\to K^*\mu\mu$, namely $Q_i$ (related to the optimised observables $P_i$),
$T_i$ (related to the angular averages $S_i$) and $B_i$ (related to the angular coefficients $J_i$), measuring
in each case the differences between muon and electron modes.

We have discussed further the merits of the observables $B_5$ and $B_{6s}$ which are built from angular
coefficients exhibiting only a linear dependence on $C_{9\ell}$ at large recoil. In principle, this allows
us to disentangle the contributions coming from NP in $C_9$ and $C_{10}$, with a clean separation between
lepton-flavour dependent (NP) and lepton-flavour universal (NP or SM long-distance) contributions to $C_9$.
We have also built an observable~$\widetilde M$ which exhibits very interesting features: in the presence of
LFNU NP in $C_{9\ell}$ or $C_{10\ell}$ only, the large-recoil expression for $\widetilde M$ is independent of
long-distance LFU contributions (in particular transversity-independent charm contributions) and provides clean signals of NP. It proves also interesting to consider
$\widetilde{B}_5$ and  $\widetilde{B}_{6s}$, built from angular coefficients divided by appropriate powers
of $\beta_\ell$, thus removing some kinematic effects affecting $B_5$ and $B_{6s}$ at very low $q^2$.

We have then considered the situation for binned observables, and we have provided predictions for the SM and
for several benchmark points inspired by our recent global analysis of $b\to s\ell\ell$ transitions. We can
summarise our findings as follows. First, the $Q_i$ observables are efficient to separate several NP scenarios
where NP enter only $b\to s\mu\mu$ transitions due to very different $q^2$ dependences in the large-recoil
region. Second, the observables $B_{5,6s}$ and $\widetilde{B}_{5,6s}$ at very large and low recoils provide
further information, as NP in different muon Wilson coefficients will affect these observables significantly.
Finally, the  $\widetilde{M}$ observable at low $q^2$ proves particularly clean and efficient in identifying
and interpreting NP in muon modes, with a limited sensitivity to charm contributions.
These observables provide complementary information compared to the measurement of the ratio $R_{K^*}$ that
is expected very soon from the LHCb collaboration.

In view of these results, we are looking forward to the next measurements to be performed at LHCb and Belle-II.
We expect their analysis to be decisive in determining the exact origin of the anomalies currently observed
in $b\to s\ell\ell$ modes.

\section*{Acknowledgements}

We would like to thank Lars Hofer and Nicola Serra for interesting discussions and comments,
as well as all the participants of the Workshop ``Rare $B$ decays 2016 - Theory and Experiment"
in Barcelona for stimulating conversations on these topics.
SDG, JM and JV acknowledge financial support from Explora project FPA2014-61478-EXP. BC has been supported
by FPA2011-25948 and the grant 2014 SGR 1450, and in part by the Centro de Excelencia Severo Ochoa SEV-2012-0234. This project has received support from the European Union's Horizon 2020 research and innovation programme under the Marie Sklodowska-Curie grant agreements No 690575, No 674896 and No. 692194.  JV is funded by the DFG within research unit FOR 1873 (QFET).
 
\newpage 
 
\appendix

\section{Large-recoil expressions for $M$ and $\widetilde{M}$}\label{sec:mobservable}

Under the notation and hypotheses in Section~\ref{sec:obsBM}, we can separate the charm contributions
from the rest of the $\widetilde{M}$ observable
\begin{equation}
\widetilde{M}=\widetilde{M}_0+\mathcal{A}'\delta C_{10}\Delta C_9+\mathcal{B}'\delta C_{10}^2\Delta C_9^2
\end{equation}
with
\begin{eqnarray}
\widetilde{M}_0&=&\frac{\left(2C_7\delta C_{10}\hat{m}_b+\delta C_9C_{10}\hat{s}+\delta C_{10}( C_9+\delta C_9)\hat{s}\right)}{C_7\delta C_9C_{10}(C_{10}+\delta C_{10})\hat{m}_b(\hat{s}-1)\hat{s}}\\
&&\qquad\qquad \times \left(C_7\delta C_{10}\hat{m}_b+\delta C_9C_{10}\hat{s}+\delta C_{10}(C_7\hat{m}_b+C_9+\delta C_9)\hat{s}\right)\nonumber\\
\mathcal{A}'&=&\frac{2\delta C_9C_{10}\hat{s}+\delta C_{10}\left(2(C_9+\delta C_9)\hat{s}+C_7\hat{m}_b(3+\hat{s})\right)}{C_7\delta C_9C_{10}(C_{10}+\delta C_{10})\hat{m}_b(\hat{s}-1)}\\
\mathcal{B}'&=&\frac{\hat{s}}{C_7\delta C_9C_{10}(C_{10}+\delta C_{10})\hat{m}_b(\hat{s}-1)}
\end{eqnarray}

$M$ can be expressed in terms of $\widetilde{M}$ and considering all the lepton mass effects coming from
$\beta_\ell=\sqrt{1-4m_\ell^2/s}$ in the large recoil limit and up to leading order
\begin{equation}
M=\widetilde{M}+\Delta M+\mathcal{A}\Delta C_9+\mathcal{B}\Delta C_9^2
\end{equation}
\begin{eqnarray}
\Delta M&=&-\frac{\beta_e^2-\beta_\mu^2}{\beta_e^2\beta_\mu^2}\frac{1}{C_7\delta C_9C_{10}(C_{10}+\delta C_{10})\hat{m}_b(\hat{s}-1) \hat{s}} \nonumber\\
&& \times\Big[-C_{10}^2(2C_7\hat{m}_b+C_9\hat{s})(C_9\hat{s}+C_7\hat{m}_b(1+\hat{s}))\beta_e^2\\
&&+(C_{10}+\delta C_{10})^2\left(2C_7\hat{m}_b+(C_9+\delta C_9)\hat{s}\right)\left((C_9+\delta C_9)\hat{s}+C_7\hat{m}_b(1+\hat{s})\right)\beta_\mu^2\Big] \nonumber\\
\mathcal{A}&=&\frac{\beta_e^2-\beta_\mu^2}{\beta_e^2\beta_\mu^2}\frac{1}{C_7\delta C_9C_{10}(C_{10}+\delta C_{10})\hat{m}_b(\hat{s}-1)}\\
&& \times \left[C_{10}^2\left(2C_9\hat{s}+C_7\hat{m}_b(3+\hat{s})\right)\beta_e^2-(C_{10}+\delta C_{10})^2\left(2(C_9+\delta C_9)\hat{s}+C_7\hat{m}_b(3+\hat{s})\right)\beta_\mu^2\right]\nonumber\\
\mathcal{B}&=&\dfrac{\beta_e^2-\beta_\mu^2}{\beta_e^2\beta_\mu^2}\dfrac{\hat{s}\left(C_{10}^2\beta_e^2-(C_{10}+\delta C_{10})^2\beta_\mu^2\right)}{C_7\delta C_9C_{10}(C_{10}+\delta C_{10})\hat{m}_b(\hat{s}-1)}
\end{eqnarray}

\newpage

\section{Definition of binned observables}\label{sec:binned}

The binned observables are defined following the same rules as in Ref.~\cite{Descotes-Genon:2013vna}:
\begin{eqnarray} \langle Q_i\rangle &=&\langle P_i^{\mu} \rangle - \langle P_i^{e} \rangle \qquad \langle \hat{Q}_i\rangle =\langle \hat{P}_i^{\mu} \rangle - \langle \hat{P}_i^{e} \rangle \qquad
\langle T_i \rangle = \frac{\langle S_i^{\mu} \rangle - \langle S_i^{e} \rangle}{\langle S_i^{\mu} \rangle + \langle S_i^{e} \rangle}\\
\langle B_i \rangle &=& \frac{\langle J_i^{\mu} \rangle}{\langle J_i^{e} \rangle}-1 \qquad
 \langle\widetilde{B}_i\rangle = 
 \frac{\langle J_i^{\mu}/\beta_\mu^2 \rangle}{\langle J_i^{e}/\beta_e^2 \rangle}-1\\
\langle M\rangle &=& \frac{\left(\langle J_5^{\mu} \rangle -\langle J_5^{e} \rangle \right)\left(\langle J_{6s}^{\mu} \rangle -\langle J_{6s}^{e} \rangle \right)}{\langle J_{6s}^{\mu} \rangle \langle J_5^{e} \rangle   -\langle J_{6s}^{e} \rangle\langle J_5^{\mu} \rangle}
\\ \langle \widetilde{M}\rangle &=& \frac{\left(\langle J_5^{\mu}/\beta_\mu^2 \rangle -\langle J_5^{e}/\beta_e^2 \rangle \right)\left(\langle J_{6s}^{\mu}/\beta_\mu^2 \rangle -\langle J_{6s}^{e}/\beta_e^2 \rangle \right)}{\langle J_{6s}^{\mu}/\beta_\mu^2 \rangle \langle J_5^{e}/\beta_e^2 \rangle   -\langle J_{6s}^{e}/\beta_e^2 \rangle\langle J_5^{\mu}/\beta_\mu^2 \rangle}
\end{eqnarray}
where $\langle P_i^{\ell} \rangle$ and $\langle S_i^{\ell} \rangle$ correspond to the observables defined in
Ref.~\cite{Descotes-Genon:2013vna} with $\ell=e$ or $\mu$. Similarly, the $\langle \hat{P}_i ^{\ell} \rangle$
are obtained from Eqs.~(\ref{eq:hatstart})-(\ref{eq:hatend}), substituting
$J_i^{\ell} \to \langle J_i^{\ell} \rangle$.

\section{Predictions for the observables in various benchmark scenarios}\label{sec:pred}

Our predictions are obtained following Ref.~\cite{Descotes-Genon:2015uva}. We quote two uncertainties,
the second corresponding to the charm contributions, the first to all other sources of uncertainties. Bars denote predictions affected by a very large uncertainty (presence of a pole).

\footnotesize

\subsection{SM}

\begin{center}

{
\scriptsize
\begin{longtable}{@{}lrrrr@{}}
\toprule[1.6pt] 
 \hspace{5mm} Bin & $Q_{F_L}$\hspace{10mm} & $Q_1$\hspace{10mm} & $Q_2$\hspace{10mm} & $Q_3$\hspace{10mm} \\ 
 \midrule 
 $ [0.1,0.98] $ & $ -0.041 \pm 0.044 \pm 0.010 $ & $ -0.001 \pm 0.001 \pm 0.001 $ & $ 0.019 \pm 0.003 \pm 0.001 $ & $ 0.000 \pm 0.000 \pm 0.000 $ \\ 
 $ [1.1,2.5] $ & $ -0.027 \pm 0.014 \pm 0.001 $ & $ -0.000 \pm 0.000 \pm 0.000 $ & $ 0.007 \pm 0.000 \pm 0.000 $ & $ 0.000 \pm 0.000 \pm 0.000 $ \\ 
 $ [2.5,4.] $ & $ -0.016 \pm 0.009 \pm 0.000 $ & $ 0.000 \pm 0.000 \pm 0.000 $ & $ 0.001 \pm 0.001 \pm 0.000 $ & $ 0.000 \pm 0.000 \pm 0.000 $ \\ 
 $ [4.,6.] $ & $ -0.010 \pm 0.008 \pm 0.000 $ & $ 0.000 \pm 0.000 \pm 0.000 $ & $ -0.001 \pm 0.000 \pm 0.000 $ & $ 0.000 \pm 0.000 \pm 0.000 $ \\ 
 $ [6.,8.] $ & $ -0.006 \pm 0.006 \pm 0.000 $ & $ 0.000 \pm 0.000 \pm 0.000 $ & $ -0.001 \pm 0.000 \pm 0.000 $ & $ 0.000 \pm 0.000 \pm 0.000 $ \\ 
 $ [15.,19.] $ & $ -0.001 \pm 0.000 \pm 0.000 $ & $ -0.000 \pm 0.000 \pm 0.000 $ & $ -0.000 \pm 0.000 \pm 0.000 $ & $ 0.000 \pm 0.000 \pm 0.000 $ \\ 
\midrule[1.6pt] 
 \hspace{5mm} Bin & $Q_4$\hspace{10mm} & $Q_5$\hspace{10mm} & $Q_6$\hspace{10mm} & $Q_8$\hspace{10mm} \\ 
 \midrule 
 $ [0.1,0.98] $ & $ 0.005 \pm 0.002 \pm 0.004 $ & $ 0.047 \pm 0.003 \pm 0.008 $ & $ -0.005 \pm 0.002 \pm 0.001 $ & $ 0.001 \pm 0.000 \pm 0.000 $ \\ 
 $ [1.1,2.5] $ & $ 0.002 \pm 0.000 \pm 0.000 $ & $ 0.001 \pm 0.002 \pm 0.001 $ & $ -0.001 \pm 0.000 \pm 0.000 $ & $ 0.000 \pm 0.000 \pm 0.000 $ \\ 
 $ [2.5,4.] $ & $ 0.000 \pm 0.000 \pm 0.000 $ & $ -0.004 \pm 0.001 \pm 0.000 $ & $ -0.000 \pm 0.000 \pm 0.000 $ & $ -0.000 \pm 0.000 \pm 0.000 $ \\ 
 $ [4.,6.] $ & $ 0.000 \pm 0.000 \pm 0.000 $ & $ -0.004 \pm 0.000 \pm 0.000 $ & $ -0.000 \pm 0.000 \pm 0.000 $ & $ 0.000 \pm 0.000 \pm 0.000 $ \\ 
 $ [6.,8.] $ & $ 0.000 \pm 0.000 \pm 0.000 $ & $ -0.003 \pm 0.000 \pm 0.000 $ & $ -0.000 \pm 0.000 \pm 0.000 $ & $ 0.000 \pm 0.000 \pm 0.000 $ \\ 
 $ [15.,19.] $ & $ 0.000 \pm 0.000 \pm 0.000 $ & $ -0.001 \pm 0.000 \pm 0.000 $ & $ 0.000 \pm 0.000 \pm 0.000 $ & $ 0.000 \pm 0.000 \pm 0.000 $ \\ 
\bottomrule[1.6pt] 
\end{longtable}

\newpage

\begin{longtable}{@{}lrrrr@{}}
\toprule[1.6pt] 
 \hspace{5mm} Bin & $\hat Q_{F_L}$\hspace{10mm} & $\hat Q_1$\hspace{10mm} & $\hat Q_2$\hspace{10mm} & $\hat Q_3$\hspace{10mm} \\ 
 \midrule 
 $ [0.1,0.98] $ & $ 0.018 \pm 0.017 \pm 0.004 $ & $ -0.007 \pm 0.006 \pm 0.018 $ & $ -0.008 \pm 0.004 \pm 0.001 $ & $ 0.000 \pm 0.001 \pm 0.001 $ \\ 
 $ [1.1,2.5] $ & $ 0.014 \pm 0.002 \pm 0.000 $ & $ -0.000 \pm 0.003 \pm 0.000 $ & $ 0.013 \pm 0.032 \pm 0.002 $ & $ 0.000 \pm 0.000 \pm 0.000 $ \\ 
 $ [2.5,4.] $ & $ 0.010 \pm 0.002 \pm 0.000 $ & $ 0.000 \pm 0.003 \pm 0.000 $ & $ 0.010 \pm 0.025 \pm 0.001 $ & $ 0.000 \pm 0.001 \pm 0.000 $ \\ 
 $ [4.,6.] $ & $ 0.008 \pm 0.001 \pm 0.000 $ & $ 0.001 \pm 0.006 \pm 0.000 $ & $ -0.004 \pm 0.005 \pm 0.000 $ & $ 0.000 \pm 0.000 \pm 0.000 $ \\ 
 $ [6.,8.] $ & $ 0.006 \pm 0.002 \pm 0.000 $ & $ 0.000 \pm 0.003 \pm 0.000 $ & $ -0.004 \pm 0.007 \pm 0.000 $ & $ 0.000 \pm 0.000 \pm 0.000 $ \\ 
 $ [15.,19.] $ & $ 0.001 \pm 0.000 \pm 0.000 $ & $ 0.001 \pm 0.000 \pm 0.000 $ & $ -0.000 \pm 0.000 \pm 0.000 $ & $ 0.000 \pm 0.000 \pm 0.000 $ \\ 
\midrule[1.6pt] 
 \hspace{5mm} Bin & $\hat Q_4$\hspace{10mm} & $\hat Q_5$\hspace{10mm} & $\hat Q_6$\hspace{10mm} & $\hat Q_8$\hspace{10mm} \\ 
 \midrule 
 $ [0.1,0.98] $ & $ 0.111 \pm 0.007 \pm 0.037 $ & $ -0.097 \pm 0.013 \pm 0.019 $ & $ 0.008 \pm 0.003 \pm 0.001 $ & $ -0.004 \pm 0.004 \pm 0.003 $ \\ 
 $ [1.1,2.5] $ & $ 0.003 \pm 0.005 \pm 0.002 $ & $ -0.003 \pm 0.007 \pm 0.001 $ & $ 0.001 \pm 0.003 \pm 0.000 $ & $ -0.001 \pm 0.002 \pm 0.000 $ \\ 
 $ [2.5,4.] $ & $ 0.001 \pm 0.016 \pm 0.001 $ & $ -0.005 \pm 0.017 \pm 0.001 $ & $ -0.001 \pm 0.003 \pm 0.000 $ & $ 0.000 \pm 0.002 \pm 0.000 $ \\ 
 $ [4.,6.] $ & $ -0.002 \pm 0.015 \pm 0.000 $ & $ -0.002 \pm 0.017 \pm 0.000 $ & $ -0.000 \pm 0.001 \pm 0.000 $ & $ -0.000 \pm 0.001 \pm 0.000 $ \\ 
 $ [6.,8.] $ & $ -0.005 \pm 0.009 \pm 0.001 $ & $ 0.002 \pm 0.010 \pm 0.000 $ & $ 0.000 \pm 0.000 \pm 0.000 $ & $ -0.000 \pm 0.000 \pm 0.000 $ \\ 
 $ [15.,19.] $ & $ -0.003 \pm 0.000 \pm 0.000 $ & $ 0.001 \pm 0.000 \pm 0.000 $ & $ 0.000 \pm 0.000 \pm 0.000 $ & $ 0.000 \pm 0.000 \pm 0.000 $ \\ 
\bottomrule[1.6pt] 
\end{longtable}
}

\begin{longtable}{@{}lrrr@{}}
\toprule[1.6pt] 
 \hspace{5mm} Bin & $T_3$\hspace{14mm} & $T_4$\hspace{14mm} & $T_5$\hspace{14mm}  \\ 
 \midrule 
 $ [0.1,0.98] $ & $--$\hspace{14mm} & $ -0.116 \pm 0.002 \pm 0.005 $ & $ -0.075 \pm 0.003 \pm 0.001 $ \\ 
 $ [1.1,2.5] $ & $--$\hspace{14mm} & $--$\hspace{14mm} & $ -0.017 \pm 0.004 \pm 0.001 $ \\ 
 $ [2.5,4.] $ & $--$\hspace{14mm} & $ -0.010 \pm 0.003 \pm 0.000 $ & $ -0.006 \pm 0.003 \pm 0.000 $ \\ 
 $ [4.,6.] $ & $ -0.007 \pm 0.006 \pm 0.000 $ & $ -0.007 \pm 0.003 \pm 0.000 $ & $ -0.004 \pm 0.003 \pm 0.000 $ \\ 
 $ [6.,8.] $ & $ -0.005 \pm 0.004 \pm 0.060 $ & $ -0.005 \pm 0.002 \pm 0.000 $ & $ -0.003 \pm 0.002 \pm 0.000 $ \\ 
 $ [15.,19.] $ & $ -0.001 \pm 0.000 \pm 0.000 $ & $ -0.001 \pm 0.000 \pm 0.000 $ & $ -0.000 \pm 0.000 \pm 0.000 $ \\ 
\midrule[1.6pt] 
 \hspace{5mm} Bin & $T_7$\hspace{14mm} & $T_8$\hspace{14mm} & $T_9$\hspace{14mm}  \\ 
 \midrule 
 $ [0.1,0.98] $ & $ -0.067 \pm 0.003 \pm 0.000 $ & $ -0.081 \pm 0.025 \pm 0.051 $ & $--$\hspace{14mm} \\ 
 $ [1.1,2.5] $ & $ -0.013 \pm 0.003 \pm 0.000 $ & $ -0.020 \pm 0.003 \pm 0.000 $ & $--$\hspace{14mm} \\ 
 $ [2.5,4.] $ & $ -0.007 \pm 0.003 \pm 0.000 $ & $ -0.010 \pm 0.003 \pm 0.000 $ & $ -0.010 \pm 0.027 \pm 0.000 $ \\ 
 $ [4.,6.] $ & $ -0.005 \pm 0.003 \pm 0.000 $ & $ -0.007 \pm 0.003 \pm 0.000 $ & $ -0.007 \pm 0.003 \pm 0.000 $ \\ 
 $ [6.,8.] $ & $ -0.003 \pm 0.002 \pm 0.000 $ & $ -0.005 \pm 0.002 \pm 0.000 $ & $ -0.005 \pm 0.004 \pm 0.000 $ \\ 
 $ [15.,19.] $ & $ -0.000 \pm 0.000 \pm 0.000 $ & $ -0.001 \pm 0.001 \pm 0.004 $ & $ -0.001 \pm 0.002 \pm 0.001 $ \\ 
\bottomrule[1.6pt] 
\end{longtable}

\begin{longtable}{@{}lrrr@{}}
\toprule[1.6pt] 
 \hspace{5mm} Bin & $B_5$\hspace{14mm} & $B_{6s}$\hspace{14mm} & $M$\hspace{14mm}  \\ 
 \midrule 
 $ [0.1,0.98] $ & $ -0.155 \pm 0.002 \pm 0.002 $ & $ -0.121 \pm 0.001 \pm 0.000 $ & $ 0.548 \pm 0.021 \pm 0.024 $ \\ 
 $ [1.1,2.5] $ & $ -0.034 \pm 0.005 \pm 0.002 $ & $ -0.027 \pm 0.000 \pm 0.000 $ & $ 0.150 \pm 0.071 \pm 0.037 $ \\ 
 $ [2.5,4.] $ & $ -0.013 \pm 0.000 \pm 0.000 $ & $ -0.015 \pm 0.001 \pm 0.000 $ & $ -0.095 \pm 0.033 \pm 0.007 $ \\ 
 $ [4.,6.] $ & $ -0.009 \pm 0.000 \pm 0.000 $ & $ -0.008 \pm 0.021 \pm 0.000 $ & $ 0.149 \pm 0.122 \pm 0.019 $ \\ 
 $ [6.,8.] $ & $ -0.006 \pm 0.000 \pm 0.000 $ & $ -0.006 \pm 0.000 \pm 0.000 $ & $ 0.617 \pm 0.253 \pm 0.204 $ \\ 
 $ [15.,19.] $ & $ -0.003 \pm 0.000 \pm 0.000 $ & $ -0.003 \pm 0.000 \pm 0.000 $ & $--$\hspace{14mm} \\ 
\bottomrule[1.6pt] 
\end{longtable}

\begin{longtable}{@{}lrrr@{}}
\toprule[1.6pt] 
 \hspace{5mm} Bin & $\widetilde B_5$\hspace{14mm} & $\widetilde B_{6s}$\hspace{14mm} & $\widetilde M$\hspace{14mm} \\ 
 \midrule 
 $ [0.1,0.98] $ & $ 0.000 \pm 0.000 \pm 0.000 $ & $ 0.000 \pm 0.000 \pm 0.000 $ & $ 0.000 \pm 0.000 \pm 0.000 $ \\ 
 $ [1.1,2.5] $ & $ 0.000 \pm 0.000 \pm 0.000 $ & $ 0.000 \pm 0.000 \pm 0.000 $ & $ 0.000 \pm 0.000 \pm 0.000 $ \\ 
 $ [2.5,4.] $ & $ 0.000 \pm 0.000 \pm 0.000 $ & $ 0.000 \pm 0.000 \pm 0.000 $ & $ 0.000 \pm 0.000 \pm 0.000 $ \\ 
 $ [4.,6.] $ & $ 0.000 \pm 0.000 \pm 0.000 $ & $ 0.000 \pm 0.000 \pm 0.000 $ & $ 0.000 \pm 0.000 \pm 0.000 $ \\ 
 $ [6.,8.] $ & $ 0.000 \pm 0.000 \pm 0.000 $ & $ 0.000 \pm 0.000 \pm 0.000 $ & $ 0.000 \pm 0.000 \pm 0.000 $ \\ 
 $ [15.,19.] $ & $ 0.000 \pm 0.000 \pm 0.000 $ & $ 0.000 \pm 0.000 \pm 0.000 $ & $ 0.000 \pm 0.000 \pm 0.000 $ \\ 
\bottomrule[1.6pt] 
\end{longtable}

\end{center}

\subsection{Scenario 1: $C_{9\mu}^{\rm NP} = -1.11$}

\begin{center}

{
\scriptsize
\begin{longtable}{@{}lrrrr@{}}
\toprule[1.6pt] 
 \hspace{5mm} Bin & $Q_{F_L}$\hspace{10mm} & $Q_1$\hspace{10mm} & $Q_2$\hspace{10mm} & $Q_3$\hspace{10mm} \\ 
 \midrule 
 $ [0.1,0.98] $ & $ -0.085 \pm 0.073 \pm 0.021 $ & $ -0.001 \pm 0.002 \pm 0.003 $ & $ 0.017 \pm 0.002 \pm 0.001 $ & $ 0.000 \pm 0.000 \pm 0.000 $ \\ 
 $ [1.1,2.5] $ & $ -0.122 \pm 0.032 \pm 0.001 $ & $ 0.001 \pm 0.008 \pm 0.003 $ & $ -0.008 \pm 0.010 \pm 0.001 $ & $ -0.000 \pm 0.001 \pm 0.000 $ \\ 
 $ [2.5,4.] $ & $ -0.086 \pm 0.037 \pm 0.002 $ & $ -0.013 \pm 0.026 \pm 0.007 $ & $ 0.174 \pm 0.058 \pm 0.006 $ & $ -0.001 \pm 0.002 \pm 0.000 $ \\ 
 $ [4.,6.] $ & $ -0.051 \pm 0.016 \pm 0.002 $ & $ -0.022 \pm 0.038 \pm 0.010 $ & $ 0.246 \pm 0.009 \pm 0.002 $ & $ -0.000 \pm 0.001 \pm 0.000 $ \\ 
 $ [6.,8.] $ & $ -0.027 \pm 0.008 \pm 0.003 $ & $ -0.017 \pm 0.028 \pm 0.009 $ & $ 0.184 \pm 0.036 \pm 0.009 $ & $ 0.000 \pm 0.000 \pm 0.000 $ \\ 
 $ [15.,19.] $ & $ -0.002 \pm 0.000 \pm 0.003 $ & $ 0.002 \pm 0.001 \pm 0.004 $ & $ 0.051 \pm 0.004 \pm 0.010 $ & $ 0.000 \pm 0.000 \pm 0.003 $ \\ 
\midrule[1.6pt] 
 \hspace{5mm} Bin & $Q_4$\hspace{10mm} & $Q_5$\hspace{10mm} & $Q_6$\hspace{10mm} & $Q_8$\hspace{10mm} \\ 
 \midrule 
 $ [0.1,0.98] $ & $ 0.136 \pm 0.011 \pm 0.049 $ & $ 0.172 \pm 0.004 \pm 0.016 $ & $ -0.011 \pm 0.004 \pm 0.001 $ & $ -0.012 \pm 0.004 \pm 0.003 $ \\ 
 $ [1.1,2.5] $ & $ 0.087 \pm 0.033 \pm 0.019 $ & $ 0.241 \pm 0.021 \pm 0.013 $ & $ -0.002 \pm 0.001 \pm 0.000 $ & $ -0.018 \pm 0.007 \pm 0.001 $ \\ 
 $ [2.5,4.] $ & $ -0.037 \pm 0.035 \pm 0.010 $ & $ 0.370 \pm 0.017 \pm 0.014 $ & $ -0.003 \pm 0.001 \pm 0.000 $ & $ -0.014 \pm 0.007 \pm 0.001 $ \\ 
 $ [4.,6.] $ & $ -0.041 \pm 0.008 \pm 0.008 $ & $ 0.312 \pm 0.044 \pm 0.017 $ & $ -0.006 \pm 0.002 \pm 0.000 $ & $ -0.006 \pm 0.004 \pm 0.000 $ \\ 
 $ [6.,8.] $ & $ -0.020 \pm 0.005 \pm 0.010 $ & $ 0.212 \pm 0.056 \pm 0.029 $ & $ -0.004 \pm 0.003 \pm 0.000 $ & $ -0.002 \pm 0.002 \pm 0.001 $ \\ 
 $ [15.,19.] $ & $ -0.001 \pm 0.000 \pm 0.002 $ & $ 0.073 \pm 0.007 \pm 0.013 $ & $ -0.001 \pm 0.000 \pm 0.020 $ & $ -0.001 \pm 0.000 \pm 0.004 $ \\ 
\bottomrule[1.6pt] 
\end{longtable}

\begin{longtable}{@{}lrrrr@{}}
\toprule[1.6pt] 
 \hspace{5mm} Bin & $\hat Q_{F_L}$\hspace{10mm} & $\hat Q_1$\hspace{10mm} & $\hat Q_2$\hspace{10mm} & $\hat Q_3$\hspace{10mm} \\ 
 \midrule 
 $ [0.1,0.98] $ & $ -0.037 \pm 0.022 \pm 0.011 $ & $ -0.007 \pm 0.007 \pm 0.019 $ & $ -0.009 \pm 0.003 \pm 0.000 $ & $ 0.000 \pm 0.001 \pm 0.001 $ \\ 
 $ [1.1,2.5] $ & $ -0.086 \pm 0.049 \pm 0.001 $ & $ 0.001 \pm 0.008 \pm 0.003 $ & $ -0.010 \pm 0.019 \pm 0.002 $ & $ -0.000 \pm 0.001 \pm 0.000 $ \\ 
 $ [2.5,4.] $ & $ -0.060 \pm 0.046 \pm 0.002 $ & $ -0.014 \pm 0.026 \pm 0.007 $ & $ 0.183 \pm 0.048 \pm 0.006 $ & $ -0.001 \pm 0.002 \pm 0.000 $ \\ 
 $ [4.,6.] $ & $ -0.033 \pm 0.021 \pm 0.002 $ & $ -0.021 \pm 0.036 \pm 0.011 $ & $ 0.247 \pm 0.011 \pm 0.002 $ & $ -0.000 \pm 0.001 \pm 0.000 $ \\ 
 $ [6.,8.] $ & $ -0.015 \pm 0.008 \pm 0.003 $ & $ -0.017 \pm 0.026 \pm 0.009 $ & $ 0.182 \pm 0.035 \pm 0.009 $ & $ 0.000 \pm 0.000 \pm 0.000 $ \\ 
 $ [15.,19.] $ & $ -0.001 \pm 0.000 \pm 0.002 $ & $ 0.002 \pm 0.001 \pm 0.004 $ & $ 0.051 \pm 0.004 \pm 0.010 $ & $ 0.000 \pm 0.000 \pm 0.003 $ \\ 
\midrule[1.6pt] 
 \hspace{5mm} Bin & $\hat Q_4$\hspace{10mm} & $\hat Q_5$\hspace{10mm} & $\hat Q_6$\hspace{10mm} & $\hat Q_8$\hspace{10mm} \\ 
 \midrule 
 $ [0.1,0.98] $ & $ 0.214 \pm 0.008 \pm 0.010 $ & $ -0.000 \pm 0.011 \pm 0.014 $ & $ 0.003 \pm 0.001 \pm 0.001 $ & $ -0.014 \pm 0.007 \pm 0.001 $ \\ 
 $ [1.1,2.5] $ & $ 0.086 \pm 0.035 \pm 0.016 $ & $ 0.227 \pm 0.021 \pm 0.010 $ & $ 0.000 \pm 0.002 \pm 0.001 $ & $ -0.019 \pm 0.007 \pm 0.001 $ \\ 
 $ [2.5,4.] $ & $ -0.040 \pm 0.042 \pm 0.009 $ & $ 0.370 \pm 0.017 \pm 0.013 $ & $ -0.003 \pm 0.002 \pm 0.000 $ & $ -0.014 \pm 0.006 \pm 0.001 $ \\ 
 $ [4.,6.] $ & $ -0.045 \pm 0.016 \pm 0.008 $ & $ 0.314 \pm 0.043 \pm 0.017 $ & $ -0.005 \pm 0.003 \pm 0.000 $ & $ -0.006 \pm 0.004 \pm 0.000 $ \\ 
 $ [6.,8.] $ & $ -0.025 \pm 0.007 \pm 0.009 $ & $ 0.216 \pm 0.054 \pm 0.029 $ & $ -0.004 \pm 0.003 \pm 0.000 $ & $ -0.002 \pm 0.002 \pm 0.001 $ \\ 
 $ [15.,19.] $ & $ -0.003 \pm 0.000 \pm 0.002 $ & $ 0.074 \pm 0.007 \pm 0.013 $ & $ -0.001 \pm 0.000 \pm 0.020 $ & $ -0.001 \pm 0.000 \pm 0.004 $ \\ 
\bottomrule[1.6pt] 
\end{longtable}
}

\newpage

\begin{longtable}{@{}lrrr@{}}
\toprule[1.6pt] 
 \hspace{5mm} Bin & $T_3$\hspace{14mm} & $T_4$\hspace{14mm} & $T_5$\hspace{14mm}  \\ 
 \midrule 
 $ [0.1,0.98] $ & $--$\hspace{14mm} & $--$\hspace{14mm} & $ -0.026 \pm 0.038 \pm 0.011 $ \\ 
 $ [1.1,2.5] $ & $--$\hspace{14mm} & $--$\hspace{14mm} & $ 0.402 \pm 0.152 \pm 0.076 $ \\ 
 $ [2.5,4.] $ & $--$\hspace{14mm} & $ 0.005 \pm 0.072 \pm 0.008 $ & $ -0.608 \pm 0.295 \pm 0.121 $ \\ 
 $ [4.,6.] $ & $--$\hspace{14mm} & $ -0.010 \pm 0.031 \pm 0.004 $ & $ -0.224 \pm 0.061 \pm 0.026 $ \\ 
 $ [6.,8.] $ & $--$\hspace{14mm} & $ -0.009 \pm 0.014 \pm 0.004 $ & $ -0.126 \pm 0.042 \pm 0.025 $ \\ 
 $ [15.,19.] $ & $ -0.001 \pm 0.001 \pm 0.004 $ & $ -0.002 \pm 0.000 \pm 0.001 $ & $ -0.069 \pm 0.006 \pm 0.015 $ \\ 
\midrule[1.6pt] 
 \hspace{5mm} Bin & $T_7$\hspace{14mm} & $T_8$\hspace{14mm} & $T_9$\hspace{14mm}  \\ 
 \midrule 
 $ [0.1,0.98] $ & $ -0.056 \pm 0.038 \pm 0.011 $ & $--$\hspace{14mm} & $--$\hspace{14mm} \\ 
 $ [1.1,2.5] $ & $ 0.029 \pm 0.071 \pm 0.010 $ & $ -0.244 \pm 0.137 \pm 0.073 $ & $--$\hspace{14mm} \\ 
 $ [2.5,4.] $ & $ 0.065 \pm 0.050 \pm 0.005 $ & $ -0.143 \pm 0.075 \pm 0.023 $ & $--$\hspace{14mm} \\ 
 $ [4.,6.] $ & $ 0.087 \pm 0.028 \pm 0.003 $ & $ -0.091 \pm 0.050 \pm 0.016 $ & $--$\hspace{14mm} \\ 
 $ [6.,8.] $ & $ 0.102 \pm 0.015 \pm 0.004 $ & $ -0.067 \pm 0.083 \pm 0.025 $ & $--$\hspace{14mm} \\ 
 $ [15.,19.] $ & $ 0.118 \pm 0.001 \pm 0.003 $ & $--$\hspace{14mm} & $--$\hspace{14mm} \\ 
\bottomrule[1.6pt] 
\end{longtable}

\begin{longtable}{@{}lrrr@{}}
\toprule[1.6pt] 
 \hspace{5mm} Bin & $B_5$\hspace{14mm} & $B_{6s}$\hspace{14mm} & $M$\hspace{14mm}  \\ 
 \midrule 
 $ [0.1,0.98] $ & $ -0.087 \pm 0.008 \pm 0.004 $ & $ -0.084 \pm 0.005 \pm 0.001 $ & $--$\hspace{14mm} \\ 
 $ [1.1,2.5] $ & $--$\hspace{14mm} & $ 0.172 \pm 0.047 \pm 0.006 $ & $ -0.203 \pm 0.049 \pm 0.012 $ \\ 
 $ [2.5,4.] $ & $ -0.785 \pm 0.181 \pm 0.078 $ & $--$\hspace{14mm} & $ -0.459 \pm 0.106 \pm 0.026 $ \\ 
 $ [4.,6.] $ & $ -0.472 \pm 0.051 \pm 0.026 $ & $--$\hspace{14mm} & $ -0.736 \pm 0.188 \pm 0.062 $ \\ 
 $ [6.,8.] $ & $ -0.372 \pm 0.040 \pm 0.027 $ & $ -0.569 \pm 0.150 \pm 0.032 $ & $ -1.101 \pm 0.328 \pm 0.242 $ \\ 
 $ [15.,19.] $ & $ -0.316 \pm 0.007 \pm 0.018 $ & $ -0.324 \pm 0.008 \pm 0.019 $ & $--$\hspace{14mm} \\ 
\bottomrule[1.6pt] 
\end{longtable}

\begin{longtable}{@{}lrrr@{}}
\toprule[1.6pt] 
 \hspace{5mm} Bin & $\widetilde B_5$\hspace{14mm} & $\widetilde B_{6s}$\hspace{14mm} & $\widetilde M$\hspace{14mm} \\ 
 \midrule 
 $ [0.1,0.98] $ & $ 0.075 \pm 0.010 \pm 0.006 $ & $ 0.040 \pm 0.006 \pm 0.001 $ & $ -0.083 \pm 0.017 \pm 0.006 $ \\ 
 $ [1.1,2.5] $ & $--$\hspace{14mm} & $ 0.204 \pm 0.048 \pm 0.006 $ & $ -0.247 \pm 0.049 \pm 0.015 $ \\ 
 $ [2.5,4.] $ & $ -0.783 \pm 0.184 \pm 0.079 $ & $--$\hspace{14mm} & $ -0.463 \pm 0.102 \pm 0.027 $ \\ 
 $ [4.,6.] $ & $ -0.467 \pm 0.051 \pm 0.026 $ & $--$\hspace{14mm} & $ -0.723 \pm 0.182 \pm 0.061 $ \\ 
 $ [6.,8.] $ & $ -0.368 \pm 0.040 \pm 0.027 $ & $ -0.566 \pm 0.151 \pm 0.032 $ & $ -1.077 \pm 0.319 \pm 0.238 $ \\ 
 $ [15.,19.] $ & $ -0.314 \pm 0.007 \pm 0.018 $ & $ -0.322 \pm 0.008 \pm 0.019 $ & $--$\hspace{14mm} \\ 
\bottomrule[1.6pt] 
\end{longtable}

\end{center}

\newpage

\subsection{Scenario 2: $C_{9\mu}^{\rm NP} = -\C{10\mu}^{\rm NP} = -0.65$}

\begin{center}

{
\scriptsize
\begin{longtable}{@{}lrrrr@{}}
\toprule[1.6pt] 
 \hspace{5mm} Bin & $Q_{F_L}$\hspace{10mm} & $Q_1$\hspace{10mm} & $Q_2$\hspace{10mm} & $Q_3$\hspace{10mm} \\ 
 \midrule 
 $ [0.1,0.98] $ & $ -0.096 \pm 0.081 \pm 0.013 $ & $ -0.001 \pm 0.001 \pm 0.002 $ & $ 0.001 \pm 0.000 \pm 0.000 $ & $ 0.000 \pm 0.000 \pm 0.000 $ \\ 
 $ [1.1,2.5] $ & $ -0.107 \pm 0.027 \pm 0.007 $ & $ -0.002 \pm 0.008 \pm 0.002 $ & $ -0.032 \pm 0.015 \pm 0.002 $ & $ -0.000 \pm 0.001 \pm 0.000 $ \\ 
 $ [2.5,4.] $ & $ -0.043 \pm 0.014 \pm 0.003 $ & $ -0.017 \pm 0.039 \pm 0.008 $ & $ 0.148 \pm 0.037 \pm 0.003 $ & $ 0.000 \pm 0.001 \pm 0.000 $ \\ 
 $ [4.,6.] $ & $ -0.009 \pm 0.012 \pm 0.002 $ & $ -0.011 \pm 0.027 \pm 0.005 $ & $ 0.134 \pm 0.029 \pm 0.006 $ & $ 0.001 \pm 0.001 \pm 0.000 $ \\ 
 $ [6.,8.] $ & $ 0.003 \pm 0.011 \pm 0.003 $ & $ -0.001 \pm 0.008 \pm 0.001 $ & $ 0.059 \pm 0.029 \pm 0.007 $ & $ 0.001 \pm 0.001 \pm 0.000 $ \\ 
 $ [15.,19.] $ & $ 0.001 \pm 0.000 \pm 0.003 $ & $ -0.002 \pm 0.001 \pm 0.005 $ & $ 0.005 \pm 0.001 \pm 0.003 $ & $ 0.000 \pm 0.000 \pm 0.003 $ \\ 
\midrule[1.6pt] 
 \hspace{5mm} Bin & $Q_4$\hspace{10mm} & $Q_5$\hspace{10mm} & $Q_6$\hspace{10mm} & $Q_8$\hspace{10mm} \\ 
 \midrule 
 $ [0.1,0.98] $ & $ -0.003 \pm 0.007 \pm 0.027 $ & $ 0.078 \pm 0.007 \pm 0.029 $ & $ -0.005 \pm 0.002 \pm 0.002 $ & $ -0.005 \pm 0.001 \pm 0.003 $ \\ 
 $ [1.1,2.5] $ & $ -0.102 \pm 0.028 \pm 0.014 $ & $ 0.136 \pm 0.017 \pm 0.012 $ & $ -0.000 \pm 0.001 \pm 0.001 $ & $ -0.005 \pm 0.002 \pm 0.001 $ \\ 
 $ [2.5,4.] $ & $ -0.152 \pm 0.013 \pm 0.010 $ & $ 0.188 \pm 0.021 \pm 0.010 $ & $ -0.007 \pm 0.002 \pm 0.001 $ & $ 0.002 \pm 0.003 \pm 0.001 $ \\ 
 $ [4.,6.] $ & $ -0.078 \pm 0.031 \pm 0.009 $ & $ 0.096 \pm 0.032 \pm 0.010 $ & $ -0.008 \pm 0.004 \pm 0.000 $ & $ 0.005 \pm 0.004 \pm 0.000 $ \\ 
 $ [6.,8.] $ & $ -0.031 \pm 0.021 \pm 0.009 $ & $ 0.033 \pm 0.021 \pm 0.011 $ & $ -0.004 \pm 0.003 \pm 0.000 $ & $ 0.004 \pm 0.003 \pm 0.001 $ \\ 
 $ [15.,19.] $ & $ 0.000 \pm 0.000 \pm 0.002 $ & $ 0.007 \pm 0.001 \pm 0.006 $ & $ -0.001 \pm 0.000 \pm 0.015 $ & $ -0.001 \pm 0.001 \pm 0.005 $ \\ 
\bottomrule[1.6pt] 
\end{longtable}

\begin{longtable}{@{}lrrrr@{}}
\toprule[1.6pt] 
 \hspace{5mm} Bin & $\hat Q_{F_L}$\hspace{10mm} & $\hat Q_1$\hspace{10mm} & $\hat Q_2$\hspace{10mm} & $\hat Q_3$\hspace{10mm} \\ 
 \midrule 
 $ [0.1,0.98] $ & $ -0.051 \pm 0.031 \pm 0.003 $ & $ -0.007 \pm 0.006 \pm 0.019 $ & $ -0.022 \pm 0.004 \pm 0.001 $ & $ 0.000 \pm 0.001 \pm 0.001 $ \\ 
 $ [1.1,2.5] $ & $ -0.071 \pm 0.043 \pm 0.008 $ & $ -0.002 \pm 0.008 \pm 0.002 $ & $ -0.034 \pm 0.020 \pm 0.003 $ & $ -0.000 \pm 0.001 \pm 0.000 $ \\ 
 $ [2.5,4.] $ & $ -0.017 \pm 0.020 \pm 0.003 $ & $ -0.017 \pm 0.040 \pm 0.008 $ & $ 0.159 \pm 0.028 \pm 0.003 $ & $ 0.000 \pm 0.001 \pm 0.000 $ \\ 
 $ [4.,6.] $ & $ 0.009 \pm 0.007 \pm 0.002 $ & $ -0.011 \pm 0.024 \pm 0.005 $ & $ 0.133 \pm 0.032 \pm 0.006 $ & $ 0.001 \pm 0.002 \pm 0.000 $ \\ 
 $ [6.,8.] $ & $ 0.016 \pm 0.006 \pm 0.003 $ & $ -0.000 \pm 0.005 \pm 0.001 $ & $ 0.056 \pm 0.027 \pm 0.007 $ & $ 0.001 \pm 0.002 \pm 0.000 $ \\ 
 $ [15.,19.] $ & $ 0.002 \pm 0.001 \pm 0.004 $ & $ -0.001 \pm 0.001 \pm 0.005 $ & $ 0.006 \pm 0.001 \pm 0.003 $ & $ 0.000 \pm 0.000 \pm 0.003 $ \\ 
\midrule[1.6pt] 
 \hspace{5mm} Bin & $\hat Q_4$\hspace{10mm} & $\hat Q_5$\hspace{10mm} & $\hat Q_6$\hspace{10mm} & $\hat Q_8$\hspace{10mm} \\ 
 \midrule 
 $ [0.1,0.98] $ & $ 0.107 \pm 0.007 \pm 0.015 $ & $ -0.075 \pm 0.008 \pm 0.005 $ & $ 0.008 \pm 0.003 \pm 0.000 $ & $ -0.008 \pm 0.004 \pm 0.001 $ \\ 
 $ [1.1,2.5] $ & $ -0.097 \pm 0.030 \pm 0.012 $ & $ 0.126 \pm 0.017 \pm 0.009 $ & $ 0.002 \pm 0.002 \pm 0.000 $ & $ -0.006 \pm 0.002 \pm 0.001 $ \\ 
 $ [2.5,4.] $ & $ -0.154 \pm 0.009 \pm 0.010 $ & $ 0.189 \pm 0.022 \pm 0.010 $ & $ -0.007 \pm 0.003 \pm 0.001 $ & $ 0.002 \pm 0.003 \pm 0.001 $ \\ 
 $ [4.,6.] $ & $ -0.079 \pm 0.023 \pm 0.008 $ & $ 0.098 \pm 0.030 \pm 0.010 $ & $ -0.008 \pm 0.004 \pm 0.000 $ & $ 0.005 \pm 0.004 \pm 0.000 $ \\ 
 $ [6.,8.] $ & $ -0.035 \pm 0.015 \pm 0.008 $ & $ 0.037 \pm 0.021 \pm 0.011 $ & $ -0.004 \pm 0.003 \pm 0.000 $ & $ 0.004 \pm 0.003 \pm 0.001 $ \\ 
 $ [15.,19.] $ & $ -0.003 \pm 0.000 \pm 0.002 $ & $ 0.009 \pm 0.001 \pm 0.006 $ & $ -0.001 \pm 0.000 \pm 0.015 $ & $ -0.001 \pm 0.001 \pm 0.005 $ \\ 
\bottomrule[1.6pt] 
\end{longtable}
}

\newpage

\begin{longtable}{@{}lrrr@{}}
\toprule[1.6pt] 
 \hspace{5mm} Bin & $T_3$\hspace{14mm} & $T_4$\hspace{14mm} & $T_5$\hspace{14mm}  \\ 
 \midrule 
 $ [0.1,0.98] $ & $--$\hspace{14mm} & $ -0.158 \pm 0.050 \pm 0.043 $ & $ -0.101 \pm 0.046 \pm 0.005 $ \\ 
 $ [1.1,2.5] $ & $--$\hspace{14mm} & $--$\hspace{14mm} & $ 0.276 \pm 0.131 \pm 0.056 $ \\ 
 $ [2.5,4.] $ & $--$\hspace{14mm} & $ -0.156 \pm 0.118 \pm 0.023 $ & $ -0.234 \pm 0.100 \pm 0.039 $ \\ 
 $ [4.,6.] $ & $--$\hspace{14mm} & $ -0.057 \pm 0.033 \pm 0.008 $ & $ -0.070 \pm 0.022 \pm 0.008 $ \\ 
 $ [6.,8.] $ & $--$\hspace{14mm} & $ -0.026 \pm 0.023 \pm 0.007 $ & $ -0.026 \pm 0.015 \pm 0.005 $ \\ 
 $ [15.,19.] $ & $ -0.001 \pm 0.001 \pm 0.005 $ & $ -0.000 \pm 0.000 \pm 0.001 $ & $ -0.007 \pm 0.002 \pm 0.006 $ \\ 
\midrule[1.6pt] 
 \hspace{5mm} Bin & $T_7$\hspace{14mm} & $T_8$\hspace{14mm} & $T_9$\hspace{14mm}  \\ 
 \midrule 
 $ [0.1,0.98] $ & $ -0.116 \pm 0.047 \pm 0.005 $ & $--$\hspace{14mm} & $--$\hspace{14mm} \\ 
 $ [1.1,2.5] $ & $ 0.015 \pm 0.056 \pm 0.002 $ & $ -0.050 \pm 0.084 \pm 0.029 $ & $--$\hspace{14mm} \\ 
 $ [2.5,4.] $ & $ 0.069 \pm 0.014 \pm 0.003 $ & $ 0.037 \pm 0.029 \pm 0.006 $ & $--$\hspace{14mm} \\ 
 $ [4.,6.] $ & $ 0.089 \pm 0.008 \pm 0.003 $ & $ 0.073 \pm 0.022 \pm 0.003 $ & $--$\hspace{14mm} \\ 
 $ [6.,8.] $ & $ 0.095 \pm 0.012 \pm 0.006 $ & $ 0.138 \pm 0.042 \pm 0.005 $ & $--$\hspace{14mm} \\ 
 $ [15.,19.] $ & $ 0.094 \pm 0.002 \pm 0.004 $ & $--$\hspace{14mm} & $--$\hspace{14mm} \\ 
\bottomrule[1.6pt] 
\end{longtable}

\begin{longtable}{@{}lrrr@{}}
\toprule[1.6pt] 
 \hspace{5mm} Bin & $B_5$\hspace{14mm} & $B_{6s}$\hspace{14mm} & $M$\hspace{14mm}  \\ 
 \midrule 
 $ [0.1,0.98] $ & $ -0.248 \pm 0.003 \pm 0.002 $ & $ -0.235 \pm 0.002 \pm 0.001 $ & $--$\hspace{14mm} \\ 
 $ [1.1,2.5] $ & $--$\hspace{14mm} & $ -0.075 \pm 0.023 \pm 0.003 $ & $ 0.062 \pm 0.011 \pm 0.004 $ \\ 
 $ [2.5,4.] $ & $ -0.546 \pm 0.090 \pm 0.039 $ & $--$\hspace{14mm} & $ -0.231 \pm 0.126 \pm 0.015 $ \\ 
 $ [4.,6.] $ & $ -0.389 \pm 0.025 \pm 0.013 $ & $--$\hspace{14mm} & $ -0.750 \pm 0.280 \pm 0.061 $ \\ 
 $ [6.,8.] $ & $ -0.338 \pm 0.020 \pm 0.013 $ & $ -0.436 \pm 0.074 \pm 0.016 $ & $ -1.550 \pm 0.570 \pm 0.305 $ \\ 
 $ [15.,19.] $ & $ -0.309 \pm 0.003 \pm 0.009 $ & $ -0.313 \pm 0.004 \pm 0.009 $ & $--$\hspace{14mm} \\ 
\bottomrule[1.6pt] 
\end{longtable}

\begin{longtable}{@{}lrrr@{}}
\toprule[1.6pt] 
 \hspace{5mm} Bin & $\widetilde B_5$\hspace{14mm} & $\widetilde B_{6s}$\hspace{14mm} & $\widetilde M$\hspace{14mm} \\ 
 \midrule 
 $ [0.1,0.98] $ & $ -0.113 \pm 0.005 \pm 0.003 $ & $ -0.131 \pm 0.003 \pm 0.001 $ & $ -0.845 \pm 0.182 \pm 0.136 $ \\ 
 $ [1.1,2.5] $ & $--$\hspace{14mm} & $ -0.049 \pm 0.024 \pm 0.003 $ & $ 0.044 \pm 0.016 \pm 0.002 $ \\ 
 $ [2.5,4.] $ & $ -0.540 \pm 0.091 \pm 0.039 $ & $--$\hspace{14mm} & $ -0.236 \pm 0.120 \pm 0.014 $ \\ 
 $ [4.,6.] $ & $ -0.383 \pm 0.025 \pm 0.013 $ & $--$\hspace{14mm} & $ -0.731 \pm 0.269 \pm 0.059 $ \\ 
 $ [6.,8.] $ & $ -0.334 \pm 0.020 \pm 0.013 $ & $ -0.432 \pm 0.075 \pm 0.016 $ & $ -1.508 \pm 0.551 \pm 0.297 $ \\ 
 $ [15.,19.] $ & $ -0.307 \pm 0.003 \pm 0.009 $ & $ -0.311 \pm 0.004 \pm 0.009 $ & $--$\hspace{14mm} \\ 
\bottomrule[1.6pt] 
\end{longtable}

\end{center}

\newpage

\subsection{Scenario 3: $C_{9\mu}^{\rm NP} = -C_{9'\mu}^{\rm NP} = -1.07$}

\begin{center}
{
\scriptsize
\begin{longtable}{@{}lrrrr@{}}
\toprule[1.6pt] 
 \hspace{5mm} Bin & $Q_{F_L}$\hspace{10mm} & $Q_1$\hspace{10mm} & $Q_2$\hspace{10mm} & $Q_3$\hspace{10mm} \\ 
 \midrule 
 $ [0.1,0.98] $ & $ -0.109 \pm 0.094 \pm 0.034 $ & $ -0.055 \pm 0.009 \pm 0.003 $ & $ 0.017 \pm 0.002 \pm 0.001 $ & $ 0.002 \pm 0.001 \pm 0.000 $ \\ 
 $ [1.1,2.5] $ & $ -0.164 \pm 0.044 \pm 0.007 $ & $ -0.204 \pm 0.024 \pm 0.005 $ & $ -0.014 \pm 0.007 \pm 0.002 $ & $ 0.009 \pm 0.004 \pm 0.001 $ \\ 
 $ [2.5,4.] $ & $ -0.133 \pm 0.060 \pm 0.003 $ & $ -0.186 \pm 0.050 \pm 0.005 $ & $ 0.148 \pm 0.062 \pm 0.006 $ & $ 0.013 \pm 0.006 \pm 0.000 $ \\ 
 $ [4.,6.] $ & $ -0.106 \pm 0.037 \pm 0.004 $ & $ -0.045 \pm 0.083 \pm 0.012 $ & $ 0.232 \pm 0.011 \pm 0.001 $ & $ 0.011 \pm 0.006 \pm 0.000 $ \\ 
 $ [6.,8.] $ & $ -0.089 \pm 0.021 \pm 0.007 $ & $ 0.074 \pm 0.072 \pm 0.015 $ & $ 0.190 \pm 0.032 \pm 0.008 $ & $ 0.007 \pm 0.005 \pm 0.000 $ \\ 
 $ [15.,19.] $ & $ -0.022 \pm 0.003 \pm 0.009 $ & $ 0.136 \pm 0.013 \pm 0.007 $ & $ 0.016 \pm 0.007 \pm 0.016 $ & $ -0.017 \pm 0.007 \pm 0.007 $ \\ 
\midrule[1.6pt] 
 \hspace{5mm} Bin & $Q_4$\hspace{10mm} & $Q_5$\hspace{10mm} & $Q_6$\hspace{10mm} & $Q_8$\hspace{10mm} \\ 
 \midrule 
 $ [0.1,0.98] $ & $ 0.295 \pm 0.023 \pm 0.107 $ & $ 0.246 \pm 0.003 \pm 0.017 $ & $ -0.017 \pm 0.007 \pm 0.002 $ & $ -0.025 \pm 0.009 \pm 0.006 $ \\ 
 $ [1.1,2.5] $ & $ 0.233 \pm 0.050 \pm 0.045 $ & $ 0.271 \pm 0.016 \pm 0.013 $ & $ -0.008 \pm 0.004 \pm 0.001 $ & $ -0.030 \pm 0.012 \pm 0.003 $ \\ 
 $ [2.5,4.] $ & $ 0.031 \pm 0.068 \pm 0.021 $ & $ 0.347 \pm 0.021 \pm 0.017 $ & $ -0.007 \pm 0.003 \pm 0.001 $ & $ -0.025 \pm 0.013 \pm 0.001 $ \\ 
 $ [4.,6.] $ & $ -0.052 \pm 0.035 \pm 0.014 $ & $ 0.267 \pm 0.054 \pm 0.021 $ & $ -0.008 \pm 0.003 \pm 0.000 $ & $ -0.015 \pm 0.010 \pm 0.001 $ \\ 
 $ [6.,8.] $ & $ -0.082 \pm 0.022 \pm 0.017 $ & $ 0.153 \pm 0.071 \pm 0.038 $ & $ -0.006 \pm 0.003 \pm 0.000 $ & $ -0.008 \pm 0.008 \pm 0.001 $ \\ 
 $ [15.,19.] $ & $ -0.055 \pm 0.006 \pm 0.003 $ & $ -0.009 \pm 0.008 \pm 0.021 $ & $ -0.002 \pm 0.001 \pm 0.034 $ & $ 0.027 \pm 0.011 \pm 0.011 $ \\ 
\bottomrule[1.6pt] 
\end{longtable}

\begin{longtable}{@{}lrrrr@{}}
\toprule[1.6pt] 
 \hspace{5mm} Bin & $\hat Q_{F_L}$\hspace{10mm} & $\hat Q_1$\hspace{10mm} & $\hat Q_2$\hspace{10mm} & $\hat Q_3$\hspace{10mm} \\ 
 \midrule 
 $ [0.1,0.98] $ & $ -0.067 \pm 0.046 \pm 0.027 $ & $ -0.048 \pm 0.011 \pm 0.019 $ & $ -0.010 \pm 0.003 \pm 0.000 $ & $ 0.002 \pm 0.001 \pm 0.001 $ \\ 
 $ [1.1,2.5] $ & $ -0.130 \pm 0.062 \pm 0.006 $ & $ -0.202 \pm 0.021 \pm 0.005 $ & $ -0.018 \pm 0.015 \pm 0.002 $ & $ 0.009 \pm 0.004 \pm 0.001 $ \\ 
 $ [2.5,4.] $ & $ -0.108 \pm 0.070 \pm 0.003 $ & $ -0.189 \pm 0.055 \pm 0.005 $ & $ 0.154 \pm 0.053 \pm 0.006 $ & $ 0.013 \pm 0.006 \pm 0.000 $ \\ 
 $ [4.,6.] $ & $ -0.089 \pm 0.045 \pm 0.004 $ & $ -0.045 \pm 0.083 \pm 0.012 $ & $ 0.233 \pm 0.008 \pm 0.001 $ & $ 0.011 \pm 0.006 \pm 0.000 $ \\ 
 $ [6.,8.] $ & $ -0.076 \pm 0.025 \pm 0.007 $ & $ 0.075 \pm 0.071 \pm 0.015 $ & $ 0.189 \pm 0.031 \pm 0.008 $ & $ 0.007 \pm 0.006 \pm 0.000 $ \\ 
 $ [15.,19.] $ & $ -0.022 \pm 0.003 \pm 0.008 $ & $ 0.136 \pm 0.013 \pm 0.007 $ & $ 0.016 \pm 0.007 \pm 0.016 $ & $ -0.017 \pm 0.007 \pm 0.007 $ \\ 
\midrule[1.6pt] 
 \hspace{5mm} Bin & $\hat Q_4$\hspace{10mm} & $\hat Q_5$\hspace{10mm} & $\hat Q_6$\hspace{10mm} & $\hat Q_8$\hspace{10mm} \\ 
 \midrule 
 $ [0.1,0.98] $ & $ 0.340 \pm 0.015 \pm 0.052 $ & $ 0.056 \pm 0.012 \pm 0.031 $ & $ -0.002 \pm 0.002 \pm 0.003 $ & $ -0.024 \pm 0.011 \pm 0.002 $ \\ 
 $ [1.1,2.5] $ & $ 0.227 \pm 0.053 \pm 0.041 $ & $ 0.255 \pm 0.015 \pm 0.010 $ & $ -0.006 \pm 0.004 \pm 0.001 $ & $ -0.031 \pm 0.013 \pm 0.003 $ \\ 
 $ [2.5,4.] $ & $ 0.025 \pm 0.074 \pm 0.020 $ & $ 0.348 \pm 0.021 \pm 0.017 $ & $ -0.007 \pm 0.003 \pm 0.001 $ & $ -0.025 \pm 0.013 \pm 0.001 $ \\ 
 $ [4.,6.] $ & $ -0.058 \pm 0.040 \pm 0.014 $ & $ 0.271 \pm 0.052 \pm 0.021 $ & $ -0.008 \pm 0.003 \pm 0.000 $ & $ -0.015 \pm 0.010 \pm 0.001 $ \\ 
 $ [6.,8.] $ & $ -0.089 \pm 0.024 \pm 0.016 $ & $ 0.159 \pm 0.069 \pm 0.038 $ & $ -0.006 \pm 0.003 \pm 0.000 $ & $ -0.009 \pm 0.008 \pm 0.001 $ \\ 
 $ [15.,19.] $ & $ -0.057 \pm 0.006 \pm 0.003 $ & $ -0.008 \pm 0.008 \pm 0.021 $ & $ -0.002 \pm 0.001 \pm 0.033 $ & $ 0.027 \pm 0.011 \pm 0.011 $ \\ 
\bottomrule[1.6pt] 
\end{longtable}
}

\newpage

\begin{longtable}{@{}lrrr@{}}
\toprule[1.6pt] 
 \hspace{5mm} Bin & $T_3$\hspace{14mm} & $T_4$\hspace{14mm} & $T_5$\hspace{14mm}  \\ 
 \midrule 
 $ [0.1,0.98] $ & $--$\hspace{14mm} & $--$\hspace{14mm} & $ -0.007 \pm 0.061 \pm 0.021 $ \\ 
 $ [1.1,2.5] $ & $--$\hspace{14mm} & $--$\hspace{14mm} & $ 0.436 \pm 0.158 \pm 0.080 $ \\ 
 $ [2.5,4.] $ & $--$\hspace{14mm} & $ 0.091 \pm 0.149 \pm 0.012 $ & $ -0.528 \pm 0.296 \pm 0.122 $ \\ 
 $ [4.,6.] $ & $--$\hspace{14mm} & $ 0.004 \pm 0.087 \pm 0.006 $ & $ -0.161 \pm 0.090 \pm 0.029 $ \\ 
 $ [6.,8.] $ & $--$\hspace{14mm} & $ -0.031 \pm 0.066 \pm 0.006 $ & $ -0.074 \pm 0.075 \pm 0.032 $ \\ 
 $ [15.,19.] $ & $ -0.103 \pm 0.021 \pm 0.011 $ & $ -0.031 \pm 0.004 \pm 0.003 $ & $ -0.002 \pm 0.008 \pm 0.017 $ \\ 
\midrule[1.6pt] 
 \hspace{5mm} Bin & $T_7$\hspace{14mm} & $T_8$\hspace{14mm} & $T_9$\hspace{14mm}  \\ 
 \midrule 
 $ [0.1,0.98] $ & $ -0.036 \pm 0.062 \pm 0.021 $ & $--$\hspace{14mm} & $--$\hspace{14mm} \\ 
 $ [1.1,2.5] $ & $ 0.081 \pm 0.105 \pm 0.021 $ & $ -0.514 \pm 0.246 \pm 0.198 $ & $--$\hspace{14mm} \\ 
 $ [2.5,4.] $ & $ 0.121 \pm 0.086 \pm 0.011 $ & $ -0.322 \pm 0.117 \pm 0.059 $ & $ 0.830 \pm 0.290 \pm 0.082 $ \\ 
 $ [4.,6.] $ & $ 0.136 \pm 0.069 \pm 0.008 $ & $ -0.283 \pm 0.112 \pm 0.045 $ & $ 0.791 \pm 0.276 \pm 0.080 $ \\ 
 $ [6.,8.] $ & $ 0.144 \pm 0.058 \pm 0.012 $ & $ -0.304 \pm 0.312 \pm 0.100 $ & $--$\hspace{14mm} \\ 
 $ [15.,19.] $ & $ 0.177 \pm 0.005 \pm 0.009 $ & $--$\hspace{14mm} & $--$\hspace{14mm} \\ 
\bottomrule[1.6pt] 
\end{longtable}

\begin{longtable}{@{}lrrr@{}}
\toprule[1.6pt] 
 \hspace{5mm} Bin & $B_5$\hspace{14mm} & $B_{6s}$\hspace{14mm} & $M$\hspace{14mm}  \\ 
 \midrule 
 $ [0.1,0.98] $ & $ -0.089 \pm 0.008 \pm 0.004 $ & $ -0.086 \pm 0.005 \pm 0.001 $ & $--$\hspace{14mm} \\ 
 $ [1.1,2.5] $ & $--$\hspace{14mm} & $ 0.165 \pm 0.045 \pm 0.006 $ & $ -0.194 \pm 0.047 \pm 0.011 $ \\ 
 $ [2.5,4.] $ & $ -0.758 \pm 0.175 \pm 0.075 $ & $--$\hspace{14mm} & $ -0.443 \pm 0.102 \pm 0.026 $ \\ 
 $ [4.,6.] $ & $ -0.455 \pm 0.049 \pm 0.025 $ & $--$\hspace{14mm} & $ -0.710 \pm 0.182 \pm 0.060 $ \\ 
 $ [6.,8.] $ & $ -0.359 \pm 0.039 \pm 0.026 $ & $ -0.549 \pm 0.145 \pm 0.031 $ & $ -1.063 \pm 0.317 \pm 0.234 $ \\ 
 $ [15.,19.] $ & $ -0.305 \pm 0.007 \pm 0.017 $ & $ -0.312 \pm 0.007 \pm 0.018 $ & $--$\hspace{14mm} \\ 
\bottomrule[1.6pt] 
\end{longtable}

\begin{longtable}{@{}lrrr@{}}
\toprule[1.6pt] 
 \hspace{5mm} Bin & $\widetilde B_5$\hspace{14mm} & $\widetilde B_{6s}$\hspace{14mm} & $\widetilde M$\hspace{14mm} \\ 
 \midrule 
 $ [0.1,0.98] $ & $ 0.072 \pm 0.010 \pm 0.005 $ & $ 0.038 \pm 0.006 \pm 0.001 $ & $ -0.080 \pm 0.016 \pm 0.006 $ \\ 
 $ [1.1,2.5] $ & $--$\hspace{14mm} & $ 0.197 \pm 0.046 \pm 0.006 $ & $ -0.238 \pm 0.047 \pm 0.015 $ \\ 
 $ [2.5,4.] $ & $ -0.755 \pm 0.177 \pm 0.076 $ & $--$\hspace{14mm} & $ -0.447 \pm 0.098 \pm 0.026 $ \\ 
 $ [4.,6.] $ & $ -0.450 \pm 0.050 \pm 0.025 $ & $--$\hspace{14mm} & $ -0.697 \pm 0.176 \pm 0.059 $ \\ 
 $ [6.,8.] $ & $ -0.355 \pm 0.039 \pm 0.026 $ & $ -0.546 \pm 0.146 \pm 0.031 $ & $ -1.038 \pm 0.308 \pm 0.229 $ \\ 
 $ [15.,19.] $ & $ -0.303 \pm 0.007 \pm 0.018 $ & $ -0.310 \pm 0.007 \pm 0.018 $ & $--$\hspace{14mm} \\ 
\bottomrule[1.6pt] 
\end{longtable}

\end{center}

\newpage

\subsection{Scenario 4: $C_{9\mu}^{\rm NP}= -C_{9'\mu}^{\rm NP} = -1.18\ ,\ C_{10\mu}^{\rm NP}= C_{10'\mu}^{\rm NP} = 0.38$}

\begin{center}
{
\scriptsize
\begin{longtable}{@{}lrrrr@{}}
\toprule[1.6pt] 
 \hspace{5mm} Bin & $Q_{F_L}$\hspace{10mm} & $Q_1$\hspace{10mm} & $Q_2$\hspace{10mm} & $Q_3$\hspace{10mm} \\ 
 \midrule 
 $ [0.1,0.98] $ & $ -0.113 \pm 0.097 \pm 0.037 $ & $ -0.063 \pm 0.010 \pm 0.004 $ & $ 0.006 \pm 0.001 \pm 0.001 $ & $ 0.002 \pm 0.001 \pm 0.000 $ \\ 
 $ [1.1,2.5] $ & $ -0.167 \pm 0.044 \pm 0.009 $ & $ -0.280 \pm 0.037 \pm 0.006 $ & $ -0.044 \pm 0.009 \pm 0.003 $ & $ 0.010 \pm 0.004 \pm 0.001 $ \\ 
 $ [2.5,4.] $ & $ -0.120 \pm 0.052 \pm 0.004 $ & $ -0.371 \pm 0.045 \pm 0.005 $ & $ 0.146 \pm 0.071 \pm 0.007 $ & $ 0.016 \pm 0.007 \pm 0.000 $ \\ 
 $ [4.,6.] $ & $ -0.084 \pm 0.027 \pm 0.005 $ & $ -0.236 \pm 0.092 \pm 0.013 $ & $ 0.230 \pm 0.014 \pm 0.004 $ & $ 0.014 \pm 0.008 \pm 0.000 $ \\ 
 $ [6.,8.] $ & $ -0.064 \pm 0.014 \pm 0.009 $ & $ -0.078 \pm 0.087 \pm 0.018 $ & $ 0.175 \pm 0.033 \pm 0.008 $ & $ 0.009 \pm 0.007 \pm 0.000 $ \\ 
 $ [15.,19.] $ & $ -0.013 \pm 0.002 \pm 0.010 $ & $ 0.068 \pm 0.008 \pm 0.011 $ & $ 0.024 \pm 0.006 \pm 0.015 $ & $ -0.020 \pm 0.009 \pm 0.008 $ \\ 
\midrule[1.6pt] 
 \hspace{5mm} Bin & $Q_4$\hspace{10mm} & $Q_5$\hspace{10mm} & $Q_6$\hspace{10mm} & $Q_8$\hspace{10mm} \\ 
 \midrule 
 $ [0.1,0.98] $ & $ 0.336 \pm 0.025 \pm 0.118 $ & $ 0.271 \pm 0.005 \pm 0.026 $ & $ -0.018 \pm 0.007 \pm 0.003 $ & $ -0.028 \pm 0.010 \pm 0.007 $ \\ 
 $ [1.1,2.5] $ & $ 0.276 \pm 0.052 \pm 0.052 $ & $ 0.337 \pm 0.022 \pm 0.006 $ & $ -0.011 \pm 0.005 \pm 0.002 $ & $ -0.034 \pm 0.014 \pm 0.003 $ \\ 
 $ [2.5,4.] $ & $ 0.089 \pm 0.066 \pm 0.025 $ & $ 0.430 \pm 0.021 \pm 0.013 $ & $ -0.012 \pm 0.004 \pm 0.001 $ & $ -0.026 \pm 0.014 \pm 0.002 $ \\ 
 $ [4.,6.] $ & $ 0.018 \pm 0.035 \pm 0.017 $ & $ 0.324 \pm 0.059 \pm 0.019 $ & $ -0.012 \pm 0.005 \pm 0.000 $ & $ -0.016 \pm 0.011 \pm 0.001 $ \\ 
 $ [6.,8.] $ & $ -0.016 \pm 0.028 \pm 0.021 $ & $ 0.187 \pm 0.074 \pm 0.035 $ & $ -0.008 \pm 0.005 \pm 0.000 $ & $ -0.009 \pm 0.009 \pm 0.001 $ \\ 
 $ [15.,19.] $ & $ -0.027 \pm 0.004 \pm 0.004 $ & $ 0.017 \pm 0.008 \pm 0.020 $ & $ -0.002 \pm 0.001 \pm 0.039 $ & $ 0.031 \pm 0.013 \pm 0.013 $ \\ 
\bottomrule[1.6pt] 
\end{longtable}

\begin{longtable}{@{}lrrrr@{}}
\toprule[1.6pt] 
 \hspace{5mm} Bin & $\hat Q_{F_L}$\hspace{10mm} & $\hat Q_1$\hspace{10mm} & $\hat Q_2$\hspace{10mm} & $\hat Q_3$\hspace{10mm} \\ 
 \midrule 
 $ [0.1,0.98] $ & $ -0.072 \pm 0.051 \pm 0.031 $ & $ -0.055 \pm 0.012 \pm 0.020 $ & $ -0.018 \pm 0.003 \pm 0.001 $ & $ 0.002 \pm 0.001 \pm 0.001 $ \\ 
 $ [1.1,2.5] $ & $ -0.133 \pm 0.062 \pm 0.009 $ & $ -0.277 \pm 0.034 \pm 0.006 $ & $ -0.048 \pm 0.014 \pm 0.003 $ & $ 0.010 \pm 0.004 \pm 0.001 $ \\ 
 $ [2.5,4.] $ & $ -0.094 \pm 0.062 \pm 0.004 $ & $ -0.378 \pm 0.054 \pm 0.005 $ & $ 0.153 \pm 0.062 \pm 0.007 $ & $ 0.016 \pm 0.008 \pm 0.000 $ \\ 
 $ [4.,6.] $ & $ -0.065 \pm 0.034 \pm 0.005 $ & $ -0.239 \pm 0.097 \pm 0.013 $ & $ 0.231 \pm 0.010 \pm 0.004 $ & $ 0.014 \pm 0.008 \pm 0.000 $ \\ 
 $ [6.,8.] $ & $ -0.051 \pm 0.017 \pm 0.009 $ & $ -0.079 \pm 0.087 \pm 0.018 $ & $ 0.173 \pm 0.032 \pm 0.008 $ & $ 0.009 \pm 0.007 \pm 0.000 $ \\ 
 $ [15.,19.] $ & $ -0.013 \pm 0.002 \pm 0.010 $ & $ 0.068 \pm 0.009 \pm 0.011 $ & $ 0.024 \pm 0.006 \pm 0.015 $ & $ -0.020 \pm 0.009 \pm 0.008 $ \\ 
\midrule[1.6pt] 
 \hspace{5mm} Bin & $\hat Q_4$\hspace{10mm} & $\hat Q_5$\hspace{10mm} & $\hat Q_6$\hspace{10mm} & $\hat Q_8$\hspace{10mm} \\ 
 \midrule 
 $ [0.1,0.98] $ & $ 0.372 \pm 0.016 \pm 0.060 $ & $ 0.076 \pm 0.014 \pm 0.041 $ & $ -0.002 \pm 0.002 \pm 0.003 $ & $ -0.027 \pm 0.012 \pm 0.003 $ \\ 
 $ [1.1,2.5] $ & $ 0.269 \pm 0.055 \pm 0.047 $ & $ 0.319 \pm 0.023 \pm 0.006 $ & $ -0.008 \pm 0.005 \pm 0.002 $ & $ -0.034 \pm 0.014 \pm 0.003 $ \\ 
 $ [2.5,4.] $ & $ 0.083 \pm 0.072 \pm 0.024 $ & $ 0.431 \pm 0.020 \pm 0.013 $ & $ -0.011 \pm 0.005 \pm 0.001 $ & $ -0.027 \pm 0.014 \pm 0.002 $ \\ 
 $ [4.,6.] $ & $ 0.012 \pm 0.040 \pm 0.017 $ & $ 0.327 \pm 0.056 \pm 0.019 $ & $ -0.012 \pm 0.005 \pm 0.000 $ & $ -0.016 \pm 0.011 \pm 0.001 $ \\ 
 $ [6.,8.] $ & $ -0.023 \pm 0.030 \pm 0.021 $ & $ 0.192 \pm 0.072 \pm 0.034 $ & $ -0.008 \pm 0.005 \pm 0.000 $ & $ -0.009 \pm 0.009 \pm 0.001 $ \\ 
 $ [15.,19.] $ & $ -0.029 \pm 0.004 \pm 0.004 $ & $ 0.018 \pm 0.008 \pm 0.020 $ & $ -0.002 \pm 0.001 \pm 0.039 $ & $ 0.031 \pm 0.013 \pm 0.013 $ \\ 
\bottomrule[1.6pt] 
\end{longtable}
}

\newpage

\begin{longtable}{@{}lrrr@{}}
\toprule[1.6pt] 
 \hspace{5mm} Bin & $T_3$\hspace{14mm} & $T_4$\hspace{14mm} & $T_5$\hspace{14mm}  \\ 
 \midrule 
 $ [0.1,0.98] $ & $--$\hspace{14mm} & $--$\hspace{14mm} & $ 0.002 \pm 0.065 \pm 0.027 $ \\ 
 $ [1.1,2.5] $ & $ 0.991 \pm 0.188 \pm 0.182 $ & $--$\hspace{14mm} & $ 0.488 \pm 0.162 \pm 0.090 $ \\ 
 $ [2.5,4.] $ & $ 1.010 \pm 0.231 \pm 0.028 $ & $ 0.133 \pm 0.149 \pm 0.012 $ & $ -0.809 \pm 0.524 \pm 0.177 $ \\ 
 $ [4.,6.] $ & $--$\hspace{14mm} & $ 0.040 \pm 0.074 \pm 0.007 $ & $ -0.222 \pm 0.085 \pm 0.032 $ \\ 
 $ [6.,8.] $ & $--$\hspace{14mm} & $ 0.002 \pm 0.050 \pm 0.008 $ & $ -0.101 \pm 0.063 \pm 0.030 $ \\ 
 $ [15.,19.] $ & $ -0.047 \pm 0.010 \pm 0.014 $ & $ -0.016 \pm 0.002 \pm 0.004 $ & $ -0.021 \pm 0.007 \pm 0.017 $ \\ 
\midrule[1.6pt] 
 \hspace{5mm} Bin & $T_7$\hspace{14mm} & $T_8$\hspace{14mm} & $T_9$\hspace{14mm}  \\ 
 \midrule 
 $ [0.1,0.98] $ & $ -0.034 \pm 0.066 \pm 0.023 $ & $--$\hspace{14mm} & $--$\hspace{14mm} \\ 
 $ [1.1,2.5] $ & $ 0.094 \pm 0.107 \pm 0.023 $ & $ -0.614 \pm 0.296 \pm 0.250 $ & $ 0.974 \pm 0.486 \pm 0.234 $ \\ 
 $ [2.5,4.] $ & $ 0.146 \pm 0.076 \pm 0.012 $ & $ -0.371 \pm 0.120 \pm 0.072 $ & $ 0.849 \pm 0.264 \pm 0.074 $ \\ 
 $ [4.,6.] $ & $ 0.170 \pm 0.053 \pm 0.009 $ & $ -0.319 \pm 0.112 \pm 0.055 $ & $ 0.817 \pm 0.252 \pm 0.071 $ \\ 
 $ [6.,8.] $ & $ 0.183 \pm 0.040 \pm 0.013 $ & $ -0.346 \pm 0.371 \pm 0.126 $ & $--$\hspace{14mm} \\ 
 $ [15.,19.] $ & $ 0.205 \pm 0.004 \pm 0.011 $ & $--$\hspace{14mm} & $--$\hspace{14mm} \\ 
\bottomrule[1.6pt] 
\end{longtable}

\begin{longtable}{@{}lrrr@{}}
\toprule[1.6pt] 
 \hspace{5mm} Bin & $B_5$\hspace{14mm} & $B_{6s}$\hspace{14mm} & $M$\hspace{14mm}  \\ 
 \midrule 
 $ [0.1,0.98] $ & $ -0.075 \pm 0.010 \pm 0.009 $ & $ -0.166 \pm 0.009 \pm 0.003 $ & $ -0.138 \pm 0.031 \pm 0.031 $ \\ 
 $ [1.1,2.5] $ & $--$\hspace{14mm} & $ 0.059 \pm 0.048 \pm 0.005 $ & $ -0.062 \pm 0.051 \pm 0.006 $ \\ 
 $ [2.5,4.] $ & $ -0.916 \pm 0.202 \pm 0.077 $ & $--$\hspace{14mm} & $ -0.446 \pm 0.163 \pm 0.022 $ \\ 
 $ [4.,6.] $ & $ -0.552 \pm 0.052 \pm 0.024 $ & $--$\hspace{14mm} & $ -1.009 \pm 0.337 \pm 0.079 $ \\ 
 $ [6.,8.] $ & $ -0.439 \pm 0.038 \pm 0.021 $ & $ -0.577 \pm 0.119 \pm 0.028 $ & $ -1.888 \pm 0.668 \pm 0.376 $ \\ 
 $ [15.,19.] $ & $ -0.369 \pm 0.007 \pm 0.016 $ & $ -0.374 \pm 0.007 \pm 0.017 $ & $--$\hspace{14mm} \\ 
\bottomrule[1.6pt] 
\end{longtable}

\begin{longtable}{@{}lrrr@{}}
\toprule[1.6pt] 
 \hspace{5mm} Bin & $\widetilde B_5$\hspace{14mm} & $\widetilde B_{6s}$\hspace{14mm} & $\widetilde M$\hspace{14mm} \\ 
 \midrule 
 $ [0.1,0.98] $ & $ 0.088 \pm 0.013 \pm 0.012 $ & $ -0.054 \pm 0.011 \pm 0.003 $ & $ 0.033 \pm 0.003 \pm 0.002 $ \\ 
 $ [1.1,2.5] $ & $--$\hspace{14mm} & $ 0.088 \pm 0.050 \pm 0.006 $ & $ -0.094 \pm 0.054 \pm 0.007 $ \\ 
 $ [2.5,4.] $ & $ -0.916 \pm 0.205 \pm 0.078 $ & $--$\hspace{14mm} & $ -0.453 \pm 0.159 \pm 0.023 $ \\ 
 $ [4.,6.] $ & $ -0.548 \pm 0.053 \pm 0.024 $ & $--$\hspace{14mm} & $ -0.994 \pm 0.328 \pm 0.078 $ \\ 
 $ [6.,8.] $ & $ -0.436 \pm 0.038 \pm 0.021 $ & $ -0.575 \pm 0.119 \pm 0.028 $ & $ -1.851 \pm 0.651 \pm 0.369 $ \\ 
 $ [15.,19.] $ & $ -0.367 \pm 0.007 \pm 0.016 $ & $ -0.372 \pm 0.007 \pm 0.017 $ & $--$\hspace{14mm} \\ 
\bottomrule[1.6pt] 
\end{longtable}

\end{center}

\newpage

\subsection{$R_{K^*}$}

\begin{center}
\begin{longtable}{@{}lrrrr@{}}
\toprule[1.6pt] 
 \hspace{5mm} $R_{K^*}$ &&&&  \\ 
 \midrule[1.1pt] 
 Bin & $ [0.1,2] $\hspace{10mm} & $ [2,4.3] $\hspace{10mm} & $ [4.3,8.68] $\hspace{10mm} & $ [16.,19.] $\hspace{10mm}  \\ 
 \midrule 
 SM & $ 0.988 \pm 0.007 \pm 0.001 $ & $ 1.000 \pm 0.006 \pm 0.000 $ & $ 1.000 \pm 0.005 \pm 0.000 $ & $ 0.998 \pm 0.000 \pm 0.000 $\\ 
 Scen.1 & $ 0.951 \pm 0.096 \pm 0.021 $ & $ 0.871 \pm 0.093 \pm 0.009 $ & $ 0.813 \pm 0.026 \pm 0.029 $ & $ 0.786 \pm 0.001 \pm 0.004 $\\ 
 Scen.2 & $ 0.889 \pm 0.102 \pm 0.008 $ & $ 0.737 \pm 0.028 \pm 0.005 $ & $ 0.701 \pm 0.016 \pm 0.045 $ & $ 0.701 \pm 0.003 \pm 0.006 $\\ 
 Scen.3 & $ 0.898 \pm 0.142 \pm 0.039 $ & $ 0.780 \pm 0.142 \pm 0.018 $ & $ 0.747 \pm 0.090 \pm 0.045 $ & $ 0.692 \pm 0.006 \pm 0.013 $\\
 Scen.4 & $ 0.890 \pm 0.149 \pm 0.043 $ & $ 0.742 \pm 0.123 \pm 0.019 $ & $ 0.690 \pm 0.059 \pm 0.052 $ & $ 0.655 \pm 0.005 \pm 0.015 $\\ 
\bottomrule[1.6pt] 
\end{longtable}
\end{center}


\normalsize


\begin{thebibliography}{99}


\bibitem{Descotes-Genon:2013wba} 
  S.~Descotes-Genon, J.~Matias and J.~Virto,
  ``Understanding the $B\to K^*\mu^+\mu^-$ Anomaly,''
  Phys.\ Rev.\ D {\bf 88}, 074002 (2013)
  [arXiv:1307.5683 [hep-ph]].


\bibitem{DescotesGenon:2012zf} 
  S.~Descotes-Genon, J.Matias, M.Ramon, J.Virto,
  ``Implications from clean observables for the binned analysis of $B \to K^*\mu^+\mu^-$ at large recoil,''
  JHEP {\bf 1301}, 048 (2013)
  [arXiv:1207.2753 [hep-ph]].


\bibitem{Aaij:2013qta} 
  {\bf LHCb} collaboration,
  ``Measurement of Form-Factor-Independent Observables in the Decay $B^{0} \to K^{*0} \mu^+ \mu^-$,''
  Phys.\ Rev.\ Lett.\  {\bf 111}, 191801 (2013)
  [arXiv:1308.1707 [hep-ex]].


\bibitem{Aaij:2015oid} 
  {\bf LHCb} collaboration,
  ``Angular analysis of the $B^{0} \to K^{*0} \mu^{+} \mu^{-}$ decay using 3 fb$^{-1}$ of integrated luminosity,''
  JHEP {\bf 1602}, 104 (2016)
  [arXiv:1512.04442 [hep-ex]].


\bibitem{Abdesselam:2016llu} 
  {\bf Belle} collaboration,
  ``Angular analysis of $B^0 \to K^\ast(892)^0 \ell^+ \ell^-$,''
  arXiv:1604.04042 [hep-ex].


\bibitem{Aaij:2013iag} 
  {\bf LHCb} collaboration,
  ``Differential branching fraction and angular analysis of the decay $B^{0} \to K^{*0} \mu^{+}\mu^{-}$,''
  JHEP {\bf 1308}, 131 (2013)
  [arXiv:1304.6325, arXiv:1304.6325 [hep-ex]].


\bibitem{Aaij:2015esa} 
  {\bf LHCb} collaboration,
  ``Angular analysis and differential branching fraction of the decay $B^0_s\to\phi\mu^+\mu^-$,''
  JHEP {\bf 1509}, 179 (2015)
  [arXiv:1506.08777 [hep-ex]].


\bibitem{Aaij:2014tfa} 
  {\bf LHCb} collaboration,
  ``Angular analysis of charged and neutral $B \to K \mu^+\mu^-$  decays,''
  JHEP {\bf 1405}, 082 (2014)
  [arXiv:1403.8045 [hep-ex]].


\bibitem{Aaij:2014ora} 
  {\bf LHCb} collaboration,
  ``Test of lepton universality using $B^{+}\rightarrow K^{+}\ell^{+}\ell^{-}$ decays,''
  Phys.\ Rev.\ Lett.\  {\bf 113}, 151601 (2014)
  [arXiv:1406.6482 [hep-ex]].


\bibitem{Lees:2013uzd} 
  {\bf BaBar} collaboration,
  ``Measurement of an Excess of $\bar{B} \to D^{(*)}\tau^- \bar{\nu}_\tau$ Decays and Implications for Charged Higgs Bosons,''
  Phys.\ Rev.\ D {\bf 88}, no. 7, 072012 (2013)
  [arXiv:1303.0571 [hep-ex]].


\bibitem{Huschle:2015rga} 
  {\bf Belle} collaboration,
  ``Measurement of the branching ratio of $\bar{B} \to D^{(\ast)} \tau^- \bar{\nu}_\tau$ relative to $\bar{B} \to D^{(\ast)} \ell^- \bar{\nu}_\ell$ decays with hadronic tagging at Belle,''
  Phys.\ Rev.\ D {\bf 92}, no. 7, 072014 (2015)
  [arXiv:1507.03233 [hep-ex]].


\bibitem{Aaij:2015yra} 
  {\bf LHCb} collaboration,
  Phys.\ Rev.\ Lett.\  {\bf 115}, no. 11, 111803 (2015)
  Addendum: [Phys.\ Rev.\ Lett.\  {\bf 115}, no. 15, 159901 (2015)]
  [arXiv:1506.08614 [hep-ex]].


\bibitem{Abdesselam:2016yvr} 
  A.~Abdesselam {\it et al.} {\bf Belle} collaboration,
  ``Measurement of the Branching Fraction and $CP$ Asymmetry in Radiative $D^0 \to V \gamma$ Decays,''
  arXiv:1603.03257 [hep-ex].


\bibitem{Becirevic:2011bp} 
  D.~Becirevic and E.~Schneider,
  ``On transverse asymmetries in $B \to K^* \ell^+\ell^-$,''
  Nucl.\ Phys.\ B {\bf 854}, 321 (2012)
  [arXiv:1106.3283 [hep-ph]].


\bibitem{Matias:2012xw} 
  J.~Matias, F.~Mescia, M.~Ramon and J.~Virto,
  ``Complete Anatomy of $\bar{B}_d \to \bar{K}^{* 0} (\to K \pi)\ell^+\ell^-$ and its angular distribution,''
  JHEP {\bf 1204}, 104 (2012)
  [arXiv:1202.4266 [hep-ph]].


\bibitem{Lees:2015ymt} 
  J.~P.~Lees {\it et al.} {\bf BaBar} collaboration,
  ``Measurement of angular asymmetries in the decays $B \to K^*\ell^+\ell^-$,''
  Phys.\ Rev.\ D {\bf 93}, no. 5, 052015 (2016)
  [arXiv:1508.07960 [hep-ex]].


\bibitem{Beneke:2001at} 
  M.~Beneke, T.~Feldmann and D.~Seidel,
  ``Systematic approach to exclusive $B \to V \ell^+ \ell^-$, $V \gamma$ decays,''
  Nucl.\ Phys.\ B {\bf 612}, 25 (2001)
  [hep-ph/0106067].


\bibitem{Altmannshofer:2013foa} 
  W.~Altmannshofer and D.~M.~Straub,
  ``New physics in $B \to K^*\mu\mu$?,''
  Eur.\ Phys.\ J.\ C {\bf 73}, 2646 (2013)
  [arXiv:1308.1501 [hep-ph]].


\bibitem{Beaujean:2013soa} 
  F.~Beaujean, C.~Bobeth and D.~van Dyk,
  ``Comprehensive Bayesian analysis of rare (semi)leptonic and radiative $B$ decays,''
  Eur.\ Phys.\ J.\ C {\bf 74}, 2897 (2014)
  Erratum: [Eur.\ Phys.\ J.\ C {\bf 74}, 3179 (2014)]
  [arXiv:1310.2478 [hep-ph]].


\bibitem{Altmannshofer:2014rta} 
  W.~Altmannshofer and D.~M.~Straub,
  ``New physics in $b\rightarrow s$ transitions after LHC run 1,''
  Eur.\ Phys.\ J.\ C {\bf 75}, no. 8, 382 (2015)
  [arXiv:1411.3161 [hep-ph]].


\bibitem{Descotes-Genon:2015uva} 
  S.~Descotes-Genon, L.~Hofer, J.~Matias and J.~Virto,
  ``Global analysis of $b\to s\ell\ell$ anomalies,''
  arXiv:1510.04239 [hep-ph].


\bibitem{Altmannshofer:2015sma} 
  W.~Altmannshofer and D.~M.~Straub,
  ``Implications of $b\to s$ measurements,''
  arXiv:1503.06199 [hep-ph].


\bibitem{Hurth:2016fbr} 
  T.~Hurth, F.~Mahmoudi and S.~Neshatpour,
  ``On the anomalies in the latest LHCb data,''
  arXiv:1603.00865 [hep-ph].


\bibitem{Aaij:2015dea} 
  {\bf LHCb} collaboration,
  ``Angular analysis of the $B^{0} \to K^{*0} e^{+} e^{−}$ decay in the low-$q^{2}$ region,''
  JHEP {\bf 1504}, 064 (2015)
  [arXiv:1501.03038 [hep-ex]].


\bibitem{1408.1627} 
  G.~Hiller and M.~Schmaltz,
  ``$R_K$ and future $b \to s \ell \ell$ physics beyond the standard model opportunities,''
  Phys.\ Rev.\ D {\bf 90}, 054014 (2014)
  [arXiv:1408.1627 [hep-ph]].


\bibitem{1408.4097} 
  D.~Ghosh, M.~Nardecchia and S.~A.~Renner,
  ``Hint of Lepton Flavour Non-Universality in $B$ Meson Decays,''
  JHEP {\bf 1412}, 131 (2014)
  [arXiv:1408.4097 [hep-ph]].


\bibitem{1411.4773} 
  G.~Hiller and M.~Schmaltz,
  ``Diagnosing lepton-nonuniversality in $b \to s \ell \ell$,''
  JHEP {\bf 1502}, 055 (2015)
  [arXiv:1411.4773 [hep-ph]].


\bibitem{1412.7164} 
  B.~Bhattacharya, A.~Datta, D.~London and S.~Shivashankara,
  ``Simultaneous Explanation of the $R_K$ and $R(D^{(*)})$ Puzzles,''
  Phys.\ Lett.\ B {\bf 742}, 370 (2015)
  [arXiv:1412.7164 [hep-ph]].


\bibitem{1503.09024} 
  D.~Becirevic, S.~Fajfer and N.~Košnik,
  ``Lepton flavor nonuniversality in $b\to s\ell\ell$ processes,''
  Phys.\ Rev.\ D {\bf 92}, no. 1, 014016 (2015)
  [arXiv:1503.09024 [hep-ph]].


\bibitem{1503.06077} 
  D.~Aristizabal Sierra, F.~Staub and A.~Vicente,
  ``Shedding light on the $b\to s$ anomalies with a dark sector,''
  Phys.\ Rev.\ D {\bf 92}, no. 1, 015001 (2015)
  [arXiv:1503.06077 [hep-ph]].
  
\bibitem{1505.03079} 
  A.~Celis, J.~Fuentes-Martin, M.~Jung and H.~Serodio,
  ``Family nonuniversal $Z'$ models with protected flavor-changing interactions,''
  Phys.\ Rev.\ D {\bf 92}, no. 1, 015007 (2015)
  [arXiv:1505.03079 [hep-ph]].
  
\bibitem{1505.05164} 
  R.~Alonso, B.~Grinstein and J.~Martin Camalich,
  ``Lepton universality violation and lepton flavor conservation in $B$-meson decays,''
  JHEP {\bf 1510}, 184 (2015)
  [arXiv:1505.05164 [hep-ph]].


\bibitem{1506.01705} 
  A.~Greljo, G.~Isidori and D.~Marzocca,
  ``On the breaking of Lepton Flavor Universality in B decays,''
  JHEP {\bf 1507}, 142 (2015)
  [arXiv:1506.01705 [hep-ph]].


\bibitem{1506.02661} 
  L.~Calibbi, A.~Crivellin and T.~Ota,
  ``Effective Field Theory Approach to $b \to s \ell \ell^{(\prime)}$, $B\to K^{(*)}\nu \overline{\nu}$ and
  $B\to D^{(*)}\tau\nu$ with Third Generation Couplings,''
  Phys.\ Rev.\ Lett.\  {\bf 115}, 181801 (2015)
  [arXiv:1506.02661 [hep-ph]].


\bibitem{1508.07009} 
  W.~Altmannshofer and I.~Yavin,
  ``Predictions for lepton flavor universality violation in rare B decays in models with gauged $L_\mu - L_\tau$,''
  Phys.\ Rev.\ D {\bf 92}, no. 7, 075022 (2015)
  [arXiv:1508.07009 [hep-ph]].


\bibitem{1511.01900} 
  M.~Bauer and M.~Neubert,
  ``Minimal Leptoquark Explanation for the R$_{D^{(*)}}$ , R$_K$ , and $(g-2)_g$ Anomalies,''
  Phys.\ Rev.\ Lett.\  {\bf 116}, no. 14, 141802 (2016)
  [arXiv:1511.01900 [hep-ph]].


\bibitem{1511.06024} 
  S.~Fajfer and N.~Košnik,
  ``Vector leptoquark resolution of $R_K$ and $R_{D^{(*)}}$ puzzles,''
  Phys.\ Lett.\ B {\bf 755}, 270 (2016)
  [arXiv:1511.06024 [hep-ph]].


\bibitem{Boucenna:2016wpr} 
  S.~M.~Boucenna, A.~Celis, J.~Fuentes-Martin, A.~Vicente and J.~Virto,
  ``Non-abelian gauge extensions for B-decay anomalies,''
  arXiv:1604.03088 [hep-ph].


\bibitem{1604.03940} 
  D.~Buttazzo, A.~Greljo, G.~Isidori and D.~Marzocca,
  ``Toward a coherent solution of diphoton and flavor anomalies,''
  arXiv:1604.03940 [hep-ph].


\bibitem{Khodjamirian:2010vf} 
  A.~Khodjamirian, T.~Mannel, A.~A.~Pivovarov and Y.-M.~Wang,
  ``Charm-loop effect in $B \to K^{(*)} \ell^{+} \ell^{-}$ and $B\to K^*\gamma$,''
  JHEP {\bf 1009}, 089 (2010)
  [arXiv:1006.4945 [hep-ph]].


\bibitem{Beylich:2011aq} 
  M.~Beylich, G.~Buchalla and T.~Feldmann,
  ``Theory of $B \to K^{(*)}\ell^+ \ell^-$ decays at high $q^2$: OPE and quark-hadron duality,''
  Eur.\ Phys.\ J.\ C {\bf 71}, 1635 (2011)
  [arXiv:1101.5118 [hep-ph]].
  
\bibitem{0404250} 
  B.~Grinstein and D.~Pirjol,
  ``Exclusive rare $B \to K^*\ell^+\ell^-$ decays at low recoil: Controlling the long-distance effects,''
  Phys.\ Rev.\ D {\bf 70}, 114005 (2004)
  [hep-ph/0404250].


\bibitem{Khodjamirian:2012rm} 
  A.~Khodjamirian, T.~Mannel and Y.~M.~Wang,
  ``$B \to K \ell^{+}\ell^{-}$ decay at large hadronic recoil,''
  JHEP {\bf 1302}, 010 (2013)
  [arXiv:1211.0234 [hep-ph]].


\bibitem{Lyon:2014hpa} 
  J.~Lyon and R.~Zwicky,
  ``Resonances gone topsy turvy - the charm of QCD or new physics in $b \to s \ell^+ \ell^-$?,''
  arXiv:1406.0566 [hep-ph].


\bibitem{Ciuchini:2015qxb} 
  M.~Ciuchini, M.~Fedele, E.~Franco, S.~Mishima, A.~Paul, L.~Silvestrini and M.~Valli,
  ``$B\to K^* \ell^+ \ell^-$ decays at large recoil in the Standard Model: a theoretical reappraisal,''
  arXiv:1512.07157 [hep-ph].


\bibitem{Descotes-Genon:2013vna} 
  S.~Descotes-Genon, T.~Hurth, J.~Matias and J.~Virto,
  ``Optimizing the basis of $B\to K^*\ell\ell$ observables in the full kinematic range,''
  JHEP {\bf 1305}, 137 (2013)
  [arXiv:1303.5794 [hep-ph]].


\bibitem{Descotes-Genon:2014uoa} 
  S.~Descotes-Genon, L.~Hofer, J.~Matias and J.~Virto,
  ``On the impact of power corrections in the prediction of $B \to K^*\mu^+\mu^-$ observables,''
  JHEP {\bf 1412}, 125 (2014)
  [arXiv:1407.8526 [hep-ph]].


\bibitem{Altmannshofer:2008dz} 
  W.~Altmannshofer, P.~Ball, A.~Bharucha, A.~J.~Buras, D.~M.~Straub and M.~Wick,
  ``Symmetries and Asymmetries of $B \to K^{*} \mu^{+} \mu^{-}$ Decays in the Standard Model and Beyond,''
  JHEP {\bf 0901}, 019 (2009)
  [arXiv:0811.1214 [hep-ph]].


\bibitem{Jager:2012uw} 
  S.~J\"ager and J.~Martin Camalich,
  ``On $B \to V \ell \ell$ at small dilepton invariant mass, power corrections, and new physics,''
  JHEP {\bf 1305}, 043 (2013)
  [arXiv:1212.2263 [hep-ph]].


\bibitem{Jager:2014rwa} 
  S.~J\"ager and J.~Martin Camalich,
  ``Reassessing the discovery potential of the $B \to K^{*} \ell^+\ell^-$ decays in the large-recoil region: SM challenges and BSM opportunities,''
  Phys.\ Rev.\ D {\bf 93}, no. 1, 014028 (2016)
  [arXiv:1412.3183 [hep-ph]].


\bibitem{Becirevic:2012dp} 
  D.~Becirevic and A.~Tayduganov,
  ``Impact of $B\to K^\ast_0 \ell^+\ell^-$ on the New Physics search in $B\to K^\ast \ell^+\ell^-$ decay,''
  Nucl.\ Phys.\ B {\bf 868}, 368 (2013)
  [arXiv:1207.4004 [hep-ph]].


\bibitem{Matias:2012qz} 
  J.~Matias,
  ``On the S-wave pollution of $B\to K^* \ell^+\ell^-$ observables,''
  Phys.\ Rev.\ D {\bf 86}, 094024 (2012)
  [arXiv:1209.1525 [hep-ph]].


\bibitem{Blake:2012mb} 
  T.~Blake, U.~Egede and A.~Shires,
  ``The effect of S-wave interference on the $B^0 \to K^{\ast 0}\ell^+\ell^-$ angular observables,''
  JHEP {\bf 1303}, 027 (2013)
  [arXiv:1210.5279 [hep-ph]].


\bibitem{Kruger:2005ep} 
  F.~Kruger and J.~Matias,
  ``Probing new physics via the transverse amplitudes of $B \to K^* (\to K^- \pi^+) \ell^+\ell^-$ at large recoil,''
  Phys.\ Rev.\ D {\bf 71}, 094009 (2005)
  [hep-ph/0502060].


\bibitem{nico} Nicola Serra, private communication.

\bibitem{Egede:2015kha} 
  U.~Egede, M.~Patel and K.~A.~Petridis,
  ``Method for an unbinned measurement of the $q^{2}$ dependent decay amplitudes of $ B^0\to {K}^{\ast 0}{\mu}^{+}{\mu}^{-} $ decays,''
  JHEP {\bf 1506}, 084 (2015)
  [arXiv:1504.00574 [hep-ph]].


\bibitem{inprep} B.Capdevila,   S.~Descotes-Genon, L.~Hofer and J.~Matias in preparation.


\bibitem{Barberio:1993qi} 
  E.~Barberio and Z.~Was,
  ``PHOTOS: A Universal Monte Carlo for QED radiative corrections. Version 2.0,''
  Comput.\ Phys.\ Commun.\  {\bf 79}, 291 (1994).

\bibitem{Stoffer:2013sfa} 
  P.~Stoffer,
  ``Isospin breaking effects in $K_{\ell 4}$ decays,''
  Eur.\ Phys.\ J.\ C {\bf 74}, 2749 (2014)
  [arXiv:1312.2066 [hep-ph]].


\bibitem{Bernard:2015vqa} 
  V.~Bernard, S.~Descotes-Genon and M.~Knecht,
  ``On some aspects of isospin breaking in the decay $K^\pm$ $\to$ $\pi^0 \pi^0 e^\pm$ $\stackrel{_{(-)}}{\nu_e}$,''
  Eur.\ Phys.\ J.\ C {\bf 75}, no. 4, 145 (2015)
  [arXiv:1501.07102 [hep-ph]].
  
\end{thebibliography}
\end{document}